\documentclass[10pt,aps,prb,longbibliography,nobibnotes,twocolumn,groupedaddress]{revtex4-1}
\usepackage[utf8]{inputenc}
\usepackage{placeins}
\usepackage{tabularx}
\usepackage{amsmath}
\usepackage{amsfonts}
\usepackage{braket}
\usepackage[colorlinks,citecolor=blue,urlcolor=blue,bookmarks=false,hypertexnames=true]{hyperref}
\usepackage{amssymb}
\usepackage{graphicx}
\usepackage{float}
\usepackage{threeparttable}
\usepackage{subcaption}
\usepackage[version=4]{mhchem}
\usepackage[dvipsnames]{xcolor}
\usepackage[shortlabels]{enumitem}
\usepackage{dcolumn} 
\newcolumntype{d}[1]{D{.}{.}{#1}}
\newcommand*{\citen}[1]{%
  \begingroup
    \romannumeral-`\x 
    \setcitestyle{numbers}%
    \cite{#1}%
  \endgroup
}

\begin{document}
\def\ie{{\it i.e.\/}}
\def\eg{{\it e.g.\/}}
\def\etal{{\it et al.\/}}
\def\cm{{cm$^{-1}$}}
\def\kJmol{kJ$\,$mol}
\def\Eh{$E_\mathrm{h}$}
\def\mEh{m$E_\mathrm{h}$}
\def\uEh{$\mu E_\mathrm{h}$}
\def\a0{$a_0$}
\def\bea{\begin{eqnarray}}
\def\eea{\end{eqnarray}}

\title{DLPNO-MP2 with Periodic Boundary Conditions}

\author{Arman Nejad}
\affiliation{University of Oxford, South Parks Road, Oxford, OX1 3QZ, UK}
\author{Andrew Zhu}
\affiliation{University of Oxford, South Parks Road, Oxford, OX1 3QZ, UK}
\author{Kesha Sorathia}
\affiliation{University of Oxford, South Parks Road, Oxford, OX1 3QZ, UK}
\author{David P. Tew}
\email[Email: ]{david.tew@chem.ox.ac.uk}
\affiliation{University of Oxford, South Parks Road, Oxford, OX1 3QZ, UK}

\date{\today}

\begin{abstract}%

We present domain-based local pair natural orbital M{\o}ller--Plesset second order perturbation theory (DLPNO-MP2) with Born--von K{\'a}rm{\'a}n boundary (BvK) conditions. The approach is based on well-localised Wannier functions in a LCAO formalism and extends the molecular DLPNO-MP2 implementation Tubromole program package to periodic systems. The PNOs are formed through a PAO-OSV-PNO cascade, using BvK projected atomic orbitals and orbital specific virtuals as intermediaries in an analogous manner to the molecular scheme. Our chargeless and surface-dipole corrected local density fitting approach is shown to be numerically stable and to ensure convergent lattice summations over the periodic images for the two- and three-index Coulomb integrals. Through careful benchmarking, we show that the DLPNO approximations in the BvK-DLPNO-MP2 methods are entirely consistent with those of molecular DLPNO-MP2 calculations, and with an alternative periodic approach Megacell-DLPNO-MP2, reported in Paper II of this series. Smooth convergence to the canonical correlation energy with tightening PNO threshold is observed. Reference MP2 correlation energies are provided for a set of 2D and 3D periodic systems using a triple-zeta basis and supercell sizes up to 11$\times$11 and 7$\times$7$\times$7.

\end{abstract}
\maketitle

\section{Introduction}
\label{sec:Intro}
Computational studies of the electronic structure of materials are currently predominantly performed using density functional theory (DFT),\cite{Kratzer_FC_2019_106, Martin_2020} due to its excellent cost-to-accuracy ratio. 
The need to surmount the uncertainties inherent in DFT predictions has motivated several research groups to explore the systematically improvable set of post-Hartree--Fock (HF) correlated wavefunction-based methods. In particular, periodic implementations of M{\o}ller--Plesset perturbation theory\cite{Pisani_JCP_2005, Pisani_PCCP_2012_7615, DelBen_JCTC_2012, Schaefer_JCP_2017_104101, Bintrim_JCTC_5374_2022} and coupled cluster methods\cite{Gruber_PRX_2018_021043, Carbone_FD_2024_586, Ye_FD_2024_628} are now available and have been applied to study the structure and stability of materials and surface processes.\cite{Pisani_PCCP_2012_7615, Maschio_PRB_2007_075101, Usvyat_PRB_2007_075102, McClain_JCTC_2017_1209, Gruber_PRX_2018_021043, Hirata_PRB_2009_085118, Hirata_JCP_2004_2581, Haritan_arxiv_2025.20482, Ayala_JCP_2001_9698, Ye_JCTC_2024_8948, Katouda_JCP_2010_184103, Booth_Nature_2013_7432, Zhang_FM_2019_123, Goldzak_JCP_2022_174112, Schaefer_JCP_2017_104101, Grueneis_JCTC_2011_2780, Grueneis_JCP_2015_102817, Shi_JACS_2023_25372, Ye_FD_2024_628, Tsatsoulis_JCP_2017_204108, Alessio_JCTC_2019_1329}

The primary challenge in using accurate wavefunction approaches for materials is the computational expense. Canonical second-order M{\o}ller--Plesset perturbation theory (MP2) scales as $\mathcal{O}(N^5)$ with system size, whilst canonical coupled cluster with singles, doubles and perturbative triples (CCSD(T)) scales as $\mathcal{O}(N^7)$. The large simulations required to converge the correlation treatment to the thermodynamic limit quickly become prohibitively expensive.
Local correlation approximations provide a solution to this problem for insulating materials. 
For example, the pioneering work by Pisani and Sch{\"u}tz and coworkers in the Cryscor program\cite{Pisani_JCC_2008_2113, Pisani_PCCP_2012_7615, Usvyat_JCP_2013_194101, Usvyat_JCP_2015_102805, Maschio_PRB_2007_075101, Usvyat_PRB_2007_075102} leverages projected atomic orbitals\cite{Saebo_ARPC_1993_213} (PAOs) and orbital specific virutals\cite{Yang_JCP_2011_044123} (OSVs) to obtain a low-scaling MP2 implementation for non-conducting systems using atom-centred Gaussian basis functions.
More recently,  Ye and Berkelbach\cite{Ye_JCTC_2024_8948} have adapted K{\'a}llay's local natural orbital (LNOs) approach\cite{Rolik_JCP_2011_104111, Rolik_JCP_2013_094105, Nagy_JCP_2017_214106, Nagy_JCTC_2018_4193} to periodic systems, at the CCSD and CCSD(T) levels of theory, also in a linear combination of atomic orbitals (LCAO) framework.  These methods have been applied successfully to obtain properties such as lattice constants, cohesive energies\cite{Ye_JCTC_2024_8948, Mueller_JCTC_2013_5590, Usvyat_PRB_2007_075102} and adsorption energies\cite{Alessio_JCTC_2019_1329, Ye_FD_2024_628, Usvyat_PRB_2012_045412, Mullan_JCP_2022_074109} for surface interactions. 

This paper concerns the extension of domain-based pair natural orbital local correlation (DLPNO) theory\cite{Neese_JCP_2009_064103, Riplinger_JCP_2013_034106} to periodic systems. DLPNO theory achieves near-linear scaling of computational effort with system size, with only modest loss in accuracy, by replacing Hamiltonian integrals and excitation amplitudes with low-rank approximations that exploit the inherent locality of electron correlation in insulators. DLPNO theory within an LCAO framework has been highly successful in the molecular setting where it has been extended to perturbative\cite{Werner_JCTC_2015_484, Schmitz_MP_2013_2463, Pinski_JCP_2015_034108, Tew_IJQC_2013_224, Haettig_JCP_2012_204105} and coupled cluster formalisms\cite{Riplinger_JCP_2013_134101, Guo_JCP_2018_011101, Schwilk_JCTC_2017_3650, Ma_JCTC_2018_198, Schmitz_JCTC_2017_2623, Schmitz_JCP_2016_234107, Liakos_JPCA_2020_90, Saitow_JCP_2017, Riplinger_JCP_2016_024109}, to explicitly correlated theory\cite{Ma_JCTC_2015_5291, Schmitz_PCCP_2014_22167, Ma_JCTC_2017_4871, Tew_IJQC_2013_224, Tew_chapter_2021}, to multireference methods\cite{Kats_JCP_2019_214107, Saitow_JCP_2020_114111, Guo_JCP_2016_094111} and to excited states,\cite{Helmich_JCP_2011_214106, Frank_JCP_2018_134102, Dutta_JCP_2016_034102} greatly extending the range of applicability of these approaches for computing energies and properties. The pair-specific nature of PNOs affords a much greater degree of compression of the correlation space than PAOs or OSVs, where domains for distant pairs are the same size as for close pairs.

We are concurrently pursuing two complementary approaches to simulating periodic systems using DLPNO correlation theory.
One where Born--von K{\'a}rm{\'a}n (BvK) boundary conditions\cite{Born_PZ_1912_297, Born_PZ_1913_15} are applied to the correlation treatment, leading to electron repulsion integrals involving lattice summations over infinite periodic images, and one that embeds a supercell correlation treatment in a much larger megacell, where integrals do not involve lattice summations and periodicity is imposed through translational invariance. The Megacell approach is presented in Paper~II of this series;\cite{paper_mega} this paper details the BvK scheme.
We provide the full working equations for DLPNO-MP2 theory, starting from periodic PAOs through OSVs, to PNOs. We detail the necessary modifications to the local density fitting method for obtaining electron repulsion integrals (ERIs) under BvK boundary conditions,\cite{Burow_JCP_2009_214101, Schuetz_chapter_2011, Ye_JCP_2021_131104} where it is essential to project out the charged contributions and to remove the spurious surface energy that arises from dipole-dipole terms.\cite{Stolarczyk_IJQC_1982_911, Makov_PRB_1995_4014, Challacombe_JCP_1997_10131, Kudin_CPL_1998_61, Burow_JCP_2009_214101, Burow_PhD_2011, Maschio_PRB_2007_075101, Pisani_JCC_2008_2113}
We present benchmark calculations validating the implementation and demonstrate that the PNO approximations are equally applicable in the molecular and periodic settings. Our implementation leverages the existing molecular DLPNO-MP2 code\cite{Schmitz_MP_2013_2463} in the Turbomole program package and is made possible due to the periodic Hartree--Fock implementation in the \verb;riper; module.\cite{Lazarski_JCTC_2015_3029, Irmler_JCTC_2018_4567} We report triple-zeta quality MP2 energies for a set of 2D and 3D periodic systems, computed using supercell sizes up to 11$\times$11 and 7$\times$7$\times$7 and involving nearly 5000 functions in the correlation treatment.

\section{Local MP2 with periodic boundary conditions}
\label{sec:mp2}

\begin{figure}[tbp]
    \centering
    \includegraphics[scale=1]{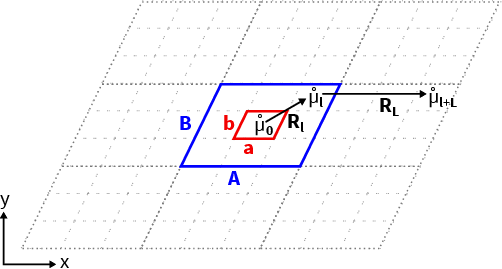}
    \caption{Definition of a 2D BvK supercell (blue box) which is composed of $\mathcal{N}=\mathcal{N}_a\cdot\mathcal{N}_b=9$ unit cells (red box).}
    \label{Fig:UC_SC_definition}
\end{figure}

\begin{table}[tbp]
    \caption{Summary on index convention.}
    \label{Tab:Conventions}
    \centering
    \begin{ruledtabular}
    \begin{tabular}{ll}
    $i$, $j$, $k$, ...                                  & Active occupied orbital \\
    $a$, $b$, $c$, ...                                  & Active virtual orbital \\
    $\tilde{\mu}$, $\tilde{\nu}$, ...                   & PAO \\
    $\tilde{a}$, $\tilde{b}$, ...                       & PNO or OSV  \\
    $\bar{a}$, $\bar{b}$, ...                           & pre-PNO \\
    $\mu$, $\nu$, $\kappa$, ...                         & AO \\
    $P$, $Q$, ...                                       & Auxiliary DF functions \\
    \\
    $\mathbf{k}$                                        & Crystal momentum \\
    $\mathbf{l}$, $\mathbf{m}$, $\mathbf{n}$            & Lattice vector index of unit cell in BvK cell \\
    $\mathbf{L}$, $\mathbf{M}$, $\mathbf{N}$            & Lattice vector index of BvK cell in crystal \\
    $\mathcal{N}=\mathcal{N}_a\mathcal{N}_b\mathcal{N}_c$ & Number of unit cells in BvK cell \\
    $\mathbf{a}$, $\mathbf{b}$, $\mathbf{c}$      & Lattice vectors of unit cell \\
    $\mathbf{A}$, $\mathbf{B}$, $\mathbf{C}$      & Lattice vectors of BvK cell \\
    $\mathbf{R}_\mathbf{l}$                 & $l_a \mathbf{a} + l_b \mathbf{b} + l_c \mathbf{c}$\\
    $\mathbf{R}_\mathbf{L}$                 & $L_a\mathbf{A} + L_b \mathbf{B} + L_c \mathbf{C}$\\
    $\mu_{\mathbf{k}}$                                  & Bloch AO \\
    $\mu_{\mathbf{l}}$                                  & BvK AO  \\
    $\mathring{\mu}_{\mathbf{l}+\mathbf{L}}$            & Individual AO\\
    \end{tabular}
    \end{ruledtabular}
\end{table}

We are working in the LCAO formulation of Hartree--Fock theory with Born--von K{\'a}rm{\'a}n boundary conditions. The basis functions are BvK AOs
\begin{align}
    \mu_\mathbf{l}(\mathbf{r})
    = \sum_{\mathbf{L}}^\infty \mathring{\mu}(\mathbf{r} - \mathbf{R}_\mathbf{l} - \mathbf{R}_\mathbf{L})
    = \sum_{\mathbf{L}}^\infty \mathring{\mu}_{\mathbf{l}+\mathbf{L}}(\mathbf{r})
    \label{eq:bvkao}
\end{align}
Here $\mu$ labels the AO within a unit cell and $\mathbf{l}$ is a vector index labelling the unit cell in the BvK supercell. The BvK AO has the periodicity of the BvK supercell since it is the infitine sum over all periodic images of the AO $\mathring{\mu}$ at $\mathbf{R}_\mathbf{L}$ where $\mathbf{L}$ is the vector index labelling the supercell within the infinite crystal. Table \ref{Tab:Conventions} and Figure \ref{Fig:UC_SC_definition} summarise the index conventions used in this paper.
Hartree--Fock calculations are performed in the basis of Bloch AOs, which are eigenfunctions of the momentum operator with eigenvalue $\mathbf{k}$
\begin{align}
    \mu_\mathbf{k}(\mathbf{r}) &= \frac{1}{\sqrt{\mathcal{N}}} \sum_\mathbf{l}^\mathcal{N} e^{i {\mathbf k} \cdot {\mathbf R}_\mathbf{l}} \mu_\mathbf{l}(\mathbf{r})
    \label{eq:blochao}
\end{align}
Here ${\mathbf R}_\mathbf{l}$ is the lattice position of unit cell $\mathbf{l}$ in the BvK supercell and $\mathcal{N}=\mathcal{N}_a\mathcal{N}_b\mathcal{N}_c$ is the number of unit cells in the BvK supercell. The BvK boundary condition imposes discretisation on the momenta $\mathbf{k}$. 
In this work we adopt the Monkhorst-Pack $k$-point grid.\cite{Monkhorst_PRB_1976_5188} The conversion between BvK AOs and Bloch AOs is in this case a generalised discrete Fourier transform (FT) and our choice of normalisation is such that the FT matrix is unitary
\begin{align}
    V^{\mathbf{k}}_{\mathbf{l}} = \frac{1}{\sqrt{\mathcal{N}}} e^{i {\mathbf k} \cdot {\mathbf R}_\mathbf{l}} 
\end{align}
The overlap and Fock matrices are block diagonal in the Bloch AO basis and
the Hartree--Fock orbitals and eigenvalues satisfy
\begin{align}
    \mathbf{F}_\mathbf{k}^{\mathbf k} \mathbf{C}_\mathbf{k}^\mathbf{k}
    &=
    \mathbf{S}_\mathbf{k}^\mathbf{k} \mathbf{C}_\mathbf{k}^\mathbf{k} {\mathcal E}_\mathbf{k}
\end{align}
where
\begin{align}
    \langle \mu_\mathbf{k} \vert \nu_{\mathbf k'} \rangle_\mathrm{BvK}
    &=
    \delta_\mathbf{k k'} S_{\mu_\mathbf{k}}^{\nu_\mathbf{k'}}
    \\
    \langle \mu_\mathbf{k} \vert \hat F \vert \nu_{\mathbf k'} \rangle_\mathrm{BvK}
    &=
    \delta_\mathbf{k k'} F_{\mu_\mathbf{k}}^{\nu_\mathbf{k'}}
\end{align}
The BvK subscript indicates integrals over the BvK supercell.
Solution of the HF equations yields the canonical molecular orbitals (MOs) $\vert p_\mathbf{k}\rangle$ and
orbital eigenvalues $\epsilon_{p_\mathbf{k}}$. The spin-free MP2 amplitude and energy equations are
\begin{align}
    0 &=
    g^{i_{\mathbf{k}_1}j_{\mathbf{k}_2}}_{a_{\mathbf{k}_3} b_{\mathbf{k}_4}}
    + (\epsilon_{a_{\mathbf{k}_3}} + \epsilon_{b_{\mathbf{k}_4}} - \epsilon_{i_{\mathbf{k}_1}} - \epsilon_{j_{\mathbf{k}_2}})
    t^{i_{\mathbf{k}_1}j_{\mathbf{k}_2}}_{a_{\mathbf{k}_3} b_{\mathbf{k}_4}} \\
    E_\mathrm{corr} &= \frac{1}{\mathcal{N}} 
    \sum_{i_{{\mathbf k}_1} j_{{\mathbf k}_2}}\sum_{a_{{\mathbf k}_3} b_{{\mathbf k}_4}}
    (2g_{i_{\mathbf{k}_1}j_{\mathbf{k}_2}}^{a_{\mathbf{k}_3} b_{\mathbf{k}_4}} - g_{i_{\mathbf{k}_1}j_{\mathbf{k}_2}}^{b_{\mathbf{k}_4} a_{\mathbf{k}_3}})
    t^{i_{\mathbf{k}_1}j_{\mathbf{k}_2}}_{a_{\mathbf{k}_3} b_{\mathbf{k}_4}}
\end{align}
where  $E_\mathrm{corr}$ is the correlation energy per unit cell and
$g_{i_{\mathbf{k}_1}j_{\mathbf{k}_2}}^{a_{\mathbf{k}_3} b_{\mathbf{k}_4}}
= \langle i_{\mathbf{k}_1} j_{\mathbf{k}_2} \vert a_{\mathbf{k}_3} b_{\mathbf{k}_4}\rangle_\mathrm{BvK}$ 
are the BvK electron repulsion integrals.

The coefficients of the canonical MOs $\vert p_\mathbf{k}\rangle$ in the BvK AO basis are the 
inverse FT of the Bloch AO coefficients in the Fock equation
\begin{align}
C_{\mu_{\mathbf{m}}}^{p_\mathbf{k}} = \sum_{\mathbf{k}'}^\mathcal{N} \delta_\mathbf{k k'} C_{\mu_{\mathbf{k}'}}^{p_\mathbf{k}} V^{\mathbf{k}'}_{\mathbf{m}}
\end{align}
These Bloch MOs are delocalised throughout the BvK supercell and are not suitable for local correlation approximations. 
To obtain functions that are localised on a unit cell a
further inverse FT is performed on the Bloch MOs to obtain Wannier functions $\vert p_\mathbf{l}\rangle$ with coefficients
\begin{align}
    C_{\mu_\mathbf{m}}^{p_\mathbf{l}}
    =
    \sum_{\mathbf{k}}^\mathcal{N} C_{\mu_\mathbf{m}}^{p_\mathbf{k}} V^{\mathbf{k}}_{\mathbf{l}}
\end{align}
Wannier functions can be further localised by rotating the Bloch MOs $\vert p_\mathbf{k}\rangle$ for each  $k$-point among themselves before wannierisation and finding the unitary matrix that maximises a chosen locality metric.
We have recently introduced a procedure for obtaining well localised Wannier functions by optimising a fourth-order Pipek--Mezey metric using atomic charges from Bloch intrinsic atomic orbitals,\cite{Zhu_JPCA_2024_8570} which we summarise in Section \ref{sec:wf}. In the Wannier basis, the MP2 energy and amplitude equations become
\begin{align}
    0
    =
    &
    g^{i_{\mathbf l_1} j_{\mathbf l_2}}_{a_{\mathbf l_3} b_{\mathbf l_4}} +
    \sum_{c_{\mathbf l_5}} (f_{a_{\mathbf l_3}}^{c_{\mathbf l_5}} t^{ i_{\mathbf l_1} j_{\mathbf l_2}}_{c_{\mathbf l_5} b_{\mathbf l_4}} + f_{b_{\mathbf l_2}}^{c_{\mathbf l_5}} t^{ i_{\mathbf l_1} j_{\mathbf l_2}}_{a_{\mathbf l_3} c_{\mathbf l_5}})
    \nonumber \\
    &
    - \sum_{k_{\mathbf l_5}} (t^{ k_{\mathbf l_5} j_{\mathbf l_2}}_{a_{\mathbf l_3} b_{\mathbf l_4}} f_{k_{\mathbf l_5}}^{i_{\mathbf l_1}} + t^{ i_{\mathbf l_1} k_{\mathbf l_5}}_{a_{\mathbf l_3} b_{\mathbf l_4}} f_{k_{\mathbf l_5}}^{j_{\mathbf l_2}})
    \\
    E_\mathrm{corr}
    &= \frac{1}{\mathcal N}
    \sum_{i_{{\mathbf l}_1} j_{{\mathbf l}_2}}\sum_{a_{{\mathbf l}_3} b_{{\mathbf l}_4}}
    (2g^{i_{\mathbf{l}_1}j_{\mathbf{l}_2}}_{a_{\mathbf{l}_3} b_{\mathbf{l}_4}} - g^{i_{\mathbf{l}_1}j_{\mathbf{l}_2}}_{b_{\mathbf{l}_4} a_{\mathbf{l}_3}})
    t^{i_{\mathbf{l}_1}j_{\mathbf{l}_2}}_{a_{\mathbf{l}_3} b_{\mathbf{l}_4}}
\end{align}
where the electron repulsion integrals in the BvK AO basis are evaluted as
\begin{align}
    \langle \mu_{\mathbf{l}_1} \nu_{\mathbf{l}_2} \vert \kappa_{\mathbf{l}_3} \lambda_{\mathbf{l}_4}\rangle_\mathrm{BvK}
    =
    \sum_{\mathbf{L}_2\mathbf{L}_3\mathbf{L}_4}^\infty \langle \mathring\mu_{\mathbf{l}_1} \mathring\nu_{\mathbf{L}_2+\mathbf{l}_2} \vert \mathring\kappa_{\mathbf{L}_3+\mathbf{l}_3} \mathring\lambda_{\mathbf{L}_4+\mathbf{l}_4} \rangle
\label{eq:eri}
\end{align}
Wannier functions are translationally invariant $C_{\mu_\mathbf{m}}^{p_\mathbf{0}} = C_{\mu_\mathbf{m+l}}^{p_\mathbf{l}}$, where it is understood that $\mathbf{m}+\mathbf{l}$ implies modulo arithmetic around the BvK supercell. All integrals and amplitudes inherit this same translational symmetry. It is therefore only necessary to compute the residual for $t^{ i_{\mathbf 0} j_{\mathbf l}}_{a_{\mathbf m} b_{\mathbf n}}$ where $\mathbf 0$ is the reference unit cell. The correlation energy per unit cell is
\begin{align}
E_\mathrm{corr} = 
\,\,\frac{2}{1+\delta_{i_\mathbf{0}j_\mathbf{ l}}}
\sum_{i_\mathbf{0}\le j_\mathbf{l}} \sum_{a_\mathbf{m} b_\mathbf{n}}
    (2g^{i_\mathbf{0}j_\mathbf{l}}_{a_\mathbf{m} b_\mathbf{n}} - g^{i_\mathbf{0}j_\mathbf{l}}_{b_\mathbf{m} a_\mathbf{n}})
t^{i_\mathbf{0}j_\mathbf{l}}_{a_\mathbf{m} b_\mathbf{n}}
\end{align}
where the restricted summation counts each pair interaction once and only includes pairs where at least one orbital is in the
reference unit cell. The Fock matrix elements in the Wannier basis can be obtained from the band energies
\begin{align}
f_{p_{\mathbf l}}^{q_{\mathbf m}} &= \sum_{\mathbf k} V_{\mathbf l}^{\mathbf k} f_{p_{\mathbf k}}^{q_{\mathbf k}} V_{\mathbf m}^{\mathbf k} \\
\mathbf{f}_{\mathbf k}^{\mathbf k} &= {\mathbf{U}_{\mathbf k}^{\mathbf k}}^\dagger \mathbf{F}_{\mathbf k}^{\mathbf k} {\mathbf{U}_{\mathbf k}^{\mathbf k}}
\end{align}
where $\mathbf U^\mathbf{k}_\mathbf{k}$ are the unitary matrices that mix the bands at each k-point to maximise the chosen locality metric.

The Wannier functions are localised at a unit cell and local approximations can be applied. 
Specifically, PAO, OSV and PNO approaches can be applied to generate an $\mathcal O(1)$ set of localised 
virtual orbitals $\{\tilde a_{i_\mathbf{0}j_{\mathbf{l}}}\}$ 
adapted to describe the correlation of pair $i_\mathbf{0}j_{\mathbf{l}}$.
In addition, distant electrons in orbitals $j_{\mathbf{l}}$ have negligible correlation with 
electrons in orbital $i_\mathbf{0}$ of the reference unit cell and the number of significant pairs 
$i_\mathbf{0}j_{\mathbf{l}}$ with pair energy greater than $\epsilon$ tends to a constant as the size of the BvK supercell is increased.

\section{Well-localised Wannier Functions}
\label{sec:wf}


\subsection{Diabatic Wannier Functions}

Bloch functions are only determined up to an arbitary phase, or gauge, and as a consequence 
the Wannier functions are not uniquely determined.  In a computer program, the
phase of a Bloch function is simply that resulting from the diagonalisation routine used to solve the Fock equation, which
is in practice arbitary. The FT of rapidly oscillating phases across the $k$-point grid results in  
Wannier functions delocalised across the supercell. Wannier functions properly localised at one unit cell can 
be obtained by fixing the gauge of the
Bloch functions relative to each other so that variations with $\mathbf{k}$ are gradual. 
The natural gauge is that where the scalar product between the coefficients of the
Bloch function at each $\mathbf{k}$ and the $\Gamma$-point $\mathbf{0}$ is real. Bloch functions in their natural
gauge $\vert p_\mathbf{k}^\text{n} \rangle$ can be obtained from those with a random gauge 
$\vert p_\mathbf{k}^\text{r} \rangle$ straightforwardly
\begin{align}
\sum_{\mu_\mathbf{l}} C_{\mu_\mathbf{l}}^{p_\mathbf{0}} C_{\mu_\mathbf{l}}^{p_\mathbf{k}} &= R e^{i\theta^{\mathbf{0k}}_p} \\
\vert p_\mathbf{k}^\text{n} \rangle &= e^{-i\theta^{\mathbf{0k}}_p} \vert p_\mathbf{k}^\text{r} \rangle 
\end{align}
Wannier functions can be further localised by rotating the Bloch functions $p$ for a given $\mathbf{k}$ among
themselves before wannierisation. In particular, due to the invariance of the HF wavefunction, we are free 
to rotate the occupied bands among themselves to obtain highly localised occupied Wannier functions by maximising a chosen
locality metric. 
To obtain a good initial guess for the optimisation routines, we construct diabatic Wannier functions. First, the Bloch orbitals of the $\Gamma$-point are localised by orthogonal transformation. 
\begin{align}
\vert i_\mathbf{0} \rangle = \sum_{j} \vert j_\mathbf{0} \rangle O_{ji}
\end{align}
The Bloch orbitals of the remaining $k$-points are chosen to be those with maximal similarity with the
$\Gamma$-point. The locality of the orbitals of the $\Gamma$-point is thus transferred diabatically across
the first Brillouin zone. The unitary matrix that transforms the unlocalised Bloch orbitals at $\mathbf{k}$ to those that best 
match $\mathbf{0}$ is obtained through a single value decomposition of the similarity metric 
$\mathbf{S}^{j_\mathbf{k}}_{i_\mathbf{0}} = \sum_{\mu_\mathbf{l}} C^{i_\mathbf 0}_{\mu_\mathbf{l}} C^{j_\mathbf k}_{\mu_\mathbf{l}}$
\begin{align}
\vert i_\mathbf{k} \rangle &= \sum_{jj'} \vert j_\mathbf{k} \rangle U_{jj'} V_{j'i} \\
\mathbf{S}^\mathbf{k}_\mathbf{0} &= \mathbf{U} \mathbf{\Sigma} \mathbf{V} 
\end{align}
In this work, we employ a convenient approximate localisation procedure for the Bloch orbitals of the $\Gamma$-point,
where we simply replace them with the Choleski vectors of the $\Gamma$-point density. In this way, well-localised 
Wannier functions can be obtained through linear algebra without any expensive optimisation steps.

\subsection{IBO Wannier Functions}

Well-localised Wannier functions have previously been obtained by optimising the Boy's metric\cite{Boys_RMP_1960_296, Foster_RMP_1960_300, Marzari_PRB_1997_12847, Pizzi_JPCM_2020_165902}, or by generalised Pipek--Mezey metrics\cite{Jonsson_JCTC_2017_460,Clement_JCTC_2021_7406,Schreder_JCP_2024_214117}. In this work we obtain well-localised Wannier functions by optimising the periodic generalisation of Knizia's intrinsic bond orbitals (IBOs),\cite{Knizia_JCTC_2013_4834} which are constructed via intrinsic Bloch atomic orbitals (IAOs) $\tau_\mathbf{l},\sigma_\mathbf{m}$. As discussed in detail in Ref.~\cite{Zhu_JPCA_2024_8570}, our chosen PM locality metric is given by
\begin{equation}
\label{PMmetric}
\braket{O}_{\textbf{PM}}=\sum_{A_\mathbf{l},i_\mathbf{0}}|Q_{i_\mathbf{0}}^{A_{\mathbf{l}}}|^4=\sum_{A_\mathbf{l},i_\mathbf{0}}\bra{i_{\mathbf{0}}}\hat{P}_{A_{\mathbf{l}}}^{\mathrm{IAO}}\ket{i_{\mathbf{0}}}^4, 
\end{equation}
where
\begin{equation}
   \label{eq:a-projector-iao}
	\hat{P}_{A_{\mathbf{l}}}^{\mathrm{IAO}}=\sum_{\tau \in A} \sum_{\sigma_\mathbf{m}} \ket{\tau_{\mathbf{l}}} (S^{-1})_{\tau_\mathbf{l}}^{\sigma_\mathbf{m}}\bra{\sigma_{\mathbf{m}}}.
\end{equation} 
and the projector $\hat{P}_{A_{\mathbf{l}}}^{\mathrm{IAO}}$ involves a restricted summation over the Bloch-IAOs belonging to the atom $A$ and $S_{\tau_\mathbf{l}}^{\sigma_\mathbf{m}}= \langle\tau_\mathbf{l}\ket{\sigma_\mathbf{m}}$. Localization of the WFs is greatly aided by our diabatic Wannierisation procedure, which serves as an initial guess by fixing the gauge the of the orbitals to vary smoothly with the Bloch functions at the $\Gamma$- point. Full implementation details, including the optimization algorithm and the explicit construction of the Bloch IAOs can be found in Ref  \citen{Zhu_JPCA_2024_8570}. The IBO Wannier functions obtained via Bloch IAOs retain the same advantages as molecular IBOs, namely that they form chemically interpretable well-localised orbitals with a well-defined basis set limit and rapidly decaying orthogonality tails.

\section{PNOs with Periodic Boundary Conditions}
\label{sec:pno}

In the domain-based PNO approach, the MP2 amplitudes for pair $i_\mathbf{0}j_{\mathbf{l}}$
are expanded in a basis of PNOs $\{\tilde a_{i_\mathbf{0}j_{\mathbf{l}}}\}$, which are specific to that pair
\begin{align}
E^{i_{{\mathbf 0}}j_{{\mathbf l}}} &= \sum_{\tilde a \tilde b \in [ i_{{\mathbf 0}}j_{{\mathbf l}} ] }
(2g^{i_{\mathbf{0}}j_{\mathbf{l}}}_{\tilde a \tilde b} - g^{i_{\mathbf{0}}j_{\mathbf{l}}}_{\tilde b \tilde a})
t^{i_{\mathbf{0}}j_{\mathbf{l}}}_{\tilde a \tilde b} \,.
\end{align}
In turn, the 
PNOs $\tilde a_{i_\mathbf{0}j_{\mathbf{l}}}$ are expanded in a domain of BvK PAOs $\{\tilde \mu_\mathbf{l}\}$ 
which is also specific to that pair
\begin{align}
\vert \tilde a_{i_\mathbf{0}j_{\mathbf{l}}} \rangle = \sum_{\mu_{\mathbf{m}} \in \mathcal{D}_{i_\mathbf{0}j_{\mathbf{l}}}} \vert \tilde{\mu}_{\mathbf{m}} \rangle C_{\mu_\mathbf{m}}^{a,i_\mathbf{0}j_{\mathbf{l}}} \,.
\end{align}
Our approach to constructing the PNOs $\tilde a_{i_\mathbf{0}j_{\mathbf{l}}}$ for the periodic case
mirrors the approach we use for the molecular case.\cite{Schmitz_MP_2013_2463, Tew_JCTC_2019_6597} First, we determine a domain of
PAOs for each occupied orbital $\vert i_\mathbf{l}\rangle$ based on integral screening thresholds. 
Within this domain, we truncate the virtual space for each $\vert i_\mathbf{l}\rangle$ to the domain of PAOs 
$\mathcal{D}_{i_\mathbf{l}}$ and set of OSVs $\{\tilde a_{i_\mathbf{l}}\}$
that contribute above a threshold to the external MP2 density for pair $\vert i_\mathbf{l}i_\mathbf{l}\rangle$. 
We use numerical Laplace integration and local density fitting to approximate the MP2 density for OSVs.
These PAO and OSV spaces are merged to pair PAO domains and pair orbital sets, within which we construct a MP2 pair density
using the semi-canonical approximation. The virtual space for each pair $\vert i_\mathbf{0}j_\mathbf{l}\rangle$ 
is then finally truncated to the domain of PAOs $\mathcal{D}_{i_\mathbf{0}j_{\mathbf{l}}}$
and set of PNOs
$\{\tilde a_{i_\mathbf{0}j_{\mathbf{l}}}\}$ that contribute to the external MP2 density for pair 
$\vert i_\mathbf{0}j_\mathbf{l}\rangle$ above a threshold. 
For the molecular case, the appropriate truncation threshold at each step has been carefully characterised and 
linked to the overall PNO truncation threshold $T_\text{PNO}$ and we use these thresholds without modification. 
In the following, we give the details of the generalisation of PAOs, OSVs, PNOs and local density fitting to 
include periodic boundary conditions.

\subsection{BvK PAOs}

The BvK PAOs are the projection of the BvK AO onto the space spanned by the virtual Wannier orbitals $a_\mathbf{l}$
\begin{align}
\vert \tilde \mu_\mathbf{0} \rangle &= \sum_{a_{\mathbf{l}}} \vert a_{\mathbf{l}} \rangle \langle a_{\mathbf{l}} \vert \mu_\mathbf{0} \rangle_\mathrm{BvK} 
= \sum_{\nu_{\mathbf{m}}} \vert \nu_{\mathbf{m}} \rangle \hat C^{\mu_\mathbf{0}}_{\nu_{\mathbf{m}}} \\
\hat C^{\mu_\mathbf{0}}_{\nu_{\mathbf{m}}} &= \sum_{ a_{\mathbf{l}} \kappa_{\mathbf{n}}} C_{\nu_{\mathbf{m}}}^{a_{\mathbf{l}}} C_{\kappa_\mathbf{n}}^{a_{\mathbf{l}}} S^{\kappa_\mathbf{n}}_{\mu_\mathbf{0}} \\
    S^{\kappa_\mathbf{n}}_{\mu_\mathbf{0}}
    &=
    \langle  \mu_{\mathbf{0}} \vert \kappa_\mathbf{n} \rangle_\mathrm{BvK}
    =
    \sum_{\mathbf{N}}^\infty \langle \mathring \mu_\mathbf{0} \vert \mathring \kappa_{\mathbf{n}+\mathbf{N}} \rangle
\end{align}
Since the Wannier functions and BvK AOs are all translational copies of the corresponding functions in the reference cell $\mathbf{0}$, the PAOs also inherit translational symmetry $\hat C^{\mu_\mathbf{l}}_{\nu_{\mathbf{m}}} = \hat C^{\mu_\mathbf{0}}_{\nu_{\mathbf{m}-\mathbf{l}}}$.  The overlap of the BvK AO functions within the BvK cell is evaluated as a sum over regular overlap integrals for the infinite periodic images of the individual AOs.
The transformation coefficients $\tilde C^{\mu_\mathbf{l}}_{\nu_{\mathbf{l}'}}$ 
for the contravariant PAOs $\vert \hat \mu_\mathbf{l} \rangle$ are the right-hand Moore--Penrose pseudo inverse of 
$\hat C^{\mu_\mathbf{l}}_{\nu_{\mathbf{m}}}$\cite{Tew_JCTC_2019_6597}.
Our approach to forming OSVs employs a numerical Laplace transformation and we 
require Laplace transformed BvK PAOs. These are defined as 
\begin{align}
\vert \tilde \mu_\mathbf{0}^z \rangle &= \sum_{a_\mathbf{k}} \vert a_{\mathbf{k}} \rangle e^{-(\epsilon_{a_\mathbf{k}}-\epsilon_F)t_z} \langle a_{\mathbf{k}} \vert \tilde \mu_\mathbf{0} \rangle_\mathrm{BvK} \\
&= \sum_{\nu_{\mathbf{m}}} \vert \nu_{\mathbf{m}} \rangle \hat C^{\mu_\mathbf{0},z}_{\nu_{\mathbf{m}}} \\
\hat C^{\mu_\mathbf{0},z}_{\nu_{\mathbf{m}}} &= \sum_{ a \mathbf{n}\mathbf{n}'\kappa_{\mathbf{m}}} C_{\nu_{\mathbf{m}}}^{a_{\mathbf{n}'}} 
L_{\mathbf{n}'\mathbf{n}}^{a,z} C_{\kappa_\mathbf{l}}^{a_\mathbf{n}} S^{\kappa_\mathbf{l}}_{\mu_\mathbf{0}}  \\
L_{\mathbf{n}'\mathbf{n}}^{a,z} &= \sum_{\mathbf{k}} (V^{\mathbf{k}}_{\mathbf{n}'})^\ast e^{-(\epsilon_{a_\mathbf{k}}-\epsilon_F)t_z}
V^{\mathbf{k}}_{\mathbf{n}}
\end{align}
where $\epsilon_F$ is the Fermi level and $t_z$ is a Laplace integration grid point. The Laplace transformed
PAOs $\vert \tilde \mu_\mathbf{l}^z \rangle$ are also all translational copies of $\vert \tilde \mu_\mathbf{0}^z \rangle$.

\subsection{OSVs}

Orbital specific virtuals are PNOs for diagonal pairs $\vert i_\mathbf{l}i_\mathbf{l}\rangle$. Our approach is to
closely follow the molecular PNO-MP2 implementation and to obtain OSVs
as an intermediary for the purpose of accelerating the determination of PNOs. We obtain OSVs for each
orbital $\vert i_\mathbf{l}\rangle$ as eigenvectors of an approximate external density matrix for pair 
$\vert i_\mathbf{l}i_\mathbf{l}\rangle$, which we construct in a prinicpal domain of 
PAOs $\tilde \mu \in \mathcal{D}_{i_\mathbf{l}}$ selected using the greedy algorithm described in Ref.~\citen{Tew_JCTC_2019_6597}.
The external density is computed from first-order amplitudes for diagonal pairs approximated using a numerical Laplace transformation.\cite{Almlof_CPL_1991_319,Haeser_JCP_1992_489,Schmitz_MP_2013_2463} The integration points $z$ and 
weights $w_z$ are determined from the orbital eigenvalues of the supercell in the same way as for molecular calculations. It is only necessary to compue the OSVs for the reference cell, since those for the other cells can be obtained trhough translational symmetry. Specifically, the OSV coefficients $C_{\tilde \nu_\mathbf{m}}^{\tilde a}$ and occupation numbers $n_{\tilde a}$ for orbital $i_\mathbf{0}$ are obtained via 
\begin{align}
t^{i_\mathbf{0}i_\mathbf{0}}_{\tilde \nu_\mathbf{l}\hat \kappa_\mathbf{n}} &= 
\sum_z w_z \langle \tilde \nu_\mathbf{l}^z \hat \kappa_\mathbf{n}^z \vert i_\mathbf{0}^z i_\mathbf{0}^z \rangle_\mathrm{BvK} \\
D_{\tilde \mu_\mathbf{l}}^{\tilde \nu_\mathbf{m}} &= \sum_{\kappa_\mathbf{n}} 4 t^{i_\mathbf{0}i_\mathbf{0}}_{\tilde \mu_\mathbf{l}\tilde \kappa_\mathbf{n}} 
 t^{i_\mathbf{0}i_\mathbf{0}}_{\tilde \nu_\mathbf{m}\hat \kappa_\mathbf{n}} \\
\sum_{\nu_\mathbf{m}} D_{\tilde \mu_\mathbf{l}}^{\tilde \nu_\mathbf{m}} C_{\tilde \nu_\mathbf{m}}^{\tilde a} &= \sum_{\nu_\mathbf{m}} S_{\tilde \mu_\mathbf{l}}^{\tilde \nu_\mathbf{m}} C_{\tilde \nu_\mathbf{m}}^{\tilde a} n_{\tilde a} 
\end{align}
Here $S_{\tilde \mu}^{\tilde \nu}$ is the overlap between the selected PAOs $S_{\tilde \mu_\mathbf{l}}^{\tilde \nu_\mathbf{m}}=\langle \tilde \mu_\mathbf{l} \vert \tilde \nu_\mathbf{m} \rangle_\mathrm{BvK}$
and $C_{\tilde \nu_\mathbf{m}}^{\tilde a}$
are the transformation coefficients from the PAOs to the OSVs $\vert \tilde a_{i_\mathbf{0}} \rangle$, 
which are particular to each orbital $\vert i_\mathbf{0}\rangle$.
Only those OSVs $\{ \tilde a_{i_\mathbf{0}} \}$ with occupation numbers $n_{\tilde a}$ greater than a user 
defined threshold are retained.
The OSVs for $\vert i_\mathbf{l}\rangle$ are simply $C_{\tilde \nu_\mathbf{m+l}}^{\tilde a}$, obtained from those of $\vert i_\mathbf{0}\rangle$
by translation.

Once the OSVs are determined, the OSV-SOS-MP2 energy can be used to compute approximate pair energies for the purpose
of discarding insignificant pairs and providing an estimate of their contribution to the total energy
\begin{align}
    E^{i_\mathbf{0} j_\mathbf{l}}_\text{SOS}
    =
    - \!\!\!\!\!\! \sum_{\tilde a \in [i_\mathbf{0}], \tilde b \in [j_\mathbf{l}]} 
    \frac{g_{\tilde a \tilde b}^{i_\mathbf{0} j_\mathbf{l}} g_{\tilde a \tilde b}^{i_\mathbf{0} j_\mathbf{l}} }
    { \varepsilon_{\tilde a} + \varepsilon_{\tilde b} - f_{i_\mathbf{0}}^{i_\mathbf{0}} - f_{j_\mathbf{l}}^{j_\mathbf{l}} } \,.
\end{align}
The OSV-SOS-MP2 energy does not contain exchange contributions and can be evaluated efficiently using asymmetric density fitting.

\subsection{PNOs}

Having obtained PAO domains and OSVs for every orbital $\vert i_\mathbf{l}\rangle$, we construct initial
PAO pair domains $\mathcal{D}_{i_\mathbf{0}j_\mathbf{l}}$ and pair specific virtuals $\{ \bar a_{i_\mathbf{0}j_\mathbf{l}} \}$
for each pair $\vert i_\mathbf{0}j_\mathbf{l} \rangle$
by merging the respective PAO and OSV domains for $\vert i_\mathbf{0}\rangle$ and $\vert j_\mathbf{l}\rangle$. 
We then form PNOs for each pair as eigenvectors of an approximate external density matrix constructed within
this pair specific subspace. The external density is computed from first-order amplitudes using the semi-canonical
approximation, where the pair specific orbitals are canonicalised to diagonalise the virtual block of the Fock
matrix for each pair $f_{\bar a}^{\bar b} = \varepsilon_{\bar a} \delta_{{\bar a} {\bar b}}$
and off-diagonal occupied Fock matrix elements are neglected. 
\begin{align}
t^{i_\mathbf{0}j_\mathbf{l}}_{\bar a\bar b} &= 
- (\varepsilon_{\bar a} - \varepsilon_{\bar b} - f_{i_\mathbf{0}}^{i_\mathbf{0}} -f_{j_\mathbf{l}}^{j_\mathbf{l}}  )^{-1}
g_{\bar a \bar b}^{i_\mathbf{0} j_\mathbf{l}}
\\
u^{i_\mathbf{0}j_\mathbf{l}}_{\bar a\bar b} &= 2t^{i_\mathbf{0}j_\mathbf{l}}_{\bar a\bar b} - t^{i_\mathbf{0}j_\mathbf{l}}_{\bar b\bar a} \\
D_{\bar a}^{\bar b} &= 2 \sum_{\bar c} ( t^{i_\mathbf{0}j_\mathbf{l}}_{\bar a\bar c} u^{i_\mathbf{0}j_\mathbf{l}}_{\bar b\bar c} + t^{i_\mathbf{0}j_\mathbf{l}}_{\bar c\bar a} u^{i_\mathbf{0}j_\mathbf{l}}_{\bar c\bar b} ) \\
D_{\bar a}^{\bar b} C_{\bar b}^{\tilde a} &= n_{\tilde a} C_{\bar a}^{\tilde a}
\end{align}
Optionally, a further truncation of the PAO domains to principle pair domains can be performed prior to 
finding the eigenvectors of the external pair density matrix. The final space of retained PNOs are those
with occupation numbers greater than the user defined PNO threshold $n_{\tilde a} > T_\text{PNO}$ and the
fully coupled first-order amplitude equations are solved in this subspace. Since the PNOs and 
corresponding integrals and amplitudes for pairs $\vert i_\mathbf{l}j_{\mathbf{m}} \rangle$
are translational copies of $\vert i_\mathbf{0}j_{\mathbf{l}} \rangle$, it is only necessary to 
compute and store those for $\vert i_\mathbf{0}j_{\mathbf{l}} \rangle$.

PNO-MP2 energy under the semi-canonical approximation (PNO-SC-MP2) can be computed as a byproduct of forming PNOs. 
\begin{align}
E^{i_\mathbf{0} j_\mathbf{l}}_\text{SC} = \sum_{\bar a \bar b \in [i_\mathbf{0}j_\mathbf{l}]} 
u_{\bar a \bar b}^{i_\mathbf{0} j_\mathbf{l}} g_{\bar a \bar b}^{i_\mathbf{0} j_\mathbf{l}} 
\end{align}
The difference between the PNO-SC-MP2 energy before and after truncation of the PNO space 
to $\{\tilde a_{i_\mathbf{0}j_\mathbf{l}}\}$ provides an estimate for the contribution of the discarded
PNOs to the total energy.

\section{PNO-MP2 with Periodic Boundary Conditions}

The final amplitude equations for periodic PNO-MP2 are analogous to those of non-periodic PNO-MP2.
\begin{align}
0 =& 
g^{i_{\mathbf 0} j_{\mathbf l}}_{\tilde a \tilde b} + 
(\varepsilon_{\tilde a} + \varepsilon_{\tilde b}) t^{i_{\mathbf 0} j_{\mathbf l}}_{\tilde a \tilde b} \nonumber \\
& - \sum_{k_{\mathbf m}} \sum_{\tilde c \tilde d \in [ k_{{\mathbf m}}j_{{\mathbf l}} ] }
S_{\tilde a, i_{\mathbf 0} j_{\mathbf l}}^{\tilde c, k_{\mathbf m} j_\mathbf{l}}
S_{\tilde b, i_{\mathbf 0} j_{\mathbf l}}^{\tilde d, k_{\mathbf m} j_\mathbf{l}}
t^{ k_{\mathbf m} j_{\mathbf l}}_{\tilde c \tilde d} f_{k_{\mathbf m}}^{i_{\mathbf 0}}  \nonumber \\
& - \sum_{k_{\mathbf m}} \sum_{\tilde c \tilde d \in [ i_{{\mathbf 0}}k_{{\mathbf m}} ] }
S_{\tilde a, i_{\mathbf 0} j_{\mathbf l}}^{\tilde c, i_{\mathbf 0} k_\mathbf{m}}
S_{\tilde b, i_{\mathbf 0} j_{\mathbf l}}^{\tilde d, i_{\mathbf 0} k_\mathbf{m}}
t^{ i_{\mathbf 0} k_{\mathbf m}}_{\tilde c \tilde d} f_{k_{\mathbf m}}^{j_{\mathbf l}} 
\end{align}
where $\tilde a \tilde b \in [ i_{{\mathbf 0}}j_{{\mathbf l}} ]$ and $S_{\tilde a, i_{\mathbf 0} j_{\mathbf l}}^{\tilde c, k_{\mathbf m} j_\mathbf{l}}$ is the overlap between PNOs for pair $i_{\mathbf 0} j_{\mathbf l}$ and pair
$k_{\mathbf m} j_\mathbf{l}$.
The amplitude equations are solved iteratively for every pair $\vert i_\mathbf{0}j_{\mathbf{l}} \rangle$ and the correlation
energy per unit cell is then evaluated as
\begin{align}
E_\mathrm{corr} = 
\,\,\frac{2}{1+\delta_{i_\mathbf{0}j_\mathbf{ l}}}
\sum_{i_\mathbf{0}\le j_\mathbf{l}} \sum_{\tilde a \tilde b \in [ i_{{\mathbf 0}}j_{{\mathbf l}} ] }
    (2g^{i_{\mathbf{0}}j_{\mathbf{l}}}_{\tilde a \tilde b} - g^{i_{\mathbf{0}}j_{\mathbf{l}}}_{\tilde b \tilde a})
    t^{i_{\mathbf{0}}j_{\mathbf{l}}}_{\tilde a \tilde b} + \Delta
\label{eq:emp2}
\end{align}
where $\Delta$ is the correction term composed of the energy estimates for the discarded pairs and PNOs.

Since the WFs and PNO domains are rigorously translationally symmetric, the amplitudes, overlaps and Fock matrix elements spanning the entire supercell can be generated from the translationally unique set of objects where at least one index resides in the reference cell. For example, $f_{k_{\mathbf m}}^{j_{\mathbf l}}=f_{k_{\mathbf m - \mathbf{l}}}^{j_{\mathbf 0}}$ and $t^{ k_{\mathbf m} j_{\mathbf l}}_{\tilde c \tilde d}=t^{ k_{\mathbf 0} j_{\mathbf l - \mathbf{m}}}_{\tilde c \tilde d} $, where it is understood that modulo arithmetic is applied to the lattice vectors.

\section{Periodic Local Density Fitting}
\label{sec:df}

PNO methods require two-electron integrals in a different virtual basis for every pair, which introduces prohibitively high costs unless the density fitting approximation is envoked to
reduce the expense of the integral transformation.\cite{Vahtras_CPL_1993, Weigend_CPL_1998_143}  Density fitting also reduces the cost of the periodic image summation, since summation is only applied to two- and three-center integrals rather than to four-center integrals.

Local density fitting approximates the ERI between two charge densities $\rho^A(\mathbf{r}_1)$ and $\rho^B(\mathbf{r}_2)$ as
\begin{align}
    (\rho^A|\rho^B)_\mathrm{BvK} &\approx \sum_{P_\mathbf{m}\in\mathcal{D}_A} \sum_{Q_l\in\mathcal{D}_B }
    C^A_{P_\mathbf{m}} V_{P_\mathbf{m}Q_\mathbf{l}} C^B_{Q_\mathbf{m}}
    \\
    V_{P_\mathbf{m}Q_\mathbf{l}} &= (P_\mathbf{m}|Q_\mathbf{l})_\mathrm{BvK} = \sum_{\mathbf{L}} (\mathring{P}_\mathbf{m}|\mathring{Q}_\mathbf{l+L})
    \\
    C^A_{P_\mathbf{m}} &= \sum_{Q_l\in\mathcal{D}_A } \gamma^A_{Q_\mathbf{l}} V_{P_\mathbf{m}Q_\mathbf{l}}^{-1} \\
\gamma^A_{Q_\mathbf{l}} &= (\rho^A | Q_\mathbf{l} )_\mathrm{BvK}  = \sum_{\mathbf{L}} (\mathring{\rho}^A | \mathring{Q}_\mathbf{l+L} )
\end{align}
Here $\mathring{\rho}$ is the charge density arising from the individual AOs within the BvK supercell, without image contributions, and, similarly, $\mathring{Q}$ are individual DF functions.
In DLPNO-MP2, the relevant charge densities are $\rho = i_\mathbf{l} \tilde{a}$. For the SOS-MP2 energy estimates, asymmetric density fitting is used,\cite{Tew_JCP_2018_011102}
where the density fitting domains are local to $A$ or $B$, but for all other integrals the usual robust
formulation is used,\cite{Dunlap_PCCP_2000_2113} where $\mathcal{D}_A = \mathcal{D}_B$ and the equations simplify to $\boldsymbol{\gamma}^A \mathbf{V}^{-1} \boldsymbol{\gamma}^B$.

To apply density fitting in periodic calculations, consideration must be given to the convergence of the lattice summations over periodic images for the two- and three-center integrals $\mathbf{V}$ and $\boldsymbol{\gamma}$.

\subsection{Coulomb lattice summation}

The Coulomb interaction between a charge density $\mathring{\rho}^A(\mathbf{r}_1)$ and an infinite lattice of charge densities $\mathring{\rho}^B_\mathbf{L}(\mathbf{r}_2)$ is not necessarily absolutely convergent, and for non-neutral charge densities, even divergent. This property manifests differently\cite{Makov_PRB_1995_4014, Nijboer_Physica_1958_422, Challacombe_JCP_1997_10131} in direct or reciprocal space representations and has profound implications for solid state simulations. A century after the seminal works by Born and von K{\'a}rm{\'a}n, Madelung, Ewald and others,\cite{Born_PZ_1912_297, Born_PZ_1913_15, Madelung_PZ_1918_524, Ewald_AP_1921} the appropriate treatment of periodic lattice sums in the context of efficient electronic structure theory remains an active research field.

Table \ref{Tab:r12_convergence} lists the convergence properties of lattice summations of the multipolar contributions to a charge distribution constiting of charges, dipoles, quadrupoles etc. which are denoted by $s$, $p$, $d$, etc. Since the relevant densities $\rho = i_\mathbf{l} \tilde{a}$ in MP2 are chargeless, divergent terms can be avoided through chargeless density fitting. The conditional convergence of the dipole-dipole terms in 3D arises from a residual surface energy, whose value depends on the shape of the surface of the lattice sum as it expands to infinity. To obtain the correct integrals, this surface energy must be removed.

The conditional convergence of dipole-dipole lattices has been investigated by many researchers. An instructive textbook example that illustrates the geometric nature of this effect is the Madelung constant of an NaCl. The cubic unit cell arrangement leads to a vanishing cell dipole moment, resulting in a fast absolutely convergent lattice sum, whereas the rhombohedral unit cell only yields the correct value if the surface effect due to the cell dipole moment is properly taken into account.\cite{Stuart_JCompP_1978_127, Harris_chapter_1975} Of particular relevance to the current discussion are the numerous works in the context of LCAO-based Hartree--Fock,\cite{Harris_chapter_1975, Stolarczyk_IJQC_1982_911, Saunders_MP_1992_629, Burow_JCP_2009_214101} and MP2,\cite{Ayala_JCP_2001_9698, Maschio_PRB_2007_075101, Pisani_JCC_2008_2113, Bintrim_JCTC_5374_2022} periodic density fitting\cite{Burow_JCP_2009_214101, Schuetz_chapter_2011, Ye_JCP_2021_131104} and the multipole method.\cite{DeWette_Physica_1958_1105, Challacombe_JCP_1997_10131, Kudin_CPL_1998_61, Kudin_CPL_1998_611, Kudin_JCP_2004_2886, Burow_JCP_2009_214101, Lazarski_JCTC_2015_3029} In addition, see also Refs.~\citen{Redlack_JPCS_1975_73, Roberts_JCP_1994_5024, Herce_JCP_2007_124106, Ballenegger_JCP_2014_161102} for a more general discussion on the nature of the surface effect.

\begin{table}[tbp]
    \centering
    \caption{Convergence properties of Coulomb lattice sums between multipole contributions to charge densities $\sum_\mathbf{L} (\mathring{\rho}^A \vert \mathring{\rho}^B_{\mathbf{L}})$.}
    \label{Tab:r12_convergence}
    \begin{ruledtabular}
    \begin{tabular}{cc ccc}
    \multicolumn{1}{c}{$(\rho^A \vert \rho^B )$} & \multicolumn{1}{c}{Decay} & \multicolumn{1}{c}{1D} & \multicolumn{1}{c}{2D} & \multicolumn{1}{c}{3D} \\
    \hline
    $( s \vert s )$ & $(r_{12})^{-1}$ & divergent  & divergent   & divergent   \\[1ex]
    $( s \vert p )$ & $(r_{12})^{-2}$ & absolute   & absolute    & divergent   \\[1ex]
    $( s \vert d )$ & $(r_{12})^{-3}$ & absolute   & absolute    & conditional \\
    $( p \vert p )$ & $(r_{12})^{-3}$ & absolute   & absolute    & conditional \\[1ex]
    $( s \vert f )$ & $(r_{12})^{-4}$ & absolute   & absolute    & absolute    \\
    $( p \vert d )$ & $(r_{12})^{-4}$ & absolute   & absolute    & absolute    \\
    \end{tabular}
    \end{ruledtabular}
\end{table}

\subsection{Chargeless density fitting}

To fit a chargeless density using local density fitting, we adapt the procedure outlaid by Burow \textit{et al}.\cite{Burow_JCP_2009_214101, Burow_PhD_2011}. The linear combination of DF functions forming the approximate density can be resolved into a charged ($\parallel$) and chargeless ($\perp$) component
\begin{align}
\sum_{Q_\mathbf{l}} |\mathring{Q}_\mathbf{l}) C_{Q_\mathbf{l}} =
\sum_{Q_\mathbf{l}} |\mathring{Q}_\mathbf{l}) C^\parallel_{Q_\mathbf{l}} + \sum_{Q_\mathbf{l}} |\mathring{Q}_\mathbf{l}) C^\perp_{Q_\mathbf{l}}
\end{align}
where
\begin{align}
\mathbf{C}^\perp &= \left(\mathbf{1} - \frac{\mathbf{n}\mathbf{n}^\intercal}{|\mathbf{n}|^2}\right) \mathbf{C}  = \mathbf{P}^\perp \mathbf{C}\\
n_{Q_\mathbf{l}} &= \int_{-\infty}^\infty \mathring{Q}_\mathbf{l}(\mathbf{r}) \, \mathrm{d}^3 \mathbf{r}
\end{align}
The vector $\mathbf{n}$ is the linear combination of $\mathring{Q}_\mathbf{l}$ that results in a charge and is
specific to the local density fitting domain $\mathcal{D}$. To fit a chargeless density, the charged component
of the fitting basis is projected out before solving for the fitting coefficients, which results in
\begin{align}
(\rho^A|\rho^B)_\mathrm{BvK} &\approx \mathbf{C}_A^\perp \mathbf{V}^\perp \mathbf{C}_B^\perp \\
\mathbf{V}^\perp &= \mathbf{P}_A^\perp \mathbf{V} \mathbf{P}_B^\perp \\
\mathbf{C}_A^\perp &= \boldsymbol{\gamma}_A^\perp (\mathbf{V}_A^\perp)^{-1} \\
\boldsymbol{\gamma}_A^\perp &= \boldsymbol{\gamma}_A \mathbf{P}^\perp_A
\end{align}
For robust DF, this simplies to
\begin{align}
(\rho^A|\rho^B)_\mathrm{BvK} &\approx \boldsymbol{\gamma}_A^\perp (\mathbf{V}^\perp)^{-1} \boldsymbol{\gamma}_B^\perp
\end{align}

\subsection{Surface charge cancellation}

The dipole of a charge density can be represented as a set of six point charges placed at the centers of the faces of
the BvK supercell. Periodically repeating these dipoles within a finite volume $\mathcal{M} = \mathcal{M}_a\mathcal{M}_b\mathcal{M}_c$ results in a cancellation of all point charges within the shape, but leaves uncancelled positive and negative charges on opposite faces. The electrostatic interaction of these charges with the dipole of the charge density in the reference supercell gives rise to a surface energy that vanishes in 1D and 2D, but converges to a shape-dependent value as $\mathcal{M}$ increases in 3D.
The correct integrals for BvK boundary conditions are obtained from lattice summation protocols that impose `tin foil'\cite{Herce_JCP_2007_124106} conditions at the surface, where there is no artificial surface contribution to the energy.

\begin{figure}[tbp]
    \includegraphics[scale=1]{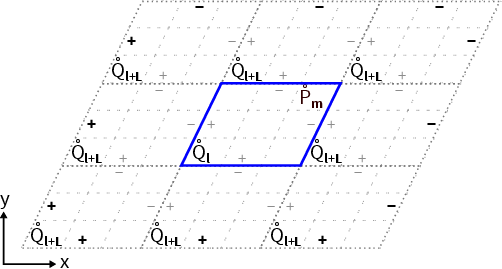}
    \caption{Direct lattice summation for
    $( P_\mathbf{m} | Q_\mathbf{l} )_\mathrm{BvK}$ with dipole correction.}
    \label{Fig:PQ_latsum_with_dipole}
\end{figure}

To eliminate the surface contribution, we follow the method by Stolarczyk and Piela,\cite{Stolarczyk_IJQC_1982_911} which has been adapted to the LCAO Hartree--Fock method in \verb:riper:\cite{Burow_JCP_2009_214101} and described in detail in Ref.~\citen{Burow_PhD_2011}. To each auxiliary basis function in the lattice summation, we add the set of six point charges located at the faces of the BvK supercell within which the function resides, that exactly cancel the dipole of the function. 
\begin{align}
(P_\mathbf{m}|Q_\mathbf{l})_\mathrm{BvK} &= \sum_{\mathbf{L}} (\mathring{P}_\mathbf{m}|\mathring{Q}_\mathbf{l+L} + \mathring{d}_\mathbf{L}^{Q_\mathbf{l}}) \\
(\rho^A | Q_\mathbf{l} )_\mathrm{BvK}  &= \sum_{\mathbf{L}} (\mathring{\rho}^A | \mathring{Q}_\mathbf{l+L} + \mathring{d}_\mathbf{L}^{Q_\mathbf{l}}) \\
\mathring{d}_\mathbf{L}^{Q_\mathbf{l}}(\mathbf{r}) &= \sum_\mathbf{I} q_\mathbf{I}^{Q_\mathbf{l}} \delta(\mathbf{r} - \tfrac{1}{2} \mathbf{R}_\mathbf{I} - \mathbf{R}_\mathbf{L})
\end{align}
Here $\mathbf{I}$ are the unit BvK supercell displacements $(\pm 1,0,0), (0,\pm 1,0), (0,0,\pm 1)$. In one and two dimensional calculations, $\mathbf{I}$ only includes the periodic lattice vectors, and two and four point charges are used, respectively, to cancel the dipole in the periodic degrees of freedom.
A 2D case is illustrated in Fig.~\ref{Fig:PQ_latsum_with_dipole}.
The Cartesian dipole components of a function $\mathring{Q}_\mathbf{l}$ are given by
\begin{align}
p_\alpha^{Q_\mathbf{l}} = -\int_{-\infty}^{\infty} \alpha \, \mathring{Q}_\mathbf{l}(\mathbf{r}) \, \mathrm{d}^3 \mathbf{r} \qquad \alpha = x,y,z
\end{align}
where the minus is due to the electronic charge.
The values of the point charges $q_\mathbf{I}^{Q_\mathbf{l}}$ that cancel the dipole are obtained by transforming $\mathbf{p}^{Q_\mathbf{l}}$ to the BvK lattice to obtain $p_a^{Q_\mathbf{l}},p_b^{Q_\mathbf{l}},p_c^{Q_\mathbf{l}}$. The point charges that cancel the dipole in lattice vector $a$, placed at supercell displacements of $(\pm0.5,0,0)$, are $\mp p_a^{Q_\mathbf{l}}/|\mathbf{A}|$, and similarly for the other lattice vectors.

In our pilot implementation, we evaluate the integrals through direct space lattice summation without applying any fast multipole acceleration techniques.\cite{Kudin_CPL_1998_61, Challacombe_JCP_1997_10131, Lazarski_JCTC_2015_3029} The lattice summation is truncated at a sufficiently large volume $\mathcal{M}$ and the integrals
are evaluated as
\begin{align}
(P_\mathbf{m}|Q_\mathbf{l})_\mathrm{BvK} &=
\sum_{\mathbf{L}}^{\mathcal{M}} (\mathring{P}_\mathbf{m}|\mathring{Q}_\mathbf{l+L}) + \sum_\mathbf{I} q_\mathbf{I}^{Q_\mathbf{l}} S_{P_\mathbf{m}}^\mathbf{I} \\
S_{P_\mathbf{m}}^\mathbf{I}  &= \sum_{\mathbf{L}\in\mathcal{F}_\mathbf{I}}
( \mathring{P}_\mathbf{m} | \delta(\mathbf{r} - \tfrac{1}{2} \mathbf{R}_\mathbf{I} - \mathbf{R}_\mathbf{L}))\\
(\mu_\mathbf{m} \nu_\mathbf{n} | Q_\mathbf{l} )_\mathrm{BvK}  &=
\sum_{\mathbf{LN}}^{\mathcal{M}} ( \mathring{\mu}_\mathbf{m} \mathring{\nu}_\mathbf{n+N} | \mathring{Q}_\mathbf{l+L} )
+ \sum_\mathbf{I} q_\mathbf{I}^{Q_\mathbf{l}} S_{\mu_\mathbf{m} \nu_\mathbf{n}}^\mathbf{I} \\
S_{\mu_\mathbf{m} \nu_\mathbf{n}}^\mathbf{I} &= \sum_{\mathbf{L}\in\mathcal{F}_\mathbf{I}}
\sum_{\mathbf{N}}^{\mathcal{M}} ( \mathring{\mu}_\mathbf{m} \mathring{\nu}_\mathbf{n+N} |\delta(\mathbf{r} - \tfrac{1}{2} \mathbf{R}_\mathbf{I} - \mathbf{R}_\mathbf{L}))
\end{align}
Here $\mathcal{F}_\mathbf{I}$ refers to the set of lattice vectors on the $\mathbf{I}^\text{th}$ face of $\mathcal{M}$. 
The surface correction term only applies to $s$-type and $p$-type DF functions, since all other functions have
a zero dipole. The net dipole arising from the $s$-type functions in the density fitting integrals
is well defined because the net charge is projected out.
For 1D and 2D calculations, the surface correction terms are analogously defined,
but are only applied in the periodic directions.
The summation over $\mathbf{N}$ in the three-center integrals is rapidly convergent since the overlaps
of $\mathring{\mu}_\mathbf{m}$ and $\mathring{\nu}_\mathbf{n+N}$ decay exponentially.
The summation over $\mathbf{L}$ is polynomially decaying, with the degree of the polynomial depending on
the basis functions and whether the calculation is 1D, 2D or 3D. In our calculations we monitor the
convergence of the integrals and resulting energies with the lattice extent $\mathcal{M}$ and our values
are accurate to the significant figures reported.

The charge projection and surface correction terms significantly complicate the exploitation of translational symmetry to obtain
computational savings. While the corrected integrals are translationally symmetric after charge projection,
the individual contributions are not. In particular, the lattice summation range and correction terms for
$s$-type DF functions must be evaluated with reference to a common origin to ensure that the surface term
is properly eliminated. We therefore compute the full set of AO integrals and correction terms in the supercell
using the centre of the supercell as the origin. Integral evaluation through naive lattice summation
is currently the primary computational bottleneck. The scaling of this step with system size is the same as the molecular code, which is  asymptotically $\mathcal{O}(N)$ when the supercell is large enough for the local DF domain to saturate and screening to be effective.

One further subtlety in the integral evaluation is that
the correction term breaks the symmetry of the Coulomb metric $V_{P_\mathbf{m}Q_\mathbf{l}}$.
Symmetry is restored in the limit of large $\mathcal{M}$. We therefore symmetrise $V_{P_\mathbf{m}Q_\mathbf{l}}$,
which we find accelerates convergence with lattice extent.

\section{Computational Details}\label{sec:compdetail}

Born--von K{\'a}rm{\'a}n domain-based local PNO-MP2 (BvK-DLPNO-MP2) has been implemented in a development version of the Turbomole program package within the \verb;pnoccsd; module, which has been used for all calculations presented here.
Periodic LCAO-based Hartree--Fock with $k$-point sampling has recently become available in the \verb;riper; module,\cite{Lazarski_JCTC_2015_3029, Lazarski_JCC_2016_2518, Burow_JCTC_2011_3097, Burow_JCP_2009_214101, Mueller_JCC_2020_2573} the output of which provides the Hartree--Fock Bloch functions and band energies required for MP2. The Bloch functions are localised for the ensuing correlation treatment through a new Wannier localization procedure which has been implemented into a development version of \verb;riper;.\cite{Zhu_JPCA_2024_8570} 

All calculations utilize the pob-TZVP and pob-TZVP-rev2 basis sets which have specifically been optimized for LCAO simulations in the solid state.\cite{Peintinger_JCC_2013_451, Laun_JCC_2022_839}
For the density fitting, we use the def2-TZVP auxiliary basis in the Hartree-Fock\cite{Weigend_PCCP_2006_1057} and MP2\cite{Weigend_CPL_1998_143} calculations. Unless otherwise indicated, the frozen-core approximation is applied. DLPNO-MP2 calculations are performed for a series of increasingly accurate PNO thresholds ($\mathcal{T}_\mathrm{PNO}=10^{-X}$, $X=6,7,8,9$). Motivated by the observation that the largest discarded amplitude is proportional to the square of the PNO threshold, the complete PNO space (CPS) limit is estimated through a square root extrapolation using the two most accurate PNO energies.\cite{Sorathia_JCP_2020_174112, Sorathia_JCTC_2024_2740}

\section{Results}
\label{sec:results}

To obtain reliable reference values for unit cell energies against which to test the performance of the BvK-DLPNO-MP2 method, we first leverage the molecular DLPNO\cite{Schmitz_MP_2013_2463, Tew_JCTC_2019_6597} and canonical RI-MP2\cite{Weigend_TCA_1997_331} infrastructure in Turbomole\cite{Balasubramani_JCP_2020_184107, Franzke_JCTC_2023_6859} to obtain unit cell energies through a molecular fragment approach. This enables us to assess the numerical stability of the BvK-DLPNO-MP2 method, to examine the rate of convergence to the thermodynamic limit, and to compare the relative accuracies of the HF and PNO approximations in the molecular and periodic settings. For future benchmarks, we provide reference data using the BvK-DLPNO-MP2 method for various 2D and 3D systems.


\subsection{Unit cell energies from molecular fragments}
\label{subsec:mol_v_bvk}

\begin{figure}[tbp]
    \centering
    \includegraphics[scale=0.6]{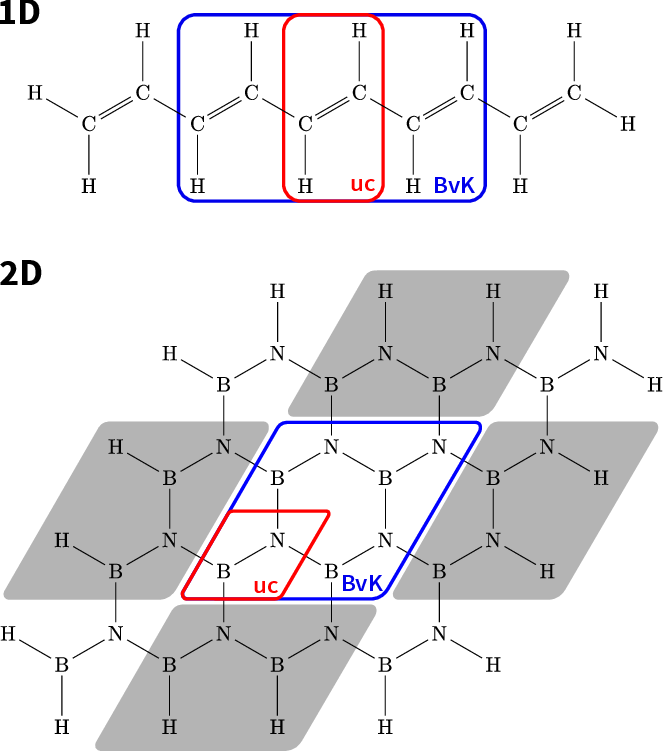}
    \caption{Molecular cluster fragments that accommodate 1D and 2D BvK supercells (dotted box). The supercells of clusters with sum formula \ce{C_{2l}H_{2l+2}} and B$_{l^2}$N$_{l^2}$H$_{4l}$ are comprised of $(l-2)$ and $(l-2)^2$ unit cells (red box), respectively.}
    \label{Fig:Lewis_C2H2_BN}
\end{figure}

\begin{table*}[tbp]
    \centering
    \caption{Hartree--Fock and MP2 correlation energies per unit cell (in Hartree) extracted from molecular cluster energies.}
    \label{Tab:Mol_dat}
    \begin{ruledtabular}
    \begin{tabular}{c d{5} d{5}d{5}d{5}d{5} d{5} d{5} }
    \multicolumn{1}{c}{effective} & & \multicolumn{5}{c}{PNO-RI-MP2} & \\
    \cline{3-7}
    \multicolumn{1}{c}{$k$-mesh} & \multicolumn{1}{c}{RI-HF\footnote{Converged BvK-RI-HF energies: $-$76.88949 (C$_2$H$_2$), $-$175.77532 (C$_2$HF), $-$75.29075 (BN).}} & \multicolumn{1}{c}{6} & \multicolumn{1}{c}{7} & \multicolumn{1}{c}{8} & \multicolumn{1}{c}{9} & \multicolumn{1}{c}{CPS} & \multicolumn{1}{c}{RI-MP2} \\
    \hline \multicolumn{7}{c}{1D C$_2$H$_2$  (pob-TZVP)} \\ \hline
             $ 3$ &  -76.88936 & -0.27609 & -0.27651 & -0.27664 & -0.27669 & -0.27671 & -0.27673 \\
             $ 5$ &  -76.88947 & -0.27619 & -0.27667 & -0.27682 & -0.27687 & -0.27689 & -0.27691 \\
             $ 7$ &  -76.88949 & -0.27622 & -0.27670 & -0.27686 & -0.27691 & -0.27693 & -0.27695 \\
             $ 9$ &  -76.88949 & -0.27622 & -0.27670 & -0.27686 & -0.27692 & -0.27695 & -0.27696 \\
             $11$ &  -76.88949 & -0.27622 & -0.27670 & -0.27686 & -0.27692 & -0.27695 & -0.27696 \\
             $13$ &  -76.88949 & -0.27622 & -0.27671 & -0.27687 & -0.27692 & -0.27694 & -0.27696 \\
             $15$ &  -76.88949 & -0.27622 & -0.27671 & -0.27686 & -0.27692 & -0.27695 & -0.27696 \\
             $17$ &  -76.88949 & -0.27622 & -0.27671 & -0.27686 & -0.27692 & -0.27695 & -0.27696 \\
             $19$ &  -76.88949 & -0.27622 & -0.27671 & -0.27686 & -0.27692 & -0.27695 & -0.27696 \\
             $21$ &  -76.88949 & -0.27622 & -0.27671 & -0.27686 & -0.27692 & -0.27695 & -0.27696 \\
             $23$ &  -76.88949 & -0.27622 & -0.27671 & -0.27686 & -0.27692 & -0.27695 & -0.27696 \\
             $25$ &  -76.88949 & -0.27622 & -0.27671 & -0.27686 & -0.27692 & -0.27695 & -0.27696 \\
             $27$ &  -76.88949 & -0.27622 & -0.27671 & -0.27686 & -0.27692 & -0.27695 & -0.27696 \\
             $37$ &  -76.88949 & -0.27622 & -0.27671 & -0.27686 & -0.27692 & -0.27695 & -0.27696 \\
    \hline \multicolumn{7}{c}{1D C$_2$HF  (pob-TZVP)} \\ \hline
             $ 3$ & -175.77440 & -0.43895 & -0.43978 & -0.44006 & -0.44016 & -0.44021 & -0.44022 \\
             $ 5$ & -175.77476 & -0.43873 & -0.43958 & -0.43988 & -0.43999 & -0.44004 & -0.44005 \\
             $ 7$ & -175.77496 & -0.43850 & -0.43936 & -0.43967 & -0.43979 & -0.43985 & -0.43985 \\
             $ 9$ & -175.77508 & -0.43831 & -0.43921 & -0.43953 & -0.43964 & -0.43969 & -0.43970 \\
             $11$ & -175.77516 & -0.43824 & -0.43912 & -0.43943 & -0.43955 & -0.43961 & -0.43961 \\
             $13$ & -175.77521 & -0.43812 & -0.43906 & -0.43938 & -0.43949 & -0.43954 & -0.43955 \\
             $15$ & -175.77524 & -0.43808 & -0.43903 & -0.43934 & -0.43945 & -0.43950 & -0.43951 \\
             $17$ & -175.77526 & -0.43811 & -0.43900 & -0.43931 & -0.43943 & -0.43949 & -0.43949 \\
             $19$ & -175.77527 & -0.43810 & -0.43899 & -0.43930 & -0.43941 & -0.43946 & -0.43947 \\
             $21$ & -175.77528 & -0.43809 & -0.43897 & -0.43929 & -0.43940 & -0.43945 & -0.43946 \\
             $23$ & -175.77529 & -0.43806 & -0.43897 & -0.43928 & -0.43939 & -0.43944 & -0.43946 \\
             $25$ & -175.77530 & -0.43807 & -0.43896 & -0.43928 & -0.43939 & -0.43944 & -0.43945 \\
             $27$ & -175.77530 & -0.43805 & -0.43896 & -0.43927 & -0.43938 & -0.43943 & -0.43945 \\
             $37$ & -175.77531 & -0.43805 & -0.43895 & -0.43926 & -0.43938 & -0.43944 & -0.43944 \\
    \hline \multicolumn{7}{c}{2D BN (pob-TZVP-rev2)} \\ \hline
    $ 5\times  5$ &  -79.29076 & -0.21773 & -0.21840 & -0.21864 & -0.21874 & -0.21879 &  -0.21880 \\
    $ 7\times  7$ &  -79.29075 & -0.21770 & -0.21838 & -0.21863 & -0.21873 & -0.21877 &  -0.21878 \\
    $ 9\times  9$ &  -79.29075 & -0.21772 & -0.21837 & -0.21862 & -0.21873 & -0.21877 &  -0.21878 \\
    $11\times 11$ &  -79.29075 & -0.21768 & -0.21837 & -0.21862 & -0.21873 & -0.21877 &  -0.21878 \\
    \end{tabular}
    \end{ruledtabular}
\end{table*}

The molecular fragment method was used to obtain benchmark unit cell energies of 1D and 2D periodic systems. 3D examples were not possible due to the high computational expense of this approach. For 1D systems, we chose the linear alkene polymers C$_2$H$_2$ and C$_2$HF, where in the latter the long-range dipole-dipole interactions due to the polar C$-$F bonds are expected to slow down the convergence with $k$-mesh. For 2D periodicity, we chose hexagonal boron nitride monolayers (lattice constant 2.501\,{\AA}; Ref.~\citen{Meng_EJIC_2010_3174}), which, unlike graphene, avoids Fermi-level degeneracies, ensuring stable periodic MP2 energies. The insulator is isoelectronic and structurally analogous to graphene but has a much higher band gap ($\sim 6$\,eV).\cite{Cassabois_NP_2016_262}  Unit cell geometries are deposited in the Supplementary Material.

In the molecular fragment approach, the energy of a unit cell is obtained by modeling the energy of a molecular fragment as a core of unit cells, plus contributions from faces, edges and corners. Starting from the periodic supercell structures, the fragments are generated by saturating the dangling bonds with hydrogens, as illustrated in Fig.~\ref{Fig:Lewis_C2H2_BN}. The energies of a series of increasing molecular fragments can then be used to subtract the surface contributions to obtain the correlation energy per unit cell, which converges to the thermodynamic limit as the size of the molecular fragments increases.

For the 1D systems, the molecular energies of C$_{2l}$H$_{2l+2}$ and C$_{2l}$H$_{l+2}$F$_l$ are modeled as 
\begin{align}
    E_{l,\text{tot}} = 2E_\mathrm{corner} + (l-2)E_\mathrm{uc}
    .
\end{align}
Using two fragment calculations, the corner energy can be subtracted to obtain an estimate of the unit cell energy
\begin{align}
    E_\mathrm{uc}^\mathrm{1D}
    = 
    \frac{E_{l,\text{tot}} - E_{l-2,\text{tot}}}{2}
    \label{eq:E_uc_mol_1D}
    .
\end{align}
The largest supercell size accommodated in the largest fragment corresponds to $k_\textit{eff}=(l-2)$. For the 2D system, the total energy may be decomposed as
\begin{align}
    E_{l,\text{tot}} = 4E_\mathrm{corner} + 4(l-2)E_\mathrm{edge}+(l-2)^2E_\mathrm{uc}
    .
\end{align}
To extract the unit cell energy for an effective mesh $k_\textit{eff}^2=(l-2)^2$, we now require three molecular clusters
\begin{align}
    E_\mathrm{uc}^\mathrm{2D}
    = 
    \frac{E_{l,\text{tot}} - E_{l-2,\text{tot}} +E_{l-4,\text{tot}}}{8}
    \label{eq:E_uc_mol_2D}
    .
\end{align}

The RI-HF, DLPNO-MP2 and canonical RI-MP2 correlation energies per unit cell obtained from this molecular fragment approach are listed in Table~\ref{Tab:Mol_dat}. Inspection of the RI-HF data demonstrates the accuracy of the fragment approach. For all three systems, the HF energy converges to the thermodynamic limit obtained from periodic RI-HF calculations using \verb:riper:, reported in the footnotes of Table~\ref{Tab:Mol_dat}. As anticipated, the C$_2$HF convergence is much slower than that of C$_2$H$_2$ and BN, which is attributed to the long range dipole-dipole interactions. A similar convergence behavior is observed for the DLPNO-MP2 correlation energies. All three systems display monotonic convergence to thermodynamic limit values for all PNO thresholds and for canonical RI-MP2.
Moreover, it is encouraging that the extrapolated CPS correlation energies match the canonical values to within $1\times10^{-5}$\,Ha, demonstrating the accuracy and numerical stability of our PAO-OSV-PNO compression cascade. The final rows in Table~\ref{Tab:Mol_dat} for each of the three systems provide the thermodynamic limit DLPNO-MP2 correlation energies for each PNO threshold. These values are accurate to $1\times10^{-5}$\,Ha.

\subsection{Validating the Accuracy of BvK-DLPNO-MP2}
\label{subsec:bvk_mp2}

\begin{figure*}[htbp]
    \centering
    \includegraphics[scale=1]{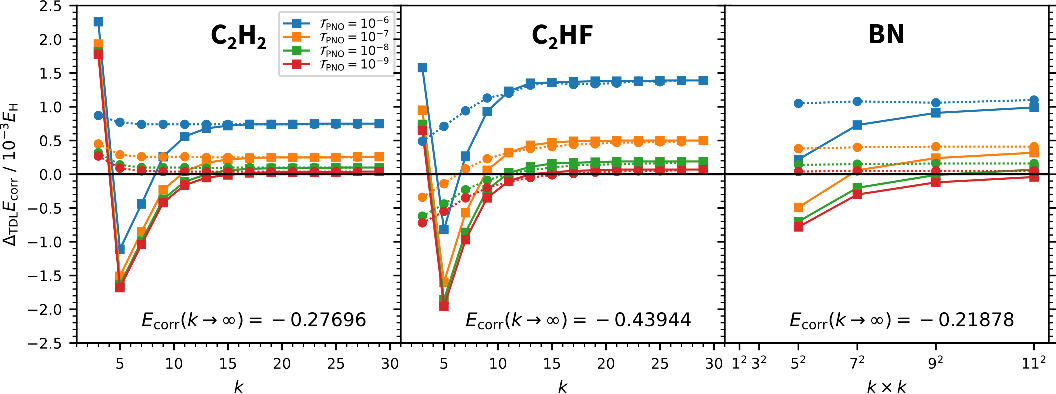}
    \caption{Deviations from the canonical thermodynamic limit of periodic BvK-DLPNO-MP2 (solid) and fragment based DLPNO-MP2 (dashed) correlation energies per unit cell for C$_2$H$_2$, C$_2$HF and  a BN monolayer}
    \label{Fig:Mol_v_BvK_MP2}
\end{figure*} 

Equipped with reference canonical MP2 unit cell energies at the thermodynamic limit for C$_2$H$_2$, C$_2$HF and BN, we are now in a position to benchmark the BvK-DLPNO-MP2 method. The convergence with respect to the PNO threshold and supercell size for the BvK-DLPNO-MP2 correlation energies is displayed in 
Fig.~\ref{Fig:Mol_v_BvK_MP2}, where we plot the deviation from the canonical thermodynamic limit. The BvK values are connected by solid lines and the corresponding molecular fragment data, which are included for comparison, by dotted lines.

We find that the molecular fragment and BvK-DLPNO-MP2 correlation energies converge to the same thermodynamic limit for each PNO threshold for all three examples tested.
For C$_2$H$_2$ and C$_2$HF the agreement is to within $10^{-5}$\,Ha at $k=29$ and for BN the agreement is better than $10^{-4}$\,Ha at $k=11$. The PNO approximation with BvK boundary conditions is therefore fully consistent with that of the molecular case, and the usual smooth convergence to the canonical limit is observed upon tightening the PNO threshold.
Fig.~\ref{Fig:Mol_v_BvK_MP2} shows a striking difference in how the molecular and BvK energies approach the thermodynamic limit. For C$_2$H$_2$ and BN, the molecular curves almost immediately plateau, whereas the BvK curves exhibit a slower convergence, necessitating comparably large supercells to achieve mHa accuracy. 

\begin{figure*}[htbp]
    \centering
    \includegraphics[scale=1]{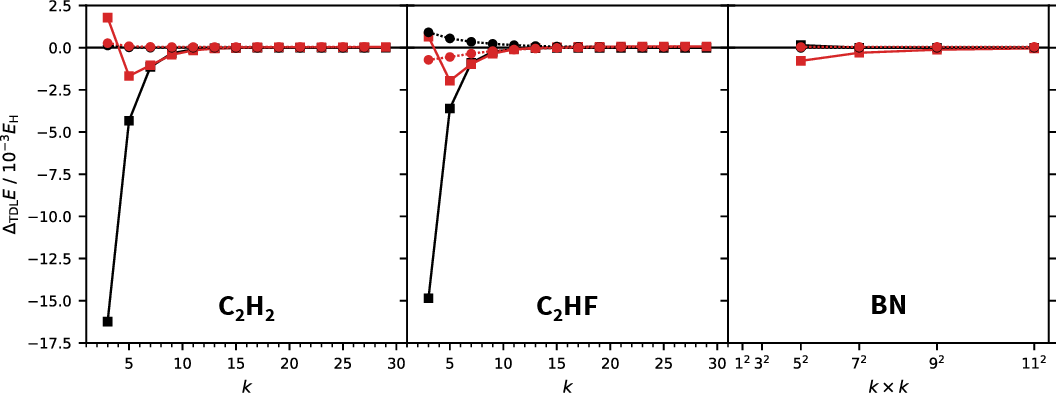}
    \caption{Molecular (dash-dot) and periodic BvK (solid) $\mathcal{T}_\mathrm{PNO}=10^{-9}$ curves from Fig.~\ref{Fig:Mol_v_BvK_MP2} (red) juxtaposed with the convergence of the Hartree–-Fock energy (black). The corresponding data are provided in the Supplementary Material.}
    \label{Fig:BvK_RHF}
\end{figure*}

To further examine the slow convergence, in Fig.~\ref{Fig:BvK_RHF} we plot the finite-size errors from BvK-DLPNO-MP2 and molecular fragment DLPNO-MP2 ($\mathcal{T}_\text{PNO}=10^{-9}$) unit cell correlation energies together with the corresponding errors in the Hartree--Fock energy.
The HF energy from the periodic calculations also displays a slower convergence to the thermodynamic limit compared to the fragment approach. These data indicate that the main source of the slow convergence of the BvK-DLPNO-MP2 correlation energy is most likely due to the finite-size errors in the underlying Hartree--Fock orbitals and band energies. Finite-size effects in the virtual HF orbitals and eigenvalues are not reflected in the HF energy, but have a direct impact on the MP2 correlation energy, which is the reason for the slow convergence of the BvK-DLPNO-MP2 energies for BN.


These findings are very interesting in the context of how the BvK-DLPNO-MP2 and Megacell-DLPNO-MP2 schemes (Paper~II) approach the thermodynamic limit. The latter embeds the supercell for the correlation treatment in a larger megacell to ensure all WFs centered in the supercell are sufficiently decayed at the boundary of the megacell. Since the underlying HF calculation is for the larger megacell, the finite-size errors for this approach will be much smaller and we anticipate a more rapid convergence to the thermodynamic limit of the Megacell-DLPNO-MP2 correlation energy with respect to the supercell size.

Since an independent benchmark for a realistic 3D case using the molecular fragment approach was too expensive, we instead verify that the dipole correction scheme is correctly removing the surface energy by computing the correlation energy of the same lattice through two different unit cells. At the thermodynamic limit, these calculations should yield the same HF and correlation energies.
The results for cubic and rhombohedral LiH unit cells are reported in Table~\ref{Tab:dipcor}. Since the cubic cell contains four LiH units and the rhombohedral only one, the former energies are divided by four.
The HF energies per LiH unit for the $3{\times} 3 {\times}3$ cubic and $5{\times} 5 {\times}5$ rhombohedral calculations agree to within $0.25$\,mHa, reducing to $10^{-5}$\,Ha for $5{\times} 5 {\times}5$ and $7{\times} 7 {\times}7$.
The DLPNO-MP2 correlation energies per LiH unit for the $3{\times} 3 {\times}3$ cubic and $5{\times} 5 {\times}5$ rhombohedral calculations agree to within $0.15$\,mHa, if the dipole correction is applied. If it is not, the energies deviate by $621$\,mHa. These data indicate that the implementation is correct.
In Paper~II of this series, we compare correlation energies computed using the BvK and Megacell schemes for 1D, 2D and 3D materials. We find that the correlation energies all converge to the same thermodynamic limit for each PNO threshold, providing further evidence that the level of accuracy of the PNO truncation scheme is entirely consistent for 1D, 2D and 3D, and molecular calculations.

\begin{table}[htbp]
    \centering
    \caption{HF and BvK-DLPNO-MP2 ($\mathcal{T}_\mathrm{PNO}=10^{-7}$) energies per LiH unit (Hartree) for a 3D LiH lattice with two different unit cells.}
    \label{Tab:dipcor}
    \begin{ruledtabular}
    \begin{tabular}{l d{5}d{5}d{5}d{5} }
    & \multicolumn{2}{c}{Cubic} & \multicolumn{2}{c}{Rhombohedral} \\
    \cline{2-3} \cline{4-5}
    $k$-mesh &
    \multicolumn{1}{c}{$3{\times} 3 {\times}3$} &
    \multicolumn{1}{c}{$5{\times} 5 {\times}5$} &
    \multicolumn{1}{c}{$5{\times} 5 {\times}5$} &
    \multicolumn{1}{c}{$7{\times} 7 {\times}7$} \\
    $n_\mathrm{atom}$ & \multicolumn{1}{c}{216} & \multicolumn{1}{c}{1000} & \multicolumn{1}{c}{250} & \multicolumn{1}{c}{686} \\
    $E_\mathrm{HF}$                & -8.06035 & -8.06058 & -8.06060 & -8.06059 \\
    $E_\mathrm{corr}$ \\
    ... dip-cor                    & -0.03060 &          & -0.03075 & -0.03092 \\
    ... no dip-cor                 & -0.077   &          & -0.698   &          \\
    \end{tabular}
    \end{ruledtabular}
\end{table}

\subsection{Illustrative 2D and 3D examples}
\label{subsec:2d_3D_numbers}

\begin{table*}[htbp]
    \centering
    \caption{Benchmark frozen-core BvK-DLPNO-MP2 correlation energies, extrapolated to the complete PNO space (CPS) for various 2D (monolayer) and 3D materials. Unit cell geometries and XYZ coordinates are in the Supplementary Material.}
    \label{Tab:Data_2D_3D}
    \begin{ruledtabular}
    \begin{tabular}{llc d{0}d{0} d{5} d{5}d{5}d{5}d{5} }
    & & & \multicolumn{2}{c}{BvK supercell} & & \multicolumn{4}{c}{PNO-RI-MP2} \\
    \cline{4-5} \cline{7-10}
    & & \multicolumn{1}{c}{$k$-mesh} & \multicolumn{1}{c}{basis func.} & \multicolumn{1}{c}{act. orb.} & \multicolumn{1}{c}{RI-HF} & \multicolumn{1}{c}{6} & \multicolumn{1}{c}{7} & \multicolumn{1}{c}{8} & \multicolumn{1}{c}{CPS} \\
    \hline
    BN\footnote{Planar honeycomb, experimental lattice constant 2.501\,{\AA}.\cite{Meng_EJIC_2010_3174}}  
                              & pob-TZVP-rev2 & $ 5{\times} 5          $ &  900 & 100 &  -79.29060 & -0.21856 & -0.21927 & -0.21948 & -0.21958 \\
                              &               & $ 7{\times} 7          $ & 1764 & 196 &  -79.29075 & -0.21805 & -0.21872 & -0.21898 & -0.21910 \\
                              &               & $ 9{\times} 9          $ & 2916 & 324 &  -79.29075 & -0.21787 & -0.21854 & -0.21879 & -0.21891 \\
                              &               & $11{\times}11          $ & 4356 & 484 &  -79.29075 & -0.21779 & -0.21846 & -0.21871 & -0.21882 \\ \hline
    SiC\footnote{Planar honeycomb, lattice constant 3.07\,{\AA}.\cite{Meng_EJIC_2010_3174}}                       
                              & pob-TZVP      & $ 3{\times} 3          $ &  360 &  36 & -326.79555 & -0.20004 & -0.20044 & -0.20058 & -0.20065 \\
                              &               & $ 5{\times} 5          $ & 1000 & 100 & -326.79905 & -0.20101 & -0.20180 & -0.20202 & -0.20212 \\
                              &               & $ 7{\times} 7          $ & 1960 & 196 & -326.79928 & -0.20089 & -0.20171 & -0.20202 & -0.20216 \\
                              &               & $ 9{\times} 9          $ & 3240 & 324 & -326.79929 & -0.20086 & -0.20164 & -0.20193 & -0.20206 \\
                              &               & $11{\times}11          $ & 4840 & 484 & -326.79929 & -0.20081 & -0.20161 & -0.20189 & -0.20203 \\ \hline
    MoS$_2$\footnote{Puckered honeycomb, lattice constant 3.190\,{\AA} (Materials Project; mp-2815);\cite{Jain_APLM_2013_011002} effective core potential for Mo.}
                              & pob-TZVP-rev2 & $ 3{\times} 3          $ &  711 & 108 & -863.92030 & -0.96718 & -0.97253 & -0.97477 &          \\
                              &               & $ 5{\times} 5          $ & 1975 & 300 & -863.92353 & -0.93369 & -0.94275 &          &          \\
                              &               & $ 7{\times} 7          $ & 3871 & 588 & -863.92352 & -0.92956 &          &          &          \\ \hline
    LiH\footnote{Rock salt type (rhombohedral), experimental cubic lattice constant 4.084\,{\AA}.\cite{Pretzel_JPCS_1960_10}}                       
                              & pob-TZVP      & $ 5{\times} 5{\times}5 $ & 1625 & 125 &   -8.06060 & -0.03066 & -0.03075 & -0.03078 & -0.03079 \\
                              &               & $ 7{\times} 7{\times}7 $ & 4459 & 343 &   -8.06059 & -0.03083 & -0.03093 & -0.03096 & -0.03098 \\ \hline
    MgO\footnote{Rock salt type (rhombohedral), cubic lattice constant 4.339\,{\AA}; frozen Ne core for Mg.}                       
                              & pob-TZVP      & $ 3{\times} 3{\times}3 $ &  999 & 108 & -274.68268 & -0.19631 & -0.19651 & -0.19660 & -0.19664 \\
                              &               & $ 5{\times} 5{\times}5 $ & 4625 & 500 & -274.68452 & -0.19853 & -0.19882 & -0.19896 & -0.19902 \\ \hline
    MgO\footnote{Frozen $1s^22s^2$ core for Mg.}
                              & pob-TZVP      & $ 3{\times} 3{\times}3 $ &  999 & 189 & -274.68268 & -0.25935 & -0.25976 & -0.25991 & -0.25998 \\
                              &               & $ 5{\times} 5{\times}5 $ & 4625 & 875 & -274.68452 & -0.26164 & -0.26219 & -0.26243 & -0.26253 \\ \hline
    Ice\footnote{Hexagonal ice (Materials Project; mp-697111).\cite{Jain_APLM_2013_011002}}                
                              & pob-TZVP      & $ 3{\times} 3{\times}3 $ & 3240 & 432 & -304.26197 & -0.74491 & -0.74646 & -0.74713 & -0.74743 \\
    \end{tabular}
    \end{ruledtabular}
\end{table*}

To further demonstrate the capabilities of our current pilot BvK-DLPNO-MP2 implementation and provide reference data for future benchmarks, Table~\ref{Tab:Data_2D_3D} shows DLPNO-MP2 unit cell energies for various 2D and 3D systems. All calculations were performed on a dual-socket workstation equipped with two Intel(R) Xeon(R) Gold 6248R CPUs (48 total cores), 386\,GB of RAM and 1.8\,TB disk within a few days of wall-time. 

In addition to hexagonal BN (see also $\mathcal{T}_\mathrm{PNO}=10^{-9}$ data in the Supplementary Material), we have computed the correlation energies for SiC and MoS$_2$ monolayers. SiC forms a planar honeycomb lattice but exhibits more polar bonds than BN. MoS$_2$ monolayers are promising candidates for future optoelectronic applications,\cite{Li_JM_2015_33, Ganatra_ACSN_2014_4074} and form puckered honeycombs.\cite{Sahin_PRB_2009_155453}
For monolayer MoS$_2$, we were only able to obtain a full PNO series for a $3\times 3$  mesh whereas tight PNO thresholds for larger supercell sizes ran out of disk space. Since the current implementation does not yet make use of translational symmetry in the integral evaluation, three-center MO integrals over the whole BvK supercell must be stored on disk, which constitutes a significant bottleneck on our machines.

LiH and MgO both crystallize in the rock salt structure and have been benchmarked extensively in the context of LCAO-HF and MP2.\cite{Irmler_JCTC_2018_4567, Usvyat_JCP_2011_214105} To diversify the benchmark set beyond materials, we also include hexagonal ice which is a prototypical example for hydrogen bonding, proton tunnelling and cooperativity in molecular crystals.\cite{Maschio_CEC_2010_2429} In all cases, the finite-size error can be reduced to within chemical accuracy except for monolayer MoS$_2$, which emerges as a promising benchmark system.

\section{Conclusion}

Wavefunction-based many-body correlation theory provides a systematically improvable hierarchy of methods that can be used to independently verify and benchmark the accuracy of density functional predictions of electronic energies and properties of molecules and materials. The steep scaling of computational cost with system size presents a severe challenge when applying correlated wavefunction theories to periodic systems, where large simulation cells are required to capture all the relevant correlation lengthscales. We are therefore developing LCAO-based local correlation theory for periodic systems as a solution to this problem, at least for insulating materials. Domain-based pair natural orbital methods (DLPNO) have transformed the range of applicability of accurate coupled-cluster methods by using local approximations to reduce the cost to scale asymptotically only linearly with system size. The goal of our work is to develop the corresponding theory for periodic systems, where we aim to leverage large parts of the existing efficient molecular implementation in the Turbomole program package.

In this contribution, we have presented the periodic generalisation of DLPNO-MP2 theory using Born--von K{\'a}rm{\'a}n periodic boundary conditions. DLPNO-MP2 is the necessary first step towards DLPNO-CCSD(T) theory and DLPNO-CC3 theory for excited states. We have provided the complete set of equations for obtaining periodic PNOs for a pair of occupied Wannier functions, using periodic PAOs and OSVs as intermediaries via the PAO-OSV-PNO cascade. This work therefore generalises the earlier periodic PAO-based OSV-MP2 method introduced by Usvyat {\it et al}.\cite{Usvyat_JCP_2015_102805}. The PNO approximations used are completely analogous to the molecular scheme and our calculations comparing unit cell energies computed from periodic calculations and molecular fragment calculations demonstrate that the PNO truncation errors are entirely equivalent. This makes it possible to straightforwardly combine results from molecular and periodic DLPNO calculations when studying molecular insertion in porous solids, or surface adsorption processes. 

 The primary difficulty for electronic structure methods under BvK conditions lies in the proper treatment of the infinite lattice sum of the electron repulsion integrals. We have used a chargeless local density fitting method that adds dipole-correction to cancel the surface energy and ensure convergent lattice sums for the two- and three-center ERIs. The method is numerically stable and the convergence with the extend of the lattice summation was sufficiently rapid that we could apply direct lattice summation without the aid of fast multipole techniques. Future incorporation of such methods would undoubtedly reduce the computational cost since the multipole lattice summation can be brought outside the loop over AOs. Integral evaluation is currently the main computational bottleneck of our current implementation. A scheme that uses maximises the use of translational symmetry for the integrals to minimise the computational effort has yet to be worked out. Nevertheless, even with our pilot implementation, we were able to apply the BvK-DLPNO-MP2 method with a pob-TZVP basis to compute the correlation energies of 3D crystals of LiH, MgO and an ice polymorph, near the thermodynamic limit.

In paper II of this series,\cite{paper_mega} we have presented an complementary strategy for treating crystalline materials, where supercell is embedded in a megacell, and rigorous translational symmetry is imposed for all Hamiltonian integrals and wavefunction parameters. This alternative approach avoids the complications arising from the lattice summation over periodic images in the integral evaluation. Encouragingly, we find that both schemes converge to the same thermodynamic limit for every PNO truncation threshold, underlining the stability of the PNO local correlation approximations. A more detailed comparison of the relative merits of the two approaches will be the subject of future work.

\section*{Acknowledgments}

We express our sincere thanks to Dr Denis Usvyat for providing benchmark thermodynamic limit MP2 energies computed using the CRYSCOR scheme.
AN gratefully acknowledges funding through a Walter Benjamin Fellowship by the Deutsche Forschungsgemeinschaft (DFG, German Research Foundation) -- 517466522. Financial support for AZ from the University of Oxford and Turbomole GmbH is gratefully acknowledged.

\bibliography{literature.bib}

\begin{thebibliography}{128}%
\makeatletter
\providecommand \@ifxundefined [1]{%
 \@ifx{#1\undefined}
}%
\providecommand \@ifnum [1]{%
 \ifnum #1\expandafter \@firstoftwo
 \else \expandafter \@secondoftwo
 \fi
}%
\providecommand \@ifx [1]{%
 \ifx #1\expandafter \@firstoftwo
 \else \expandafter \@secondoftwo
 \fi
}%
\providecommand \natexlab [1]{#1}%
\providecommand \enquote  [1]{``#1''}%
\providecommand \bibnamefont  [1]{#1}%
\providecommand \bibfnamefont [1]{#1}%
\providecommand \citenamefont [1]{#1}%
\providecommand \href@noop [0]{\@secondoftwo}%
\providecommand \href [0]{\begingroup \@sanitize@url \@href}%
\providecommand \@href[1]{\@@startlink{#1}\@@href}%
\providecommand \@@href[1]{\endgroup#1\@@endlink}%
\providecommand \@sanitize@url [0]{\catcode `\\12\catcode `\$12\catcode
  `\&12\catcode `\#12\catcode `\^12\catcode `\_12\catcode `\%12\relax}%
\providecommand \@@startlink[1]{}%
\providecommand \@@endlink[0]{}%
\providecommand \url  [0]{\begingroup\@sanitize@url \@url }%
\providecommand \@url [1]{\endgroup\@href {#1}{\urlprefix }}%
\providecommand \urlprefix  [0]{URL }%
\providecommand \Eprint [0]{\href }%
\providecommand \doibase [0]{http://dx.doi.org/}%
\providecommand \selectlanguage [0]{\@gobble}%
\providecommand \bibinfo  [0]{\@secondoftwo}%
\providecommand \bibfield  [0]{\@secondoftwo}%
\providecommand \translation [1]{[#1]}%
\providecommand \BibitemOpen [0]{}%
\providecommand \bibitemStop [0]{}%
\providecommand \bibitemNoStop [0]{.\EOS\space}%
\providecommand \EOS [0]{\spacefactor3000\relax}%
\providecommand \BibitemShut  [1]{\csname bibitem#1\endcsname}%
\let\auto@bib@innerbib\@empty
\bibitem [{\citenamefont {Kratzer}\ and\ \citenamefont
  {Neugebauer}(2019)}]{Kratzer_FC_2019_106}%
  \BibitemOpen
  \bibfield  {author} {\bibinfo {author} {\bibfnamefont {Peter}\ \bibnamefont
  {Kratzer}}\ and\ \bibinfo {author} {\bibfnamefont {J{\"o}rg}\ \bibnamefont
  {Neugebauer}},\ }\bibfield  {title} {\enquote {\bibinfo {title} {The basics
  of electronic structure theory for periodic systems},}\ }\href {\doibase
  10.3389/fchem.2019.00106} {\bibfield  {journal} {\bibinfo  {journal} {Front.
  Chem.}\ }\textbf {\bibinfo {volume} {7}},\ \bibinfo {pages} {106} (\bibinfo
  {year} {2019})}\BibitemShut {NoStop}%
\bibitem [{\citenamefont {Martin}(2020)}]{Martin_2020}%
  \BibitemOpen
  \bibfield  {author} {\bibinfo {author} {\bibfnamefont {R.M.}\ \bibnamefont
  {Martin}},\ }\href@noop {} {\emph {\bibinfo {title} {Electronic Structure:
  Basic Theory and Practical Methods}}},\ \bibinfo {edition} {2nd}\ ed.\
  (\bibinfo  {publisher} {Cambridge University Press},\ \bibinfo {address}
  {Cambridge},\ \bibinfo {year} {2020})\BibitemShut {NoStop}%
\bibitem [{\citenamefont {Pisani}\ \emph {et~al.}(2005)\citenamefont {Pisani},
  \citenamefont {Busso}, \citenamefont {Capecchi}, \citenamefont {Casassa},
  \citenamefont {Dovesi}, \citenamefont {Maschio}, \citenamefont
  {Zicovich-Wilson},\ and\ \citenamefont {Sch{\" u}tz}}]{Pisani_JCP_2005}%
  \BibitemOpen
  \bibfield  {author} {\bibinfo {author} {\bibfnamefont {C.}~\bibnamefont
  {Pisani}}, \bibinfo {author} {\bibfnamefont {M.}~\bibnamefont {Busso}},
  \bibinfo {author} {\bibfnamefont {G.}~\bibnamefont {Capecchi}}, \bibinfo
  {author} {\bibfnamefont {S.}~\bibnamefont {Casassa}}, \bibinfo {author}
  {\bibfnamefont {R.}~\bibnamefont {Dovesi}}, \bibinfo {author} {\bibfnamefont
  {L.}~\bibnamefont {Maschio}}, \bibinfo {author} {\bibfnamefont
  {C.}~\bibnamefont {Zicovich-Wilson}}, \ and\ \bibinfo {author} {\bibfnamefont
  {M.}~\bibnamefont {Sch{\" u}tz}},\ }\bibfield  {title} {\enquote {\bibinfo
  {title} {Local-mp2 electron correlation method for nonconducting crystals},}\
  }\href {\doibase 10.1063/1.1857479} {\bibfield  {journal} {\bibinfo
  {journal} {J. Chem. Phys.}\ }\textbf {\bibinfo {volume} {122}},\ \bibinfo
  {pages} {094113} (\bibinfo {year} {2005})}\BibitemShut {NoStop}%
\bibitem [{\citenamefont {Pisani}\ \emph {et~al.}(2012)\citenamefont {Pisani},
  \citenamefont {Sch{\" u}tz}, \citenamefont {Casassa}, \citenamefont {Usvyat},
  \citenamefont {Maschio}, \citenamefont {Lorenz},\ and\ \citenamefont
  {Erba}}]{Pisani_PCCP_2012_7615}%
  \BibitemOpen
  \bibfield  {author} {\bibinfo {author} {\bibfnamefont {Cesare}\ \bibnamefont
  {Pisani}}, \bibinfo {author} {\bibfnamefont {Martin}\ \bibnamefont {Sch{\"
  u}tz}}, \bibinfo {author} {\bibfnamefont {Silvia}\ \bibnamefont {Casassa}},
  \bibinfo {author} {\bibfnamefont {Denis}\ \bibnamefont {Usvyat}}, \bibinfo
  {author} {\bibfnamefont {Lorenzo}\ \bibnamefont {Maschio}}, \bibinfo {author}
  {\bibfnamefont {Marco}\ \bibnamefont {Lorenz}}, \ and\ \bibinfo {author}
  {\bibfnamefont {Alessandro}\ \bibnamefont {Erba}},\ }\bibfield  {title}
  {\enquote {\bibinfo {title} {Cryscor: a program for the post-hartree--fock
  treatment of periodic systems},}\ }\href {\doibase 10.1039/C2CP23927B}
  {\bibfield  {journal} {\bibinfo  {journal} {Phys. Chem. Chem. Phys.}\
  }\textbf {\bibinfo {volume} {14}},\ \bibinfo {pages} {7615--7628} (\bibinfo
  {year} {2012})}\BibitemShut {NoStop}%
\bibitem [{\citenamefont {{Del Ben}}\ \emph {et~al.}(2012)\citenamefont {{Del
  Ben}}, \citenamefont {Hutter},\ and\ \citenamefont
  {VandeVondele}}]{DelBen_JCTC_2012}%
  \BibitemOpen
  \bibfield  {author} {\bibinfo {author} {\bibfnamefont {Mauro}\ \bibnamefont
  {{Del Ben}}}, \bibinfo {author} {\bibfnamefont {J{\"u}rg}\ \bibnamefont
  {Hutter}}, \ and\ \bibinfo {author} {\bibfnamefont {Joost}\ \bibnamefont
  {VandeVondele}},\ }\bibfield  {title} {\enquote {\bibinfo {title}
  {Second-order m{\o}ller--plesset perturbation theory in the condensed phase:
  An efficient and massively parallel gaussian and plane waves approach},}\
  }\href {\doibase 10.1021/ct300531w} {\bibfield  {journal} {\bibinfo
  {journal} {J. Chem. Theory Comput.}\ }\textbf {\bibinfo {volume} {8}},\
  \bibinfo {pages} {4177--4188} (\bibinfo {year} {2012})}\BibitemShut {NoStop}%
\bibitem [{\citenamefont {Sch{\" a}fer}\ \emph {et~al.}(2017)\citenamefont
  {Sch{\" a}fer}, \citenamefont {Ramberger},\ and\ \citenamefont
  {Kresse}}]{Schaefer_JCP_2017_104101}%
  \BibitemOpen
  \bibfield  {author} {\bibinfo {author} {\bibfnamefont {Tobias}\ \bibnamefont
  {Sch{\" a}fer}}, \bibinfo {author} {\bibfnamefont {Benjamin}\ \bibnamefont
  {Ramberger}}, \ and\ \bibinfo {author} {\bibfnamefont {Georg}\ \bibnamefont
  {Kresse}},\ }\bibfield  {title} {\enquote {\bibinfo {title} {Quartic scaling
  {MP2} for solids: A highly parallelized algorithm in the plane wave basis},}\
  }\href {\doibase 10.1063/1.4976937} {\bibfield  {journal} {\bibinfo
  {journal} {J. Chem. Phys.}\ }\textbf {\bibinfo {volume} {146}},\ \bibinfo
  {pages} {104101} (\bibinfo {year} {2017})}\BibitemShut {NoStop}%
\bibitem [{\citenamefont {Bintrim}\ \emph {et~al.}(2022)\citenamefont
  {Bintrim}, \citenamefont {Berkelbach},\ and\ \citenamefont
  {Ye}}]{Bintrim_JCTC_5374_2022}%
  \BibitemOpen
  \bibfield  {author} {\bibinfo {author} {\bibfnamefont {Sylvia~J.}\
  \bibnamefont {Bintrim}}, \bibinfo {author} {\bibfnamefont {Timothy~C.}\
  \bibnamefont {Berkelbach}}, \ and\ \bibinfo {author} {\bibfnamefont
  {Hong-Zhou}\ \bibnamefont {Ye}},\ }\bibfield  {title} {\enquote {\bibinfo
  {title} {Integral-direct hartree--fock and m{\o}ller--{P}lesset perturbation
  theory for periodic systems with density fitting: Application to the benzene
  crystal},}\ }\href {\doibase 10.1021/acs.jctc.2c00640} {\bibfield  {journal}
  {\bibinfo  {journal} {J. Chem. Theory Comput.}\ }\textbf {\bibinfo {volume}
  {18}},\ \bibinfo {pages} {5374--5381} (\bibinfo {year} {2022})}\BibitemShut
  {NoStop}%
\bibitem [{\citenamefont {Gruber}\ \emph {et~al.}(2018)\citenamefont {Gruber},
  \citenamefont {Liao}, \citenamefont {Tsatsoulis}, \citenamefont {Hummel},\
  and\ \citenamefont {Gr{\" u}neis}}]{Gruber_PRX_2018_021043}%
  \BibitemOpen
  \bibfield  {author} {\bibinfo {author} {\bibfnamefont {Thomas}\ \bibnamefont
  {Gruber}}, \bibinfo {author} {\bibfnamefont {Ke}~\bibnamefont {Liao}},
  \bibinfo {author} {\bibfnamefont {Theodoros}\ \bibnamefont {Tsatsoulis}},
  \bibinfo {author} {\bibfnamefont {Felix}\ \bibnamefont {Hummel}}, \ and\
  \bibinfo {author} {\bibfnamefont {Andreas}\ \bibnamefont {Gr{\" u}neis}},\
  }\bibfield  {title} {\enquote {\bibinfo {title} {Applying the coupled-cluster
  ansatz to solids and surfaces in the thermodynamic limit},}\ }\href {\doibase
  10.1103/PhysRevX.8.021043} {\bibfield  {journal} {\bibinfo  {journal} {Phys.
  Rev. X}\ }\textbf {\bibinfo {volume} {8}},\ \bibinfo {pages} {021043}
  (\bibinfo {year} {2018})}\BibitemShut {NoStop}%
\bibitem [{\citenamefont {Carbone}\ \emph {et~al.}(2024)\citenamefont
  {Carbone}, \citenamefont {Irmler}, \citenamefont {Gallo}, \citenamefont
  {Sch{\"a}fer}, \citenamefont {Van~Benschoten}, \citenamefont {Shepherd},\
  and\ \citenamefont {Gr{\"u}neis}}]{Carbone_FD_2024_586}%
  \BibitemOpen
  \bibfield  {author} {\bibinfo {author} {\bibfnamefont {Johanna~P.}\
  \bibnamefont {Carbone}}, \bibinfo {author} {\bibfnamefont {Andreas}\
  \bibnamefont {Irmler}}, \bibinfo {author} {\bibfnamefont {Alejandro}\
  \bibnamefont {Gallo}}, \bibinfo {author} {\bibfnamefont {Tobias}\
  \bibnamefont {Sch{\"a}fer}}, \bibinfo {author} {\bibfnamefont {William~Z.}\
  \bibnamefont {Van~Benschoten}}, \bibinfo {author} {\bibfnamefont {James~J.}\
  \bibnamefont {Shepherd}}, \ and\ \bibinfo {author} {\bibfnamefont {Andreas}\
  \bibnamefont {Gr{\"u}neis}},\ }\bibfield  {title} {\enquote {\bibinfo {title}
  {{CO} adsorption on pt(111) studied by periodic coupled cluster theory},}\
  }\href {\doibase 10.1039/D4FD00085D} {\bibfield  {journal} {\bibinfo
  {journal} {Faraday Discuss.}\ }\textbf {\bibinfo {volume} {254}},\ \bibinfo
  {pages} {586--597} (\bibinfo {year} {2024})}\BibitemShut {NoStop}%
\bibitem [{\citenamefont {Ye}\ and\ \citenamefont
  {Berkelbach}(2024{\natexlab{a}})}]{Ye_FD_2024_628}%
  \BibitemOpen
  \bibfield  {author} {\bibinfo {author} {\bibfnamefont {Hong-Zhou}\
  \bibnamefont {Ye}}\ and\ \bibinfo {author} {\bibfnamefont {Timothy~C.}\
  \bibnamefont {Berkelbach}},\ }\bibfield  {title} {\enquote {\bibinfo {title}
  {Adsorption and vibrational spectroscopy of {CO} on the surface of {MgO} from
  periodic local coupled-cluster theory},}\ }\href {\doibase
  10.1039/D4FD00041B} {\bibfield  {journal} {\bibinfo  {journal} {Faraday
  Discuss.}\ }\textbf {\bibinfo {volume} {254}},\ \bibinfo {pages} {628--640}
  (\bibinfo {year} {2024}{\natexlab{a}})}\BibitemShut {NoStop}%
\bibitem [{\citenamefont {Maschio}\ \emph {et~al.}(2007)\citenamefont
  {Maschio}, \citenamefont {Usvyat}, \citenamefont {Manby}, \citenamefont
  {Casassa}, \citenamefont {Pisani},\ and\ \citenamefont
  {Sch{\"u}tz}}]{Maschio_PRB_2007_075101}%
  \BibitemOpen
  \bibfield  {author} {\bibinfo {author} {\bibfnamefont {Lorenzo}\ \bibnamefont
  {Maschio}}, \bibinfo {author} {\bibfnamefont {Denis}\ \bibnamefont {Usvyat}},
  \bibinfo {author} {\bibfnamefont {Frederick~R.}\ \bibnamefont {Manby}},
  \bibinfo {author} {\bibfnamefont {Silvia}\ \bibnamefont {Casassa}}, \bibinfo
  {author} {\bibfnamefont {Cesare}\ \bibnamefont {Pisani}}, \ and\ \bibinfo
  {author} {\bibfnamefont {Martin}\ \bibnamefont {Sch{\"u}tz}},\ }\bibfield
  {title} {\enquote {\bibinfo {title} {Fast local-{MP2} method with
  density-fitting for crystals. {I}. theory and algorithms},}\ }\href {\doibase
  10.1103/PhysRevB.76.075101} {\bibfield  {journal} {\bibinfo  {journal} {Phys.
  Rev. B}\ }\textbf {\bibinfo {volume} {76}},\ \bibinfo {pages} {075101}
  (\bibinfo {year} {2007})}\BibitemShut {NoStop}%
\bibitem [{\citenamefont {Usvyat}\ \emph {et~al.}(2007)\citenamefont {Usvyat},
  \citenamefont {Maschio}, \citenamefont {Manby}, \citenamefont {Casassa},
  \citenamefont {Sch{\"u}tz},\ and\ \citenamefont
  {Pisani}}]{Usvyat_PRB_2007_075102}%
  \BibitemOpen
  \bibfield  {author} {\bibinfo {author} {\bibfnamefont {Denis}\ \bibnamefont
  {Usvyat}}, \bibinfo {author} {\bibfnamefont {Lorenzo}\ \bibnamefont
  {Maschio}}, \bibinfo {author} {\bibfnamefont {Frederick~R.}\ \bibnamefont
  {Manby}}, \bibinfo {author} {\bibfnamefont {Silvia}\ \bibnamefont {Casassa}},
  \bibinfo {author} {\bibfnamefont {Martin}\ \bibnamefont {Sch{\"u}tz}}, \ and\
  \bibinfo {author} {\bibfnamefont {Cesare}\ \bibnamefont {Pisani}},\
  }\bibfield  {title} {\enquote {\bibinfo {title} {Fast local-{MP2} method with
  density-fitting for crystals. {II}. test calculations and application to the
  carbon dioxide crystal},}\ }\href {\doibase 10.1103/PhysRevB.76.075102}
  {\bibfield  {journal} {\bibinfo  {journal} {Phys. Rev. B}\ }\textbf {\bibinfo
  {volume} {76}},\ \bibinfo {pages} {075102} (\bibinfo {year}
  {2007})}\BibitemShut {NoStop}%
\bibitem [{\citenamefont {McClain}\ \emph {et~al.}(2017)\citenamefont
  {McClain}, \citenamefont {Sun}, \citenamefont {Chan},\ and\ \citenamefont
  {Berkelbach}}]{McClain_JCTC_2017_1209}%
  \BibitemOpen
  \bibfield  {author} {\bibinfo {author} {\bibfnamefont {James}\ \bibnamefont
  {McClain}}, \bibinfo {author} {\bibfnamefont {Qiming}\ \bibnamefont {Sun}},
  \bibinfo {author} {\bibfnamefont {Garnet Kin-Lic}\ \bibnamefont {Chan}}, \
  and\ \bibinfo {author} {\bibfnamefont {Timothy~C.}\ \bibnamefont
  {Berkelbach}},\ }\bibfield  {title} {\enquote {\bibinfo {title}
  {Gaussian-based coupled-cluster theory for the ground-state and band
  structure of solids},}\ }\href {\doibase 10.1021/acs.jctc.7b00049} {\bibfield
   {journal} {\bibinfo  {journal} {J. Chem. Theory Comput.}\ }\textbf {\bibinfo
  {volume} {13}},\ \bibinfo {pages} {1209--1218} (\bibinfo {year}
  {2017})}\BibitemShut {NoStop}%
\bibitem [{\citenamefont {Hirata}\ and\ \citenamefont
  {Shimazaki}(2009)}]{Hirata_PRB_2009_085118}%
  \BibitemOpen
  \bibfield  {author} {\bibinfo {author} {\bibfnamefont {So}~\bibnamefont
  {Hirata}}\ and\ \bibinfo {author} {\bibfnamefont {Tomomi}\ \bibnamefont
  {Shimazaki}},\ }\bibfield  {title} {\enquote {\bibinfo {title} {Fast
  second-order many-body perturbation method for extended systems},}\ }\href
  {\doibase 10.1103/PhysRevB.80.085118} {\bibfield  {journal} {\bibinfo
  {journal} {Phys. Rev. B}\ }\textbf {\bibinfo {volume} {80}},\ \bibinfo
  {pages} {085118} (\bibinfo {year} {2009})}\BibitemShut {NoStop}%
\bibitem [{\citenamefont {Hirata}\ \emph {et~al.}(2004)\citenamefont {Hirata},
  \citenamefont {Podeszwa}, \citenamefont {Tobita},\ and\ \citenamefont
  {Bartlett}}]{Hirata_JCP_2004_2581}%
  \BibitemOpen
  \bibfield  {author} {\bibinfo {author} {\bibfnamefont {So}~\bibnamefont
  {Hirata}}, \bibinfo {author} {\bibfnamefont {Rafa{\l}}\ \bibnamefont
  {Podeszwa}}, \bibinfo {author} {\bibfnamefont {Motoi}\ \bibnamefont
  {Tobita}}, \ and\ \bibinfo {author} {\bibfnamefont {Rodney~J.}\ \bibnamefont
  {Bartlett}},\ }\bibfield  {title} {\enquote {\bibinfo {title}
  {Coupled-cluster singles and doubles for extended systems},}\ }\href
  {\doibase 10.1063/1.1637577} {\bibfield  {journal} {\bibinfo  {journal} {J.
  Chem. Phys.}\ }\textbf {\bibinfo {volume} {120}},\ \bibinfo {pages}
  {2581--2592} (\bibinfo {year} {2004})}\BibitemShut {NoStop}%
\bibitem [{\citenamefont {Haritan}\ \emph {et~al.}(2025)\citenamefont
  {Haritan}, \citenamefont {Wang},\ and\ \citenamefont
  {Goldzak}}]{Haritan_arxiv_2025.20482}%
  \BibitemOpen
  \bibfield  {author} {\bibinfo {author} {\bibfnamefont {Idan}\ \bibnamefont
  {Haritan}}, \bibinfo {author} {\bibfnamefont {Xiao}\ \bibnamefont {Wang}}, \
  and\ \bibinfo {author} {\bibfnamefont {Tamar}\ \bibnamefont {Goldzak}},\
  }\href {https://arxiv.org/abs/2503.20482} {\enquote {\bibinfo {title} {An
  efficient scaled opposite-spin {MP2} method for periodic systems},}\ }
  (\bibinfo {year} {2025}),\ \Eprint {http://arxiv.org/abs/2503.20482}
  {arXiv:2503.20482 [physics.chem-ph]} \BibitemShut {NoStop}%
\bibitem [{\citenamefont {Ayala}\ \emph {et~al.}(2001)\citenamefont {Ayala},
  \citenamefont {Kudin},\ and\ \citenamefont {Scuseria}}]{Ayala_JCP_2001_9698}%
  \BibitemOpen
  \bibfield  {author} {\bibinfo {author} {\bibfnamefont {Philippe~Y.}\
  \bibnamefont {Ayala}}, \bibinfo {author} {\bibfnamefont {Konstantin~N.}\
  \bibnamefont {Kudin}}, \ and\ \bibinfo {author} {\bibfnamefont {Gustavo~E.}\
  \bibnamefont {Scuseria}},\ }\bibfield  {title} {\enquote {\bibinfo {title}
  {Atomic orbital {L}aplace-transformed second-order {M}{\o}ller-{P}lesset
  theory for periodic systems},}\ }\href {\doibase 10.1063/1.1414369}
  {\bibfield  {journal} {\bibinfo  {journal} {J. Chem. Phys.}\ }\textbf
  {\bibinfo {volume} {115}},\ \bibinfo {pages} {9698--9707} (\bibinfo {year}
  {2001})}\BibitemShut {NoStop}%
\bibitem [{\citenamefont {Ye}\ and\ \citenamefont
  {Berkelbach}(2024{\natexlab{b}})}]{Ye_JCTC_2024_8948}%
  \BibitemOpen
  \bibfield  {author} {\bibinfo {author} {\bibfnamefont {Hong-Zhou}\
  \bibnamefont {Ye}}\ and\ \bibinfo {author} {\bibfnamefont {Timothy~C.}\
  \bibnamefont {Berkelbach}},\ }\bibfield  {title} {\enquote {\bibinfo {title}
  {Periodic local {C}oupled-{C}luster theory for insulators and metals},}\
  }\href {\doibase 10.1021/acs.jctc.4c00936} {\bibfield  {journal} {\bibinfo
  {journal} {J. Chemi. Theory Comput.}\ }\textbf {\bibinfo {volume} {20}},\
  \bibinfo {pages} {8948--8959} (\bibinfo {year}
  {2024}{\natexlab{b}})}\BibitemShut {NoStop}%
\bibitem [{\citenamefont {Katouda}\ and\ \citenamefont
  {Nagase}(2010)}]{Katouda_JCP_2010_184103}%
  \BibitemOpen
  \bibfield  {author} {\bibinfo {author} {\bibfnamefont {Michio}\ \bibnamefont
  {Katouda}}\ and\ \bibinfo {author} {\bibfnamefont {Shigeru}\ \bibnamefont
  {Nagase}},\ }\bibfield  {title} {\enquote {\bibinfo {title} {Application of
  second-order m{\o}ller-plesset perturbation theory with
  resolution-of-identity approximation to periodic systems},}\ }\href {\doibase
  10.1063/1.3503153} {\bibfield  {journal} {\bibinfo  {journal} {The Journal of
  Chemical Physics}\ }\textbf {\bibinfo {volume} {133}},\ \bibinfo {pages}
  {184103} (\bibinfo {year} {2010})}\BibitemShut {NoStop}%
\bibitem [{\citenamefont {Booth}\ \emph {et~al.}(2013)\citenamefont {Booth},
  \citenamefont {Gr{\"u}neis}, \citenamefont {Kresse},\ and\ \citenamefont
  {Alavi}}]{Booth_Nature_2013_7432}%
  \BibitemOpen
  \bibfield  {author} {\bibinfo {author} {\bibfnamefont {George~H}\
  \bibnamefont {Booth}}, \bibinfo {author} {\bibfnamefont {Andreas}\
  \bibnamefont {Gr{\"u}neis}}, \bibinfo {author} {\bibfnamefont {Georg}\
  \bibnamefont {Kresse}}, \ and\ \bibinfo {author} {\bibfnamefont {Ali}\
  \bibnamefont {Alavi}},\ }\bibfield  {title} {\enquote {\bibinfo {title}
  {Towards an exact description of electronic wavefunctions in real solids},}\
  }\href {\doibase 10.1038/nature11770} {\bibfield  {journal} {\bibinfo
  {journal} {Nature}\ }\textbf {\bibinfo {volume} {493}},\ \bibinfo {pages}
  {365--370} (\bibinfo {year} {2013})}\BibitemShut {NoStop}%
\bibitem [{\citenamefont {Zhang}\ and\ \citenamefont
  {Gr{\"u}neis}(2019)}]{Zhang_FM_2019_123}%
  \BibitemOpen
  \bibfield  {author} {\bibinfo {author} {\bibfnamefont {Igor~Ying}\
  \bibnamefont {Zhang}}\ and\ \bibinfo {author} {\bibfnamefont {Andreas}\
  \bibnamefont {Gr{\"u}neis}},\ }\bibfield  {title} {\enquote {\bibinfo {title}
  {Coupled cluster theory in materials science},}\ }\href {\doibase
  10.3389/fmats.2019.00123} {\bibfield  {journal} {\bibinfo  {journal} {Front.
  Mater.}\ }\textbf {\bibinfo {volume} {6}},\ \bibinfo {pages} {123} (\bibinfo
  {year} {2019})}\BibitemShut {NoStop}%
\bibitem [{\citenamefont {Goldzak}\ \emph {et~al.}(2022)\citenamefont
  {Goldzak}, \citenamefont {Wang}, \citenamefont {Ye},\ and\ \citenamefont
  {Berkelbach}}]{Goldzak_JCP_2022_174112}%
  \BibitemOpen
  \bibfield  {author} {\bibinfo {author} {\bibfnamefont {Tamar}\ \bibnamefont
  {Goldzak}}, \bibinfo {author} {\bibfnamefont {Xiao}\ \bibnamefont {Wang}},
  \bibinfo {author} {\bibfnamefont {Hong-Zhou}\ \bibnamefont {Ye}}, \ and\
  \bibinfo {author} {\bibfnamefont {Timothy~C.}\ \bibnamefont {Berkelbach}},\
  }\bibfield  {title} {\enquote {\bibinfo {title} {Accurate thermochemistry of
  covalent and ionic solids from spin-component-scaled {MP2}},}\ }\href
  {\doibase 10.1063/5.0119633} {\bibfield  {journal} {\bibinfo  {journal} {J.
  Chem. Phys.}\ }\textbf {\bibinfo {volume} {157}},\ \bibinfo {pages} {174112}
  (\bibinfo {year} {2022})}\BibitemShut {NoStop}%
\bibitem [{\citenamefont {Gr{\"u}neis}\ \emph {et~al.}(2011)\citenamefont
  {Gr{\"u}neis}, \citenamefont {Booth}, \citenamefont {Marsman}, \citenamefont
  {Spencer}, \citenamefont {Alavi},\ and\ \citenamefont
  {Kresse}}]{Grueneis_JCTC_2011_2780}%
  \BibitemOpen
  \bibfield  {author} {\bibinfo {author} {\bibfnamefont {Andreas}\ \bibnamefont
  {Gr{\"u}neis}}, \bibinfo {author} {\bibfnamefont {George~H.}\ \bibnamefont
  {Booth}}, \bibinfo {author} {\bibfnamefont {Martijn}\ \bibnamefont
  {Marsman}}, \bibinfo {author} {\bibfnamefont {James}\ \bibnamefont
  {Spencer}}, \bibinfo {author} {\bibfnamefont {Ali}\ \bibnamefont {Alavi}}, \
  and\ \bibinfo {author} {\bibfnamefont {Georg}\ \bibnamefont {Kresse}},\
  }\bibfield  {title} {\enquote {\bibinfo {title} {Natural orbitals for wave
  function based correlated calculations using a plane wave basis set},}\
  }\href {\doibase 10.1021/ct200263g} {\bibfield  {journal} {\bibinfo
  {journal} {J. Chem. Theory Comput.}\ }\textbf {\bibinfo {volume} {7}},\
  \bibinfo {pages} {2780--2785} (\bibinfo {year} {2011})}\BibitemShut {NoStop}%
\bibitem [{\citenamefont {Gr{\"u}neis}(2015)}]{Grueneis_JCP_2015_102817}%
  \BibitemOpen
  \bibfield  {author} {\bibinfo {author} {\bibfnamefont {Andreas}\ \bibnamefont
  {Gr{\"u}neis}},\ }\bibfield  {title} {\enquote {\bibinfo {title} {A coupled
  cluster and {M}{\o}ller-{P}lesset perturbation theory study of the pressure
  induced phase transition in the {LiH} crystal},}\ }\href {\doibase
  10.1063/1.4928645} {\bibfield  {journal} {\bibinfo  {journal} {J. Chem.
  Phys.}\ }\textbf {\bibinfo {volume} {143}},\ \bibinfo {pages} {102817}
  (\bibinfo {year} {2015})}\BibitemShut {NoStop}%
\bibitem [{\citenamefont {Shi}\ \emph {et~al.}(2023)\citenamefont {Shi},
  \citenamefont {Zen}, \citenamefont {Kapil}, \citenamefont {Nagy},
  \citenamefont {Gr{\"u}neis},\ and\ \citenamefont
  {Michaelides}}]{Shi_JACS_2023_25372}%
  \BibitemOpen
  \bibfield  {author} {\bibinfo {author} {\bibfnamefont {Benjamin~X.}\
  \bibnamefont {Shi}}, \bibinfo {author} {\bibfnamefont {Andrea}\ \bibnamefont
  {Zen}}, \bibinfo {author} {\bibfnamefont {Venkat}\ \bibnamefont {Kapil}},
  \bibinfo {author} {\bibfnamefont {P{\'e}ter~R.}\ \bibnamefont {Nagy}},
  \bibinfo {author} {\bibfnamefont {Andreas}\ \bibnamefont {Gr{\"u}neis}}, \
  and\ \bibinfo {author} {\bibfnamefont {Angelos}\ \bibnamefont
  {Michaelides}},\ }\bibfield  {title} {\enquote {\bibinfo {title} {Many-body
  methods for surface chemistry come of age: Achieving consensus with
  experiments},}\ }\href {\doibase 10.1021/jacs.3c09616} {\bibfield  {journal}
  {\bibinfo  {journal} {J. Am. Chem. Soc.}\ }\textbf {\bibinfo {volume}
  {145}},\ \bibinfo {pages} {25372--25381} (\bibinfo {year}
  {2023})}\BibitemShut {NoStop}%
\bibitem [{\citenamefont {Tsatsoulis}\ \emph {et~al.}(2017)\citenamefont
  {Tsatsoulis}, \citenamefont {Hummel}, \citenamefont {Usvyat}, \citenamefont
  {Sch{\"u}tz}, \citenamefont {Booth}, \citenamefont {Binnie}, \citenamefont
  {Gillan}, \citenamefont {Alfè}, \citenamefont {Michaelides},\ and\
  \citenamefont {Gr{\"u}neis}}]{Tsatsoulis_JCP_2017_204108}%
  \BibitemOpen
  \bibfield  {author} {\bibinfo {author} {\bibfnamefont {Theodoros}\
  \bibnamefont {Tsatsoulis}}, \bibinfo {author} {\bibfnamefont {Felix}\
  \bibnamefont {Hummel}}, \bibinfo {author} {\bibfnamefont {Denis}\
  \bibnamefont {Usvyat}}, \bibinfo {author} {\bibfnamefont {Martin}\
  \bibnamefont {Sch{\"u}tz}}, \bibinfo {author} {\bibfnamefont {George~H.}\
  \bibnamefont {Booth}}, \bibinfo {author} {\bibfnamefont {Simon~S.}\
  \bibnamefont {Binnie}}, \bibinfo {author} {\bibfnamefont {Michael~J.}\
  \bibnamefont {Gillan}}, \bibinfo {author} {\bibfnamefont {Dario}\
  \bibnamefont {Alfè}}, \bibinfo {author} {\bibfnamefont {Angelos}\
  \bibnamefont {Michaelides}}, \ and\ \bibinfo {author} {\bibfnamefont
  {Andreas}\ \bibnamefont {Gr{\"u}neis}},\ }\bibfield  {title} {\enquote
  {\bibinfo {title} {A comparison between quantum chemistry and quantum {M}onte
  {C}arlo techniques for the adsorption of water on the {(001)} {LiH}
  surface},}\ }\href {\doibase 10.1063/1.4984048} {\bibfield  {journal}
  {\bibinfo  {journal} {J. Chem. Phys.}\ }\textbf {\bibinfo {volume} {146}},\
  \bibinfo {pages} {204108} (\bibinfo {year} {2017})}\BibitemShut {NoStop}%
\bibitem [{\citenamefont {Alessio}\ \emph {et~al.}(2019)\citenamefont
  {Alessio}, \citenamefont {Usvyat},\ and\ \citenamefont
  {Sauer}}]{Alessio_JCTC_2019_1329}%
  \BibitemOpen
  \bibfield  {author} {\bibinfo {author} {\bibfnamefont {Maristella}\
  \bibnamefont {Alessio}}, \bibinfo {author} {\bibfnamefont {Denis}\
  \bibnamefont {Usvyat}}, \ and\ \bibinfo {author} {\bibfnamefont {Joachim}\
  \bibnamefont {Sauer}},\ }\bibfield  {title} {\enquote {\bibinfo {title}
  {Chemically accurate adsorption energies: {CO} and {H$_2$O} on the {MgO(001)}
  surface},}\ }\href {\doibase 10.1021/acs.jctc.8b01122} {\bibfield  {journal}
  {\bibinfo  {journal} {J. Chem. Theory Comput.}\ }\textbf {\bibinfo {volume}
  {15}},\ \bibinfo {pages} {1329--1344} (\bibinfo {year} {2019})}\BibitemShut
  {NoStop}%
\bibitem [{\citenamefont {Pisani}\ \emph {et~al.}(2008)\citenamefont {Pisani},
  \citenamefont {Maschio}, \citenamefont {Casassa}, \citenamefont {Halo},
  \citenamefont {Sch{\"u}tz},\ and\ \citenamefont
  {Usvyat}}]{Pisani_JCC_2008_2113}%
  \BibitemOpen
  \bibfield  {author} {\bibinfo {author} {\bibfnamefont {Cesare}\ \bibnamefont
  {Pisani}}, \bibinfo {author} {\bibfnamefont {Lorenzo}\ \bibnamefont
  {Maschio}}, \bibinfo {author} {\bibfnamefont {Silvia}\ \bibnamefont
  {Casassa}}, \bibinfo {author} {\bibfnamefont {Migen}\ \bibnamefont {Halo}},
  \bibinfo {author} {\bibfnamefont {Martin}\ \bibnamefont {Sch{\"u}tz}}, \ and\
  \bibinfo {author} {\bibfnamefont {Denis}\ \bibnamefont {Usvyat}},\ }\bibfield
   {title} {\enquote {\bibinfo {title} {Periodic local {MP2} method for the
  study of electronic correlation in crystals: Theory and preliminary
  applications},}\ }\href {\doibase 10.1002/jcc.20975} {\bibfield  {journal}
  {\bibinfo  {journal} {J. Comput. Chem.}\ }\textbf {\bibinfo {volume} {29}},\
  \bibinfo {pages} {2113--2124} (\bibinfo {year} {2008})}\BibitemShut {NoStop}%
\bibitem [{\citenamefont {Usvyat}(2013)}]{Usvyat_JCP_2013_194101}%
  \BibitemOpen
  \bibfield  {author} {\bibinfo {author} {\bibfnamefont {Denis}\ \bibnamefont
  {Usvyat}},\ }\bibfield  {title} {\enquote {\bibinfo {title} {Linear-scaling
  explicitly correlated treatment of solids: Periodic local {MP2-F12}
  method},}\ }\href {\doibase 10.1063/1.4829898} {\bibfield  {journal}
  {\bibinfo  {journal} {J. Chem. Phys.}\ }\textbf {\bibinfo {volume} {139}},\
  \bibinfo {pages} {194101} (\bibinfo {year} {2013})}\BibitemShut {NoStop}%
\bibitem [{\citenamefont {Usvyat}\ \emph {et~al.}(2015)\citenamefont {Usvyat},
  \citenamefont {Maschio},\ and\ \citenamefont {Sch{\"
  u}tz}}]{Usvyat_JCP_2015_102805}%
  \BibitemOpen
  \bibfield  {author} {\bibinfo {author} {\bibfnamefont {Denis}\ \bibnamefont
  {Usvyat}}, \bibinfo {author} {\bibfnamefont {Lorenzo}\ \bibnamefont
  {Maschio}}, \ and\ \bibinfo {author} {\bibfnamefont {Martin}\ \bibnamefont
  {Sch{\" u}tz}},\ }\bibfield  {title} {\enquote {\bibinfo {title} {Periodic
  local {MP2} method employing orbital specific virtuals},}\ }\href {\doibase
  10.1063/1.4921301} {\bibfield  {journal} {\bibinfo  {journal} {J. Chem.
  Phys.}\ }\textbf {\bibinfo {volume} {143}},\ \bibinfo {pages} {102805}
  (\bibinfo {year} {2015})}\BibitemShut {NoStop}%
\bibitem [{\citenamefont {Saeb{\o}}\ and\ \citenamefont
  {Pulay}(1993)}]{Saebo_ARPC_1993_213}%
  \BibitemOpen
  \bibfield  {author} {\bibinfo {author} {\bibfnamefont {S}~\bibnamefont
  {Saeb{\o}}}\ and\ \bibinfo {author} {\bibfnamefont {P}~\bibnamefont
  {Pulay}},\ }\bibfield  {title} {\enquote {\bibinfo {title} {Local treatment
  of electron correlation},}\ }\href {\doibase
  10.1146/annurev.pc.44.100193.001241} {\bibfield  {journal} {\bibinfo
  {journal} {Annu. Rev. Phys. Chem.}\ }\textbf {\bibinfo {volume} {44}},\
  \bibinfo {pages} {213--236} (\bibinfo {year} {1993})}\BibitemShut {NoStop}%
\bibitem [{\citenamefont {Yang}\ \emph {et~al.}(2011)\citenamefont {Yang},
  \citenamefont {Kurashige}, \citenamefont {Manby},\ and\ \citenamefont
  {Chan}}]{Yang_JCP_2011_044123}%
  \BibitemOpen
  \bibfield  {author} {\bibinfo {author} {\bibfnamefont {Jun}\ \bibnamefont
  {Yang}}, \bibinfo {author} {\bibfnamefont {Yuki}\ \bibnamefont {Kurashige}},
  \bibinfo {author} {\bibfnamefont {Frederick~R.}\ \bibnamefont {Manby}}, \
  and\ \bibinfo {author} {\bibfnamefont {Garnet K.~L.}\ \bibnamefont {Chan}},\
  }\bibfield  {title} {\enquote {\bibinfo {title} {Tensor factorizations of
  local second-order {M}{\o}ller-{P}lesset theory},}\ }\href {\doibase
  10.1063/1.3528935} {\bibfield  {journal} {\bibinfo  {journal} {J. Chem.
  Phys.}\ }\textbf {\bibinfo {volume} {134}},\ \bibinfo {pages} {044123}
  (\bibinfo {year} {2011})}\BibitemShut {NoStop}%
\bibitem [{\citenamefont {Rolik}\ and\ \citenamefont
  {K{\'a}llay}(2011)}]{Rolik_JCP_2011_104111}%
  \BibitemOpen
  \bibfield  {author} {\bibinfo {author} {\bibfnamefont {Zolt{\'a}n}\
  \bibnamefont {Rolik}}\ and\ \bibinfo {author} {\bibfnamefont {Mih{\'a}ly}\
  \bibnamefont {K{\'a}llay}},\ }\bibfield  {title} {\enquote {\bibinfo {title}
  {A general-order local coupled-cluster method based on the
  cluster-in-molecule approach},}\ }\href {\doibase 10.1063/1.3632085}
  {\bibfield  {journal} {\bibinfo  {journal} {J. Chem. Phys.}\ }\textbf
  {\bibinfo {volume} {135}},\ \bibinfo {pages} {104111} (\bibinfo {year}
  {2011})}\BibitemShut {NoStop}%
\bibitem [{\citenamefont {Rolik}\ \emph {et~al.}(2013)\citenamefont {Rolik},
  \citenamefont {Szegedy}, \citenamefont {Ladj{\'a}nszki}, \citenamefont
  {Lad{\'o}czki},\ and\ \citenamefont {K{\'a}llay}}]{Rolik_JCP_2013_094105}%
  \BibitemOpen
  \bibfield  {author} {\bibinfo {author} {\bibfnamefont {Zolt{\'a}n}\
  \bibnamefont {Rolik}}, \bibinfo {author} {\bibfnamefont {L{\'o}r{\'a}nt}\
  \bibnamefont {Szegedy}}, \bibinfo {author} {\bibfnamefont {Istv{\'a}n}\
  \bibnamefont {Ladj{\'a}nszki}}, \bibinfo {author} {\bibfnamefont {Bence}\
  \bibnamefont {Lad{\'o}czki}}, \ and\ \bibinfo {author} {\bibfnamefont
  {Mih{\'a}ly}\ \bibnamefont {K{\'a}llay}},\ }\bibfield  {title} {\enquote
  {\bibinfo {title} {An efficient linear-scaling {CCSD(T)} method based on
  local natural orbitals},}\ }\href {\doibase 10.1063/1.4819401} {\bibfield
  {journal} {\bibinfo  {journal} {J. Chem. Phys.}\ }\textbf {\bibinfo {volume}
  {139}},\ \bibinfo {pages} {094105} (\bibinfo {year} {2013})}\BibitemShut
  {NoStop}%
\bibitem [{\citenamefont {Nagy}\ and\ \citenamefont
  {K{\'a}llay}(2017)}]{Nagy_JCP_2017_214106}%
  \BibitemOpen
  \bibfield  {author} {\bibinfo {author} {\bibfnamefont {P{\'e}ter~R.}\
  \bibnamefont {Nagy}}\ and\ \bibinfo {author} {\bibfnamefont {Mih{\'a}ly}\
  \bibnamefont {K{\'a}llay}},\ }\bibfield  {title} {\enquote {\bibinfo {title}
  {Optimization of the linear-scaling local natural orbital {CCSD(T)} method:
  Redundancy-free triples correction using {L}aplace transform},}\ }\href
  {\doibase 10.1063/1.4984322} {\bibfield  {journal} {\bibinfo  {journal} {J.
  Chem. Phys.}\ }\textbf {\bibinfo {volume} {146}},\ \bibinfo {pages} {214106}
  (\bibinfo {year} {2017})}\BibitemShut {NoStop}%
\bibitem [{\citenamefont {Nagy}\ \emph {et~al.}(2018)\citenamefont {Nagy},
  \citenamefont {Samu},\ and\ \citenamefont
  {K{\'a}llay}}]{Nagy_JCTC_2018_4193}%
  \BibitemOpen
  \bibfield  {author} {\bibinfo {author} {\bibfnamefont {P{\'e}ter~R.}\
  \bibnamefont {Nagy}}, \bibinfo {author} {\bibfnamefont {Gyula}\ \bibnamefont
  {Samu}}, \ and\ \bibinfo {author} {\bibfnamefont {Mih{\'a}ly}\ \bibnamefont
  {K{\'a}llay}},\ }\bibfield  {title} {\enquote {\bibinfo {title} {Optimization
  of the linear-scaling local natural orbital {CCSD(T)} method: Improved
  algorithm and benchmark applications},}\ }\href {\doibase
  10.1021/acs.jctc.8b00442} {\bibfield  {journal} {\bibinfo  {journal} {J.
  Chem. Theory Comput.}\ }\textbf {\bibinfo {volume} {14}},\ \bibinfo {pages}
  {4193--4215} (\bibinfo {year} {2018})}\BibitemShut {NoStop}%
\bibitem [{\citenamefont {M{\"u}ller}\ and\ \citenamefont
  {Usvyat}(2013)}]{Mueller_JCTC_2013_5590}%
  \BibitemOpen
  \bibfield  {author} {\bibinfo {author} {\bibfnamefont {Carsten}\ \bibnamefont
  {M{\"u}ller}}\ and\ \bibinfo {author} {\bibfnamefont {Denis}\ \bibnamefont
  {Usvyat}},\ }\bibfield  {title} {\enquote {\bibinfo {title} {Incrementally
  corrected periodic local {MP2} calculations: {I}. the cohesive energy of
  molecular crystals},}\ }\href {\doibase 10.1021/ct400797w} {\bibfield
  {journal} {\bibinfo  {journal} {J. Chem. Theory Comput.}\ }\textbf {\bibinfo
  {volume} {9}},\ \bibinfo {pages} {5590--5598} (\bibinfo {year}
  {2013})}\BibitemShut {NoStop}%
\bibitem [{\citenamefont {Usvyat}\ \emph {et~al.}(2012)\citenamefont {Usvyat},
  \citenamefont {Sadeghian}, \citenamefont {Maschio},\ and\ \citenamefont
  {Sch{\"u}tz}}]{Usvyat_PRB_2012_045412}%
  \BibitemOpen
  \bibfield  {author} {\bibinfo {author} {\bibfnamefont {Denis}\ \bibnamefont
  {Usvyat}}, \bibinfo {author} {\bibfnamefont {Keyarash}\ \bibnamefont
  {Sadeghian}}, \bibinfo {author} {\bibfnamefont {Lorenzo}\ \bibnamefont
  {Maschio}}, \ and\ \bibinfo {author} {\bibfnamefont {Martin}\ \bibnamefont
  {Sch{\"u}tz}},\ }\bibfield  {title} {\enquote {\bibinfo {title} {Geometrical
  frustration of an argon monolayer adsorbed on the {MgO} (100) surface: An
  accurate periodic ab initio study},}\ }\href {\doibase
  10.1103/PhysRevB.86.045412} {\bibfield  {journal} {\bibinfo  {journal} {Phys.
  Rev. B}\ }\textbf {\bibinfo {volume} {86}},\ \bibinfo {pages} {045412}
  (\bibinfo {year} {2012})}\BibitemShut {NoStop}%
\bibitem [{\citenamefont {Mullan}\ \emph {et~al.}(2022)\citenamefont {Mullan},
  \citenamefont {Maschio}, \citenamefont {Saalfrank},\ and\ \citenamefont
  {Usvyat}}]{Mullan_JCP_2022_074109}%
  \BibitemOpen
  \bibfield  {author} {\bibinfo {author} {\bibfnamefont {Thomas}\ \bibnamefont
  {Mullan}}, \bibinfo {author} {\bibfnamefont {Lorenzo}\ \bibnamefont
  {Maschio}}, \bibinfo {author} {\bibfnamefont {Peter}\ \bibnamefont
  {Saalfrank}}, \ and\ \bibinfo {author} {\bibfnamefont {Denis}\ \bibnamefont
  {Usvyat}},\ }\bibfield  {title} {\enquote {\bibinfo {title} {Reaction
  barriers on non-conducting surfaces beyond periodic local {MP2}: Diffusion of
  hydrogen on $\alpha$-{Al$_2$O$_3$}(0001) as a test case},}\ }\href {\doibase
  10.1063/5.0082805} {\bibfield  {journal} {\bibinfo  {journal} {J. Chem.
  Phys.}\ }\textbf {\bibinfo {volume} {156}},\ \bibinfo {pages} {074109}
  (\bibinfo {year} {2022})}\BibitemShut {NoStop}%
\bibitem [{\citenamefont {Neese}\ \emph {et~al.}(2009)\citenamefont {Neese},
  \citenamefont {Hansen},\ and\ \citenamefont
  {Liakos}}]{Neese_JCP_2009_064103}%
  \BibitemOpen
  \bibfield  {author} {\bibinfo {author} {\bibfnamefont {Frank}\ \bibnamefont
  {Neese}}, \bibinfo {author} {\bibfnamefont {Andreas}\ \bibnamefont {Hansen}},
  \ and\ \bibinfo {author} {\bibfnamefont {Dimitrios~G.}\ \bibnamefont
  {Liakos}},\ }\bibfield  {title} {\enquote {\bibinfo {title} {Efficient and
  accurate approximations to the local coupled cluster singles doubles method
  using a truncated pair natural orbital basis},}\ }\href {\doibase
  10.1063/1.3173827} {\bibfield  {journal} {\bibinfo  {journal} {J. Chem.
  Phys.}\ }\textbf {\bibinfo {volume} {131}},\ \bibinfo {pages} {064103}
  (\bibinfo {year} {2009})}\BibitemShut {NoStop}%
\bibitem [{\citenamefont {Riplinger}\ and\ \citenamefont
  {Neese}(2013)}]{Riplinger_JCP_2013_034106}%
  \BibitemOpen
  \bibfield  {author} {\bibinfo {author} {\bibfnamefont {Christoph}\
  \bibnamefont {Riplinger}}\ and\ \bibinfo {author} {\bibfnamefont {Frank}\
  \bibnamefont {Neese}},\ }\bibfield  {title} {\enquote {\bibinfo {title} {An
  efficient and near linear scaling pair natural orbital based local coupled
  cluster method},}\ }\href {\doibase 10.1063/1.4773581} {\bibfield  {journal}
  {\bibinfo  {journal} {J. Chem. Phys.}\ }\textbf {\bibinfo {volume} {138}},\
  \bibinfo {pages} {034106} (\bibinfo {year} {2013})}\BibitemShut {NoStop}%
\bibitem [{\citenamefont {Werner}\ \emph {et~al.}(2015)\citenamefont {Werner},
  \citenamefont {Knizia}, \citenamefont {Krause}, \citenamefont {Schwilk},\
  and\ \citenamefont {Dornbach}}]{Werner_JCTC_2015_484}%
  \BibitemOpen
  \bibfield  {author} {\bibinfo {author} {\bibfnamefont {Hans-Joachim}\
  \bibnamefont {Werner}}, \bibinfo {author} {\bibfnamefont {Gerald}\
  \bibnamefont {Knizia}}, \bibinfo {author} {\bibfnamefont {Christine}\
  \bibnamefont {Krause}}, \bibinfo {author} {\bibfnamefont {Max}\ \bibnamefont
  {Schwilk}}, \ and\ \bibinfo {author} {\bibfnamefont {Mark}\ \bibnamefont
  {Dornbach}},\ }\bibfield  {title} {\enquote {\bibinfo {title} {Scalable
  electron correlation methods {I}.: {PNO-LMP2} with linear scaling in the
  molecular size and near-inverse-linear scaling in the number of
  processors},}\ }\href {\doibase 10.1021/ct500725e} {\bibfield  {journal}
  {\bibinfo  {journal} {J. Chem. Theory Comput.}\ }\textbf {\bibinfo {volume}
  {11}},\ \bibinfo {pages} {484--507} (\bibinfo {year} {2015})}\BibitemShut
  {NoStop}%
\bibitem [{\citenamefont {Schmitz}\ \emph {et~al.}(2013)\citenamefont
  {Schmitz}, \citenamefont {Helmich},\ and\ \citenamefont
  {H{\"a}ttig}}]{Schmitz_MP_2013_2463}%
  \BibitemOpen
  \bibfield  {author} {\bibinfo {author} {\bibfnamefont {Gunnar}\ \bibnamefont
  {Schmitz}}, \bibinfo {author} {\bibfnamefont {Benjamin}\ \bibnamefont
  {Helmich}}, \ and\ \bibinfo {author} {\bibfnamefont {Christof}\ \bibnamefont
  {H{\"a}ttig}},\ }\bibfield  {title} {\enquote {\bibinfo {title} {A
  ($\mathcal{O}(\mathcal{N}^3)$ scaling pno-mp2 method using a hybrid osv-pno
  approach with an iterative direct generation of osvs},}\ }\href {\doibase
  10.1080/00268976.2013.794314} {\bibfield  {journal} {\bibinfo  {journal}
  {Mol. Phys.}\ }\textbf {\bibinfo {volume} {111}},\ \bibinfo {pages}
  {2463--2476} (\bibinfo {year} {2013})}\BibitemShut {NoStop}%
\bibitem [{\citenamefont {Pinski}\ \emph {et~al.}(2015)\citenamefont {Pinski},
  \citenamefont {Riplinger}, \citenamefont {Valeev},\ and\ \citenamefont
  {Neese}}]{Pinski_JCP_2015_034108}%
  \BibitemOpen
  \bibfield  {author} {\bibinfo {author} {\bibfnamefont {Peter}\ \bibnamefont
  {Pinski}}, \bibinfo {author} {\bibfnamefont {Christoph}\ \bibnamefont
  {Riplinger}}, \bibinfo {author} {\bibfnamefont {Edward~F.}\ \bibnamefont
  {Valeev}}, \ and\ \bibinfo {author} {\bibfnamefont {Frank}\ \bibnamefont
  {Neese}},\ }\bibfield  {title} {\enquote {\bibinfo {title} {Sparse maps--a
  systematic infrastructure for reduced-scaling electronic structure methods.
  {I}. an efficient and simple linear scaling local {MP2} method that uses an
  intermediate basis of pair natural orbitals},}\ }\href {\doibase
  10.1063/1.4926879} {\bibfield  {journal} {\bibinfo  {journal} {J. Chem.
  Phys.}\ }\textbf {\bibinfo {volume} {143}},\ \bibinfo {pages} {034108}
  (\bibinfo {year} {2015})}\BibitemShut {NoStop}%
\bibitem [{\citenamefont {Tew}\ and\ \citenamefont
  {H{\"a}ttig}(2013)}]{Tew_IJQC_2013_224}%
  \BibitemOpen
  \bibfield  {author} {\bibinfo {author} {\bibfnamefont {David~P.}\
  \bibnamefont {Tew}}\ and\ \bibinfo {author} {\bibfnamefont {Christof}\
  \bibnamefont {H{\"a}ttig}},\ }\bibfield  {title} {\enquote {\bibinfo {title}
  {Pair natural orbitals in explicitly correlated second-order
  m{\o}ller--plesset theory},}\ }\href {\doibase 10.1002/qua.24098} {\bibfield
  {journal} {\bibinfo  {journal} {Int. J. Quantum Chem.}\ }\textbf {\bibinfo
  {volume} {113}},\ \bibinfo {pages} {224--229} (\bibinfo {year}
  {2013})}\BibitemShut {NoStop}%
\bibitem [{\citenamefont {H{\"a}ttig}\ \emph {et~al.}(2012)\citenamefont
  {H{\"a}ttig}, \citenamefont {Tew},\ and\ \citenamefont
  {Helmich}}]{Haettig_JCP_2012_204105}%
  \BibitemOpen
  \bibfield  {author} {\bibinfo {author} {\bibfnamefont {Christof}\
  \bibnamefont {H{\"a}ttig}}, \bibinfo {author} {\bibfnamefont {David~P.}\
  \bibnamefont {Tew}}, \ and\ \bibinfo {author} {\bibfnamefont {Benjamin}\
  \bibnamefont {Helmich}},\ }\bibfield  {title} {\enquote {\bibinfo {title}
  {Local explicitly correlated second- and third-order {M}{\o}ller-{P}lesset
  perturbation theory with pair natural orbitals},}\ }\href {\doibase
  10.1063/1.4719981} {\bibfield  {journal} {\bibinfo  {journal} {J. Chem.
  Phys.}\ }\textbf {\bibinfo {volume} {136}},\ \bibinfo {pages} {204105}
  (\bibinfo {year} {2012})}\BibitemShut {NoStop}%
\bibitem [{\citenamefont {Riplinger}\ \emph {et~al.}(2013)\citenamefont
  {Riplinger}, \citenamefont {Sandhoefer}, \citenamefont {Hansen},\ and\
  \citenamefont {Neese}}]{Riplinger_JCP_2013_134101}%
  \BibitemOpen
  \bibfield  {author} {\bibinfo {author} {\bibfnamefont {Christoph}\
  \bibnamefont {Riplinger}}, \bibinfo {author} {\bibfnamefont {Barbara}\
  \bibnamefont {Sandhoefer}}, \bibinfo {author} {\bibfnamefont {Andreas}\
  \bibnamefont {Hansen}}, \ and\ \bibinfo {author} {\bibfnamefont {Frank}\
  \bibnamefont {Neese}},\ }\bibfield  {title} {\enquote {\bibinfo {title}
  {Natural triple excitations in local coupled cluster calculations with pair
  natural orbitals},}\ }\href {\doibase 10.1063/1.4821834} {\bibfield
  {journal} {\bibinfo  {journal} {J. Chem. Phys.}\ }\textbf {\bibinfo {volume}
  {139}},\ \bibinfo {pages} {134101} (\bibinfo {year} {2013})}\BibitemShut
  {NoStop}%
\bibitem [{\citenamefont {Guo}\ \emph {et~al.}(2018)\citenamefont {Guo},
  \citenamefont {Riplinger}, \citenamefont {Becker}, \citenamefont {Liakos},
  \citenamefont {Minenkov}, \citenamefont {Cavallo},\ and\ \citenamefont
  {Neese}}]{Guo_JCP_2018_011101}%
  \BibitemOpen
  \bibfield  {author} {\bibinfo {author} {\bibfnamefont {Yang}\ \bibnamefont
  {Guo}}, \bibinfo {author} {\bibfnamefont {Christoph}\ \bibnamefont
  {Riplinger}}, \bibinfo {author} {\bibfnamefont {Ute}\ \bibnamefont {Becker}},
  \bibinfo {author} {\bibfnamefont {Dimitrios~G.}\ \bibnamefont {Liakos}},
  \bibinfo {author} {\bibfnamefont {Yury}\ \bibnamefont {Minenkov}}, \bibinfo
  {author} {\bibfnamefont {Luigi}\ \bibnamefont {Cavallo}}, \ and\ \bibinfo
  {author} {\bibfnamefont {Frank}\ \bibnamefont {Neese}},\ }\bibfield  {title}
  {\enquote {\bibinfo {title} {Communication: An improved linear scaling
  perturbative triples correction for the domain based local pair-natural
  orbital based singles and doubles coupled cluster method
  [{DLPNO-CCSD(T)}]},}\ }\href {\doibase 10.1063/1.5011798} {\bibfield
  {journal} {\bibinfo  {journal} {J. Chem. Phys.}\ }\textbf {\bibinfo {volume}
  {148}},\ \bibinfo {pages} {011101} (\bibinfo {year} {2018})}\BibitemShut
  {NoStop}%
\bibitem [{\citenamefont {Schwilk}\ \emph {et~al.}(2017)\citenamefont
  {Schwilk}, \citenamefont {Ma}, \citenamefont {K{\"o}ppl},\ and\ \citenamefont
  {Werner}}]{Schwilk_JCTC_2017_3650}%
  \BibitemOpen
  \bibfield  {author} {\bibinfo {author} {\bibfnamefont {Max}\ \bibnamefont
  {Schwilk}}, \bibinfo {author} {\bibfnamefont {Qianli}\ \bibnamefont {Ma}},
  \bibinfo {author} {\bibfnamefont {Christoph}\ \bibnamefont {K{\"o}ppl}}, \
  and\ \bibinfo {author} {\bibfnamefont {Hans-Joachim}\ \bibnamefont
  {Werner}},\ }\bibfield  {title} {\enquote {\bibinfo {title} {Scalable
  electron correlation methods. 3. efficient and accurate parallel local
  coupled cluster with pair natural orbitals ({PNO-LCCSD})},}\ }\href {\doibase
  10.1021/acs.jctc.7b00554} {\bibfield  {journal} {\bibinfo  {journal} {J.
  Chem. Theory Comput.}\ }\textbf {\bibinfo {volume} {13}},\ \bibinfo {pages}
  {3650--3675} (\bibinfo {year} {2017})}\BibitemShut {NoStop}%
\bibitem [{\citenamefont {Ma}\ and\ \citenamefont
  {Werner}(2018)}]{Ma_JCTC_2018_198}%
  \BibitemOpen
  \bibfield  {author} {\bibinfo {author} {\bibfnamefont {Qianli}\ \bibnamefont
  {Ma}}\ and\ \bibinfo {author} {\bibfnamefont {Hans-Joachim}\ \bibnamefont
  {Werner}},\ }\bibfield  {title} {\enquote {\bibinfo {title} {Scalable
  electron correlation methods. 5. parallel perturbative triples correction for
  explicitly correlated local coupled cluster with pair natural orbitals},}\
  }\href {\doibase 10.1021/acs.jctc.7b01141} {\bibfield  {journal} {\bibinfo
  {journal} {J. Chem. Theory Comput.}\ }\textbf {\bibinfo {volume} {14}},\
  \bibinfo {pages} {198--215} (\bibinfo {year} {2018})}\BibitemShut {NoStop}%
\bibitem [{\citenamefont {Schmitz}\ and\ \citenamefont
  {H{\"a}ttig}(2017)}]{Schmitz_JCTC_2017_2623}%
  \BibitemOpen
  \bibfield  {author} {\bibinfo {author} {\bibfnamefont {Gunnar}\ \bibnamefont
  {Schmitz}}\ and\ \bibinfo {author} {\bibfnamefont {Christof}\ \bibnamefont
  {H{\"a}ttig}},\ }\bibfield  {title} {\enquote {\bibinfo {title} {Accuracy of
  explicitly correlated local {PNO-CCSD(T)}},}\ }\href {\doibase
  10.1021/acs.jctc.7b00180} {\bibfield  {journal} {\bibinfo  {journal} {J.
  Chem. Theory Comput.}\ }\textbf {\bibinfo {volume} {13}},\ \bibinfo {pages}
  {2623--2633} (\bibinfo {year} {2017})}\BibitemShut {NoStop}%
\bibitem [{\citenamefont {Schmitz}\ and\ \citenamefont
  {H{\"a}ttig}(2016)}]{Schmitz_JCP_2016_234107}%
  \BibitemOpen
  \bibfield  {author} {\bibinfo {author} {\bibfnamefont {Gunnar}\ \bibnamefont
  {Schmitz}}\ and\ \bibinfo {author} {\bibfnamefont {Christof}\ \bibnamefont
  {H{\"a}ttig}},\ }\bibfield  {title} {\enquote {\bibinfo {title} {Perturbative
  triples correction for local pair natural orbital based explicitly correlated
  {CCSD(F12$^*$)} using laplace transformation techniques},}\ }\href {\doibase
  10.1063/1.4972001} {\bibfield  {journal} {\bibinfo  {journal} {J. Chem.
  Phys.}\ }\textbf {\bibinfo {volume} {145}},\ \bibinfo {pages} {234107}
  (\bibinfo {year} {2016})}\BibitemShut {NoStop}%
\bibitem [{\citenamefont {Liakos}\ \emph {et~al.}(2020)\citenamefont {Liakos},
  \citenamefont {Guo},\ and\ \citenamefont {Neese}}]{Liakos_JPCA_2020_90}%
  \BibitemOpen
  \bibfield  {author} {\bibinfo {author} {\bibfnamefont {Dimitrios~G.}\
  \bibnamefont {Liakos}}, \bibinfo {author} {\bibfnamefont {Yang}\ \bibnamefont
  {Guo}}, \ and\ \bibinfo {author} {\bibfnamefont {Frank}\ \bibnamefont
  {Neese}},\ }\bibfield  {title} {\enquote {\bibinfo {title} {Comprehensive
  benchmark results for the domain based local pair natural orbital coupled
  cluster method ({DLPNO-CCSD(T)}) for closed- and open-shell systems},}\
  }\href {\doibase 10.1021/acs.jpca.9b05734} {\bibfield  {journal} {\bibinfo
  {journal} {J. Phys. Chem. A}\ }\textbf {\bibinfo {volume} {124}},\ \bibinfo
  {pages} {90--100} (\bibinfo {year} {2020})}\BibitemShut {NoStop}%
\bibitem [{\citenamefont {Saitow}\ \emph {et~al.}(2017)\citenamefont {Saitow},
  \citenamefont {Becker}, \citenamefont {Riplinger}, \citenamefont {Valeev},\
  and\ \citenamefont {Neese}}]{Saitow_JCP_2017}%
  \BibitemOpen
  \bibfield  {author} {\bibinfo {author} {\bibfnamefont {Masaaki}\ \bibnamefont
  {Saitow}}, \bibinfo {author} {\bibfnamefont {Ute}\ \bibnamefont {Becker}},
  \bibinfo {author} {\bibfnamefont {Christoph}\ \bibnamefont {Riplinger}},
  \bibinfo {author} {\bibfnamefont {Edward~F.}\ \bibnamefont {Valeev}}, \ and\
  \bibinfo {author} {\bibfnamefont {Frank}\ \bibnamefont {Neese}},\ }\bibfield
  {title} {\enquote {\bibinfo {title} {A new near-linear scaling, efficient and
  accurate, open-shell domain-based local pair natural orbital coupled cluster
  singles and doubles theory},}\ }\href {\doibase 10.1063/1.4981521} {\bibfield
   {journal} {\bibinfo  {journal} {J. Chem. Phys.}\ }\textbf {\bibinfo {volume}
  {146}},\ \bibinfo {pages} {164105} (\bibinfo {year} {2017})}\BibitemShut
  {NoStop}%
\bibitem [{\citenamefont {Riplinger}\ \emph {et~al.}(2016)\citenamefont
  {Riplinger}, \citenamefont {Pinski}, \citenamefont {Becker}, \citenamefont
  {Valeev},\ and\ \citenamefont {Neese}}]{Riplinger_JCP_2016_024109}%
  \BibitemOpen
  \bibfield  {author} {\bibinfo {author} {\bibfnamefont {Christoph}\
  \bibnamefont {Riplinger}}, \bibinfo {author} {\bibfnamefont {Peter}\
  \bibnamefont {Pinski}}, \bibinfo {author} {\bibfnamefont {Ute}\ \bibnamefont
  {Becker}}, \bibinfo {author} {\bibfnamefont {Edward~F.}\ \bibnamefont
  {Valeev}}, \ and\ \bibinfo {author} {\bibfnamefont {Frank}\ \bibnamefont
  {Neese}},\ }\bibfield  {title} {\enquote {\bibinfo {title} {Sparse maps--a
  systematic infrastructure for reduced-scaling electronic structure methods.
  {II}. linear scaling domain based pair natural orbital coupled cluster
  theory},}\ }\href {\doibase 10.1063/1.4939030} {\bibfield  {journal}
  {\bibinfo  {journal} {J. Chem. Phys.}\ }\textbf {\bibinfo {volume} {144}},\
  \bibinfo {pages} {024109} (\bibinfo {year} {2016})}\BibitemShut {NoStop}%
\bibitem [{\citenamefont {Ma}\ and\ \citenamefont
  {Werner}(2015)}]{Ma_JCTC_2015_5291}%
  \BibitemOpen
  \bibfield  {author} {\bibinfo {author} {\bibfnamefont {Qianli}\ \bibnamefont
  {Ma}}\ and\ \bibinfo {author} {\bibfnamefont {Hans-Joachim}\ \bibnamefont
  {Werner}},\ }\bibfield  {title} {\enquote {\bibinfo {title} {Scalable
  electron correlation methods. 2. parallel {PNO-LMP2-F12} with near linear
  scaling in the molecular size},}\ }\href {\doibase 10.1021/acs.jctc.5b00843}
  {\bibfield  {journal} {\bibinfo  {journal} {J. Chem. Theory Comput.}\
  }\textbf {\bibinfo {volume} {11}},\ \bibinfo {pages} {5291--5304} (\bibinfo
  {year} {2015})},\ \bibinfo {note} {pMID: 26574323}\BibitemShut {NoStop}%
\bibitem [{\citenamefont {Schmitz}\ \emph {et~al.}(2014)\citenamefont
  {Schmitz}, \citenamefont {H{\"a}ttig},\ and\ \citenamefont
  {Tew}}]{Schmitz_PCCP_2014_22167}%
  \BibitemOpen
  \bibfield  {author} {\bibinfo {author} {\bibfnamefont {Gunnar}\ \bibnamefont
  {Schmitz}}, \bibinfo {author} {\bibfnamefont {Christof}\ \bibnamefont
  {H{\"a}ttig}}, \ and\ \bibinfo {author} {\bibfnamefont {David~P.}\
  \bibnamefont {Tew}},\ }\bibfield  {title} {\enquote {\bibinfo {title}
  {Explicitly correlated pno-mp2 and pno-ccsd and their application to the
  {S66} set and large molecular systems},}\ }\href {\doibase
  10.1039/C4CP03502J} {\bibfield  {journal} {\bibinfo  {journal} {Phys. Chem.
  Chem. Phys.}\ }\textbf {\bibinfo {volume} {16}},\ \bibinfo {pages}
  {22167--22178} (\bibinfo {year} {2014})}\BibitemShut {NoStop}%
\bibitem [{\citenamefont {Ma}\ \emph {et~al.}(2017)\citenamefont {Ma},
  \citenamefont {Schwilk}, \citenamefont {K{\"o}ppl},\ and\ \citenamefont
  {Werner}}]{Ma_JCTC_2017_4871}%
  \BibitemOpen
  \bibfield  {author} {\bibinfo {author} {\bibfnamefont {Qianli}\ \bibnamefont
  {Ma}}, \bibinfo {author} {\bibfnamefont {Max}\ \bibnamefont {Schwilk}},
  \bibinfo {author} {\bibfnamefont {Christoph}\ \bibnamefont {K{\"o}ppl}}, \
  and\ \bibinfo {author} {\bibfnamefont {Hans-Joachim}\ \bibnamefont
  {Werner}},\ }\bibfield  {title} {\enquote {\bibinfo {title} {Scalable
  electron correlation methods. 4. parallel explicitly correlated local coupled
  cluster with pair natural orbitals ({PNO-LCCSD-F12})},}\ }\href {\doibase
  10.1021/acs.jctc.7b00799} {\bibfield  {journal} {\bibinfo  {journal} {J.
  Chem. Theory Comput.}\ }\textbf {\bibinfo {volume} {13}},\ \bibinfo {pages}
  {4871--4896} (\bibinfo {year} {2017})}\BibitemShut {NoStop}%
\bibitem [{\citenamefont {Tew}(2021)}]{Tew_chapter_2021}%
  \BibitemOpen
  \bibfield  {author} {\bibinfo {author} {\bibfnamefont {David~P.}\
  \bibnamefont {Tew}},\ }\bibfield  {title} {\enquote {\bibinfo {title}
  {Principal domains in {F12} explicitly correlated theory},}\ }in\ \href
  {\doibase 10.1016/bs.aiq.2021.06.001} {\emph {\bibinfo {booktitle} {New
  Electron Correlation Methods and their Applications, and Use of Atomic
  Orbitals with Exponential Asymptotes}}},\ \bibinfo {series} {Advances in
  Quantum Chemistry}, Vol.~\bibinfo {volume} {83},\ \bibinfo {editor} {edited
  by\ \bibinfo {editor} {\bibfnamefont {Monika}\ \bibnamefont {Musial}}\ and\
  \bibinfo {editor} {\bibfnamefont {Philip~E.}\ \bibnamefont {Hoggan}}}\
  (\bibinfo  {publisher} {Academic Press},\ \bibinfo {year} {2021})\ pp.\
  \bibinfo {pages} {83--106}\BibitemShut {NoStop}%
\bibitem [{\citenamefont {Kats}\ and\ \citenamefont
  {Werner}(2019)}]{Kats_JCP_2019_214107}%
  \BibitemOpen
  \bibfield  {author} {\bibinfo {author} {\bibfnamefont {Daniel}\ \bibnamefont
  {Kats}}\ and\ \bibinfo {author} {\bibfnamefont {Hans-Joachim}\ \bibnamefont
  {Werner}},\ }\bibfield  {title} {\enquote {\bibinfo {title} {Multi-state
  local complete active space second-order perturbation theory using pair
  natural orbitals ({PNO-MS-CASPT2})},}\ }\href {\doibase 10.1063/1.5097644}
  {\bibfield  {journal} {\bibinfo  {journal} {J. Chem. Phys.}\ }\textbf
  {\bibinfo {volume} {150}},\ \bibinfo {pages} {214107} (\bibinfo {year}
  {2019})}\BibitemShut {NoStop}%
\bibitem [{\citenamefont {Saitow}\ and\ \citenamefont
  {Yanai}(2020)}]{Saitow_JCP_2020_114111}%
  \BibitemOpen
  \bibfield  {author} {\bibinfo {author} {\bibfnamefont {Masaaki}\ \bibnamefont
  {Saitow}}\ and\ \bibinfo {author} {\bibfnamefont {Takeshi}\ \bibnamefont
  {Yanai}},\ }\bibfield  {title} {\enquote {\bibinfo {title} {A multireference
  coupled-electron pair approximation combined with complete-active space
  perturbation theory in local pair-natural orbital framework},}\ }\href
  {\doibase 10.1063/1.5142622} {\bibfield  {journal} {\bibinfo  {journal} {J.
  Chem. Phys.}\ }\textbf {\bibinfo {volume} {152}},\ \bibinfo {pages} {114111}
  (\bibinfo {year} {2020})}\BibitemShut {NoStop}%
\bibitem [{\citenamefont {Guo}\ \emph {et~al.}(2016)\citenamefont {Guo},
  \citenamefont {Sivalingam}, \citenamefont {Valeev},\ and\ \citenamefont
  {Neese}}]{Guo_JCP_2016_094111}%
  \BibitemOpen
  \bibfield  {author} {\bibinfo {author} {\bibfnamefont {Yang}\ \bibnamefont
  {Guo}}, \bibinfo {author} {\bibfnamefont {Kantharuban}\ \bibnamefont
  {Sivalingam}}, \bibinfo {author} {\bibfnamefont {Edward~F.}\ \bibnamefont
  {Valeev}}, \ and\ \bibinfo {author} {\bibfnamefont {Frank}\ \bibnamefont
  {Neese}},\ }\bibfield  {title} {\enquote {\bibinfo {title} {Sparse{M}aps--a
  systematic infrastructure for reduced-scaling electronic structure methods.
  {III}. linear-scaling multireference domain-based pair natural orbital
  {N}-electron valence perturbation theory},}\ }\href {\doibase
  10.1063/1.4942769} {\bibfield  {journal} {\bibinfo  {journal} {J. Chem.
  Phys.}\ }\textbf {\bibinfo {volume} {144}},\ \bibinfo {pages} {094111}
  (\bibinfo {year} {2016})}\BibitemShut {NoStop}%
\bibitem [{\citenamefont {Helmich}\ and\ \citenamefont
  {H{\"a}ttig}(2011)}]{Helmich_JCP_2011_214106}%
  \BibitemOpen
  \bibfield  {author} {\bibinfo {author} {\bibfnamefont {Benjamin}\
  \bibnamefont {Helmich}}\ and\ \bibinfo {author} {\bibfnamefont {Christof}\
  \bibnamefont {H{\"a}ttig}},\ }\bibfield  {title} {\enquote {\bibinfo {title}
  {Local pair natural orbitals for excited states},}\ }\href {\doibase
  10.1063/1.3664902} {\bibfield  {journal} {\bibinfo  {journal} {J. Chem.
  Phys.}\ }\textbf {\bibinfo {volume} {135}},\ \bibinfo {pages} {214106}
  (\bibinfo {year} {2011})}\BibitemShut {NoStop}%
\bibitem [{\citenamefont {Frank}\ and\ \citenamefont
  {H{\"a}ttig}(2018)}]{Frank_JCP_2018_134102}%
  \BibitemOpen
  \bibfield  {author} {\bibinfo {author} {\bibfnamefont {Marius~S.}\
  \bibnamefont {Frank}}\ and\ \bibinfo {author} {\bibfnamefont {Christof}\
  \bibnamefont {H{\"a}ttig}},\ }\bibfield  {title} {\enquote {\bibinfo {title}
  {A pair natural orbital based implementation of {CCSD} excitation energies
  within the framework of linear response theory},}\ }\href {\doibase
  10.1063/1.5018514} {\bibfield  {journal} {\bibinfo  {journal} {J. Chem.
  Phys.}\ }\textbf {\bibinfo {volume} {148}},\ \bibinfo {pages} {134102}
  (\bibinfo {year} {2018})}\BibitemShut {NoStop}%
\bibitem [{\citenamefont {Dutta}\ \emph {et~al.}(2016)\citenamefont {Dutta},
  \citenamefont {Neese},\ and\ \citenamefont
  {Izs{\'a}k}}]{Dutta_JCP_2016_034102}%
  \BibitemOpen
  \bibfield  {author} {\bibinfo {author} {\bibfnamefont {Achintya~Kumar}\
  \bibnamefont {Dutta}}, \bibinfo {author} {\bibfnamefont {Frank}\ \bibnamefont
  {Neese}}, \ and\ \bibinfo {author} {\bibfnamefont {R{\'o}bert}\ \bibnamefont
  {Izs{\'a}k}},\ }\bibfield  {title} {\enquote {\bibinfo {title} {Towards a
  pair natural orbital coupled cluster method for excited states},}\ }\href
  {\doibase 10.1063/1.4958734} {\bibfield  {journal} {\bibinfo  {journal} {J.
  Chem. Phys.}\ }\textbf {\bibinfo {volume} {145}},\ \bibinfo {pages} {034102}
  (\bibinfo {year} {2016})}\BibitemShut {NoStop}%
\bibitem [{\citenamefont {Born}\ and\ \citenamefont {{von
  K{\'a}rm{\'a}n}}(1912)}]{Born_PZ_1912_297}%
  \BibitemOpen
  \bibfield  {author} {\bibinfo {author} {\bibfnamefont {M.}~\bibnamefont
  {Born}}\ and\ \bibinfo {author} {\bibfnamefont {T.}~\bibnamefont {{von
  K{\'a}rm{\'a}n}}},\ }\bibfield  {title} {\enquote {\bibinfo {title} {{\"U}ber
  schwingungen in raumgittern},}\ }\href@noop {} {\bibfield  {journal}
  {\bibinfo  {journal} {Phys. Z.}\ }\textbf {\bibinfo {volume} {13}},\ \bibinfo
  {pages} {297--309} (\bibinfo {year} {1912})}\BibitemShut {NoStop}%
\bibitem [{\citenamefont {Born}\ and\ \citenamefont {{von
  K{\'a}rm{\'a}n}}(1913)}]{Born_PZ_1913_15}%
  \BibitemOpen
  \bibfield  {author} {\bibinfo {author} {\bibfnamefont {M.}~\bibnamefont
  {Born}}\ and\ \bibinfo {author} {\bibfnamefont {T.}~\bibnamefont {{von
  K{\'a}rm{\'a}n}}},\ }\bibfield  {title} {\enquote {\bibinfo {title} {Zur
  theorie der spezifischen w{\"a}rme},}\ }\href@noop {} {\bibfield  {journal}
  {\bibinfo  {journal} {Phys. Z.}\ }\textbf {\bibinfo {volume} {14}},\ \bibinfo
  {pages} {15--19} (\bibinfo {year} {1913})}\BibitemShut {NoStop}%
\bibitem [{\citenamefont {Zhu}\ \emph {et~al.}(2025)\citenamefont {Zhu},
  \citenamefont {Nejad}, \citenamefont {Komonvasee}, \citenamefont {Sorathia},\
  and\ \citenamefont {Tew}}]{paper_mega}%
  \BibitemOpen
  \bibfield  {author} {\bibinfo {author} {\bibfnamefont {Andrew}\ \bibnamefont
  {Zhu}}, \bibinfo {author} {\bibfnamefont {Arman}\ \bibnamefont {Nejad}},
  \bibinfo {author} {\bibfnamefont {Poramas}\ \bibnamefont {Komonvasee}},
  \bibinfo {author} {\bibfnamefont {Kesha}\ \bibnamefont {Sorathia}}, \ and\
  \bibinfo {author} {\bibfnamefont {David~P.}\ \bibnamefont {Tew}},\ }\href
  {https://arxiv.org/abs/XXXX.XXXXX} {\enquote {\bibinfo {title}
  {Megacell-{DLPNO-MP2} for periodic systems},}\ } (\bibinfo {year} {2025}),\
  \Eprint {http://arxiv.org/abs/XXXX.XXXXX} {arXiv:XXXX.XXXXX} \BibitemShut
  {NoStop}%
\bibitem [{\citenamefont {Burow}\ \emph {et~al.}(2009)\citenamefont {Burow},
  \citenamefont {Sierka},\ and\ \citenamefont
  {Mohamed}}]{Burow_JCP_2009_214101}%
  \BibitemOpen
  \bibfield  {author} {\bibinfo {author} {\bibfnamefont {Asbj{\"o}rn~M.}\
  \bibnamefont {Burow}}, \bibinfo {author} {\bibfnamefont {Marek}\ \bibnamefont
  {Sierka}}, \ and\ \bibinfo {author} {\bibfnamefont {Fawzi}\ \bibnamefont
  {Mohamed}},\ }\bibfield  {title} {\enquote {\bibinfo {title} {Resolution of
  identity approximation for the {C}oulomb term in molecular and periodic
  systems},}\ }\href {\doibase 10.1063/1.3267858} {\bibfield  {journal}
  {\bibinfo  {journal} {J. Chem. Phys.}\ }\textbf {\bibinfo {volume} {131}},\
  \bibinfo {pages} {214101} (\bibinfo {year} {2009})}\BibitemShut {NoStop}%
\bibitem [{\citenamefont {Sch{\"u}tz}\ \emph {et~al.}(2010)\citenamefont
  {Sch{\"u}tz}, \citenamefont {Usvyat}, \citenamefont {Lorenz}, \citenamefont
  {Pisani}, \citenamefont {Maschio}, \citenamefont {Casassa},\ and\
  \citenamefont {Halo}}]{Schuetz_chapter_2011}%
  \BibitemOpen
  \bibfield  {author} {\bibinfo {author} {\bibfnamefont {Martin}\ \bibnamefont
  {Sch{\"u}tz}}, \bibinfo {author} {\bibfnamefont {Denis}\ \bibnamefont
  {Usvyat}}, \bibinfo {author} {\bibfnamefont {Marco}\ \bibnamefont {Lorenz}},
  \bibinfo {author} {\bibfnamefont {Cesare}\ \bibnamefont {Pisani}}, \bibinfo
  {author} {\bibfnamefont {Lorenzo}\ \bibnamefont {Maschio}}, \bibinfo {author}
  {\bibfnamefont {Silvia}\ \bibnamefont {Casassa}}, \ and\ \bibinfo {author}
  {\bibfnamefont {Migen}\ \bibnamefont {Halo}},\ }\enquote {\bibinfo {title}
  {Density fitting for correlated calculations in periodic systems},}\ in\
  \href {\doibase 10.1201/9781439808375} {\emph {\bibinfo {booktitle} {Accurate
  Condensed-Phase Quantum Chemistry}}},\ \bibinfo {editor} {edited by\ \bibinfo
  {editor} {\bibfnamefont {F.~R.}\ \bibnamefont {Manby}}}\ (\bibinfo
  {publisher} {CRC Press},\ \bibinfo {address} {Boca Raton},\ \bibinfo {year}
  {2010})\ pp.\ \bibinfo {pages} {29--55}\BibitemShut {NoStop}%
\bibitem [{\citenamefont {Ye}\ and\ \citenamefont
  {Berkelbach}(2021)}]{Ye_JCP_2021_131104}%
  \BibitemOpen
  \bibfield  {author} {\bibinfo {author} {\bibfnamefont {Hong-Zhou}\
  \bibnamefont {Ye}}\ and\ \bibinfo {author} {\bibfnamefont {Timothy~C.}\
  \bibnamefont {Berkelbach}},\ }\bibfield  {title} {\enquote {\bibinfo {title}
  {Fast periodic {G}aussian density fitting by range separation},}\ }\href
  {\doibase 10.1063/5.0046617} {\bibfield  {journal} {\bibinfo  {journal} {J.
  Chem. Phys.}\ }\textbf {\bibinfo {volume} {154}},\ \bibinfo {pages} {131104}
  (\bibinfo {year} {2021})}\BibitemShut {NoStop}%
\bibitem [{\citenamefont {Stolarczyk}\ and\ \citenamefont
  {Piela}(1982)}]{Stolarczyk_IJQC_1982_911}%
  \BibitemOpen
  \bibfield  {author} {\bibinfo {author} {\bibfnamefont {Leszek~Z.}\
  \bibnamefont {Stolarczyk}}\ and\ \bibinfo {author} {\bibfnamefont {Lucjan}\
  \bibnamefont {Piela}},\ }\bibfield  {title} {\enquote {\bibinfo {title}
  {Direct calculation of lattice sums. a method to account for the crystal
  field effects},}\ }\href {\doibase 10.1002/qua.560220506} {\bibfield
  {journal} {\bibinfo  {journal} {Int. J. Quantum Chem.}\ }\textbf {\bibinfo
  {volume} {22}},\ \bibinfo {pages} {911--927} (\bibinfo {year}
  {1982})}\BibitemShut {NoStop}%
\bibitem [{\citenamefont {Makov}\ and\ \citenamefont
  {Payne}(1995)}]{Makov_PRB_1995_4014}%
  \BibitemOpen
  \bibfield  {author} {\bibinfo {author} {\bibfnamefont {G.}~\bibnamefont
  {Makov}}\ and\ \bibinfo {author} {\bibfnamefont {M.~C.}\ \bibnamefont
  {Payne}},\ }\bibfield  {title} {\enquote {\bibinfo {title} {Periodic boundary
  conditions in \textit{ab initio} calculations},}\ }\href {\doibase
  10.1103/PhysRevB.51.4014} {\bibfield  {journal} {\bibinfo  {journal} {Phys.
  Rev. B}\ }\textbf {\bibinfo {volume} {51}},\ \bibinfo {pages} {4014--4022}
  (\bibinfo {year} {1995})}\BibitemShut {NoStop}%
\bibitem [{\citenamefont {Challacombe}\ \emph {et~al.}(1997)\citenamefont
  {Challacombe}, \citenamefont {White},\ and\ \citenamefont
  {{Head-Gordon}}}]{Challacombe_JCP_1997_10131}%
  \BibitemOpen
  \bibfield  {author} {\bibinfo {author} {\bibfnamefont {Matt}\ \bibnamefont
  {Challacombe}}, \bibinfo {author} {\bibfnamefont {Chris}\ \bibnamefont
  {White}}, \ and\ \bibinfo {author} {\bibfnamefont {Martin}\ \bibnamefont
  {{Head-Gordon}}},\ }\bibfield  {title} {\enquote {\bibinfo {title} {Periodic
  boundary conditions and the fast multipole method},}\ }\href {\doibase
  10.1063/1.474150} {\bibfield  {journal} {\bibinfo  {journal} {J. Chem.
  Phys.}\ }\textbf {\bibinfo {volume} {107}},\ \bibinfo {pages} {10131--10140}
  (\bibinfo {year} {1997})}\BibitemShut {NoStop}%
\bibitem [{\citenamefont {Kudin}\ and\ \citenamefont
  {Scuseria}(1998{\natexlab{a}})}]{Kudin_CPL_1998_61}%
  \BibitemOpen
  \bibfield  {author} {\bibinfo {author} {\bibfnamefont {Konstantin~N}\
  \bibnamefont {Kudin}}\ and\ \bibinfo {author} {\bibfnamefont {Gustavo~E}\
  \bibnamefont {Scuseria}},\ }\bibfield  {title} {\enquote {\bibinfo {title} {A
  fast multipole method for periodic systems with arbitrary unit cell
  geometries},}\ }\href {\doibase 10.1016/S0009-2614(97)01329-8} {\bibfield
  {journal} {\bibinfo  {journal} {Chem. Phys. Lett.}\ }\textbf {\bibinfo
  {volume} {283}},\ \bibinfo {pages} {61--68} (\bibinfo {year}
  {1998}{\natexlab{a}})}\BibitemShut {NoStop}%
\bibitem [{\citenamefont {Burow}(2011)}]{Burow_PhD_2011}%
  \BibitemOpen
  \bibfield  {author} {\bibinfo {author} {\bibfnamefont {Asbj{\"o}rn~Manfred}\
  \bibnamefont {Burow}},\ }\href@noop {} {\emph {\bibinfo {title} {Methoden zur
  Beschreibung von chemischen Strukturen beliebiger Dimensionalit{\"a}t mit der
  Dichtefunktionaltheorie unter periodischen Randbedingungen}}}\ (\bibinfo
  {publisher} {Universit{\"a}tsbibliothek der Humboldt-Universit{\"a}t zu
  Berlin},\ \bibinfo {address} {Berlin},\ \bibinfo {year} {2011})\ \bibinfo
  {note} {{doi}~10.18452/16415}\BibitemShut {NoStop}%
\bibitem [{\citenamefont {{\L}azarski}\ \emph {et~al.}(2015)\citenamefont
  {{\L}azarski}, \citenamefont {Burow},\ and\ \citenamefont
  {Sierka}}]{Lazarski_JCTC_2015_3029}%
  \BibitemOpen
  \bibfield  {author} {\bibinfo {author} {\bibfnamefont {Roman}\ \bibnamefont
  {{\L}azarski}}, \bibinfo {author} {\bibfnamefont {Asbj{\"o}rn~M.}\
  \bibnamefont {Burow}}, \ and\ \bibinfo {author} {\bibfnamefont {Marek}\
  \bibnamefont {Sierka}},\ }\bibfield  {title} {\enquote {\bibinfo {title}
  {Density functional theory for molecular and periodic systems using density
  fitting and continuous fast multipole methods},}\ }\href {\doibase
  10.1021/acs.jctc.5b00252} {\bibfield  {journal} {\bibinfo  {journal} {J.
  Chem. Theory Comput.}\ }\textbf {\bibinfo {volume} {11}},\ \bibinfo {pages}
  {3029--3041} (\bibinfo {year} {2015})}\BibitemShut {NoStop}%
\bibitem [{\citenamefont {Irmler}\ \emph {et~al.}(2018)\citenamefont {Irmler},
  \citenamefont {Burow},\ and\ \citenamefont {Pauly}}]{Irmler_JCTC_2018_4567}%
  \BibitemOpen
  \bibfield  {author} {\bibinfo {author} {\bibfnamefont {Andreas}\ \bibnamefont
  {Irmler}}, \bibinfo {author} {\bibfnamefont {Asbj{\"o}rn~M.}\ \bibnamefont
  {Burow}}, \ and\ \bibinfo {author} {\bibfnamefont {Fabian}\ \bibnamefont
  {Pauly}},\ }\bibfield  {title} {\enquote {\bibinfo {title} {Robust periodic
  {F}ock exchange with atom-centered {G}aussian basis sets},}\ }\href {\doibase
  10.1021/acs.jctc.8b00122} {\bibfield  {journal} {\bibinfo  {journal} {J.
  Chem. Theory Comput.}\ }\textbf {\bibinfo {volume} {14}},\ \bibinfo {pages}
  {4567--4580} (\bibinfo {year} {2018})}\BibitemShut {NoStop}%
\bibitem [{\citenamefont {Monkhorst}\ and\ \citenamefont
  {Pack}(1976)}]{Monkhorst_PRB_1976_5188}%
  \BibitemOpen
  \bibfield  {author} {\bibinfo {author} {\bibfnamefont {Hendrik~J.}\
  \bibnamefont {Monkhorst}}\ and\ \bibinfo {author} {\bibfnamefont {James~D.}\
  \bibnamefont {Pack}},\ }\bibfield  {title} {\enquote {\bibinfo {title}
  {Special points for {B}rillouin-zone integrations},}\ }\href {\doibase
  10.1103/PhysRevB.13.5188} {\bibfield  {journal} {\bibinfo  {journal} {Phys.
  Rev. B}\ }\textbf {\bibinfo {volume} {13}},\ \bibinfo {pages} {5188--5192}
  (\bibinfo {year} {1976})}\BibitemShut {NoStop}%
\bibitem [{\citenamefont {Zhu}\ and\ \citenamefont
  {Tew}(2024)}]{Zhu_JPCA_2024_8570}%
  \BibitemOpen
  \bibfield  {author} {\bibinfo {author} {\bibfnamefont {Andrew}\ \bibnamefont
  {Zhu}}\ and\ \bibinfo {author} {\bibfnamefont {David~P.}\ \bibnamefont
  {Tew}},\ }\bibfield  {title} {\enquote {\bibinfo {title} {Wannier function
  localization using bloch intrinsic atomic orbitals},}\ }\href {\doibase
  10.1021/acs.jpca.4c04555} {\bibfield  {journal} {\bibinfo  {journal} {J.
  Phys. Chem. A}\ }\textbf {\bibinfo {volume} {128}},\ \bibinfo {pages}
  {8570--8579} (\bibinfo {year} {2024})}\BibitemShut {NoStop}%
\bibitem [{\citenamefont {Boys}(1960)}]{Boys_RMP_1960_296}%
  \BibitemOpen
  \bibfield  {author} {\bibinfo {author} {\bibfnamefont {S.~F.}\ \bibnamefont
  {Boys}},\ }\bibfield  {title} {\enquote {\bibinfo {title} {Construction of
  some molecular orbitals to be approximately invariant for changes from one
  molecule to another},}\ }\href {\doibase 10.1103/RevModPhys.32.296}
  {\bibfield  {journal} {\bibinfo  {journal} {Rev. Mod. Phys.}\ }\textbf
  {\bibinfo {volume} {32}},\ \bibinfo {pages} {296--299} (\bibinfo {year}
  {1960})}\BibitemShut {NoStop}%
\bibitem [{\citenamefont {Foster}\ and\ \citenamefont
  {Boys}(1960)}]{Foster_RMP_1960_300}%
  \BibitemOpen
  \bibfield  {author} {\bibinfo {author} {\bibfnamefont {J.~M.}\ \bibnamefont
  {Foster}}\ and\ \bibinfo {author} {\bibfnamefont {S.~F.}\ \bibnamefont
  {Boys}},\ }\bibfield  {title} {\enquote {\bibinfo {title} {Canonical
  configurational interaction procedure},}\ }\href {\doibase
  10.1103/RevModPhys.32.300} {\bibfield  {journal} {\bibinfo  {journal} {Rev.
  Mod. Phys.}\ }\textbf {\bibinfo {volume} {32}},\ \bibinfo {pages} {300--302}
  (\bibinfo {year} {1960})}\BibitemShut {NoStop}%
\bibitem [{\citenamefont {Marzari}\ and\ \citenamefont
  {Vanderbilt}(1997)}]{Marzari_PRB_1997_12847}%
  \BibitemOpen
  \bibfield  {author} {\bibinfo {author} {\bibfnamefont {Nicola}\ \bibnamefont
  {Marzari}}\ and\ \bibinfo {author} {\bibfnamefont {David}\ \bibnamefont
  {Vanderbilt}},\ }\bibfield  {title} {\enquote {\bibinfo {title} {Maximally
  localized generalized wannier functions for composite energy bands},}\ }\href
  {\doibase 10.1103/PhysRevB.56.12847} {\bibfield  {journal} {\bibinfo
  {journal} {Phys. Rev. B}\ }\textbf {\bibinfo {volume} {56}},\ \bibinfo
  {pages} {12847--12865} (\bibinfo {year} {1997})}\BibitemShut {NoStop}%
\bibitem [{\citenamefont {Pizzi}\ \emph {et~al.}(2020)\citenamefont {Pizzi},
  \citenamefont {Vitale}, \citenamefont {Arita}, \citenamefont {Bl{\"u}gel},
  \citenamefont {Freimuth}, \citenamefont {G{\'e}ranton}, \citenamefont
  {Gibertini}, \citenamefont {Gresch}, \citenamefont {Johnson}, \citenamefont
  {Koretsune}, \citenamefont {Iba\~{n}ez Azpiroz}, \citenamefont {Lee},
  \citenamefont {Lihm}, \citenamefont {Marchand}, \citenamefont {Marrazzo},
  \citenamefont {Mokrousov}, \citenamefont {Mustafa}, \citenamefont {Nohara},
  \citenamefont {Nomura}, \citenamefont {Paulatto}, \citenamefont {Ponc{\'e}},
  \citenamefont {Ponweiser}, \citenamefont {Qiao}, \citenamefont {Th{\"o}le},
  \citenamefont {Tsirkin}, \citenamefont {Wierzbowska}, \citenamefont
  {Marzari}, \citenamefont {Vanderbilt}, \citenamefont {Souza}, \citenamefont
  {Mostofi},\ and\ \citenamefont {Yates}}]{Pizzi_JPCM_2020_165902}%
  \BibitemOpen
  \bibfield  {author} {\bibinfo {author} {\bibfnamefont {Giovanni}\
  \bibnamefont {Pizzi}}, \bibinfo {author} {\bibfnamefont {Valerio}\
  \bibnamefont {Vitale}}, \bibinfo {author} {\bibfnamefont {Ryotaro}\
  \bibnamefont {Arita}}, \bibinfo {author} {\bibfnamefont {Stefan}\
  \bibnamefont {Bl{\"u}gel}}, \bibinfo {author} {\bibfnamefont {Frank}\
  \bibnamefont {Freimuth}}, \bibinfo {author} {\bibfnamefont {Guillaume}\
  \bibnamefont {G{\'e}ranton}}, \bibinfo {author} {\bibfnamefont {Marco}\
  \bibnamefont {Gibertini}}, \bibinfo {author} {\bibfnamefont {Dominik}\
  \bibnamefont {Gresch}}, \bibinfo {author} {\bibfnamefont {Charles}\
  \bibnamefont {Johnson}}, \bibinfo {author} {\bibfnamefont {Takashi}\
  \bibnamefont {Koretsune}}, \bibinfo {author} {\bibfnamefont {Julen}\
  \bibnamefont {Iba\~{n}ez Azpiroz}}, \bibinfo {author} {\bibfnamefont
  {Hyungjun}\ \bibnamefont {Lee}}, \bibinfo {author} {\bibfnamefont {Jae-Mo}\
  \bibnamefont {Lihm}}, \bibinfo {author} {\bibfnamefont {Daniel}\ \bibnamefont
  {Marchand}}, \bibinfo {author} {\bibfnamefont {Antimo}\ \bibnamefont
  {Marrazzo}}, \bibinfo {author} {\bibfnamefont {Yuriy}\ \bibnamefont
  {Mokrousov}}, \bibinfo {author} {\bibfnamefont {Jamal~I}\ \bibnamefont
  {Mustafa}}, \bibinfo {author} {\bibfnamefont {Yoshiro}\ \bibnamefont
  {Nohara}}, \bibinfo {author} {\bibfnamefont {Yusuke}\ \bibnamefont {Nomura}},
  \bibinfo {author} {\bibfnamefont {Lorenzo}\ \bibnamefont {Paulatto}},
  \bibinfo {author} {\bibfnamefont {Samuel}\ \bibnamefont {Ponc{\'e}}},
  \bibinfo {author} {\bibfnamefont {Thomas}\ \bibnamefont {Ponweiser}},
  \bibinfo {author} {\bibfnamefont {Junfeng}\ \bibnamefont {Qiao}}, \bibinfo
  {author} {\bibfnamefont {Florian}\ \bibnamefont {Th{\"o}le}}, \bibinfo
  {author} {\bibfnamefont {Stepan~S}\ \bibnamefont {Tsirkin}}, \bibinfo
  {author} {\bibfnamefont {Ma{\l}gorzata}\ \bibnamefont {Wierzbowska}},
  \bibinfo {author} {\bibfnamefont {Nicola}\ \bibnamefont {Marzari}}, \bibinfo
  {author} {\bibfnamefont {David}\ \bibnamefont {Vanderbilt}}, \bibinfo
  {author} {\bibfnamefont {Ivo}\ \bibnamefont {Souza}}, \bibinfo {author}
  {\bibfnamefont {Arash~A}\ \bibnamefont {Mostofi}}, \ and\ \bibinfo {author}
  {\bibfnamefont {Jonathan~R}\ \bibnamefont {Yates}},\ }\bibfield  {title}
  {\enquote {\bibinfo {title} {Wannier90 as a community code: new features and
  applications},}\ }\href {\doibase 10.1088/1361-648X/ab51ff} {\bibfield
  {journal} {\bibinfo  {journal} {J. Phys.: Condens. Matter}\ }\textbf
  {\bibinfo {volume} {32}},\ \bibinfo {pages} {165902} (\bibinfo {year}
  {2020})}\BibitemShut {NoStop}%
\bibitem [{\citenamefont {J{\'o}nsson}\ \emph {et~al.}(2017)\citenamefont
  {J{\'o}nsson}, \citenamefont {Lehtola}, \citenamefont {Puska},\ and\
  \citenamefont {J{\'o}nsson}}]{Jonsson_JCTC_2017_460}%
  \BibitemOpen
  \bibfield  {author} {\bibinfo {author} {\bibfnamefont {Elvar~{\"O}.}\
  \bibnamefont {J{\'o}nsson}}, \bibinfo {author} {\bibfnamefont {Susi}\
  \bibnamefont {Lehtola}}, \bibinfo {author} {\bibfnamefont {Martti}\
  \bibnamefont {Puska}}, \ and\ \bibinfo {author} {\bibfnamefont {Hannes}\
  \bibnamefont {J{\'o}nsson}},\ }\bibfield  {title} {\enquote {\bibinfo {title}
  {Theory and applications of generalized {P}ipek-{M}ezey {W}annier
  functions},}\ }\href {\doibase 10.1021/acs.jctc.6b00809} {\bibfield
  {journal} {\bibinfo  {journal} {J. Chem. Theory Comput.}\ }\textbf {\bibinfo
  {volume} {13}},\ \bibinfo {pages} {460--474} (\bibinfo {year}
  {2017})}\BibitemShut {NoStop}%
\bibitem [{\citenamefont {Clement}\ \emph {et~al.}(2021)\citenamefont
  {Clement}, \citenamefont {Wang},\ and\ \citenamefont
  {Valeev}}]{Clement_JCTC_2021_7406}%
  \BibitemOpen
  \bibfield  {author} {\bibinfo {author} {\bibfnamefont {Marjory~C.}\
  \bibnamefont {Clement}}, \bibinfo {author} {\bibfnamefont {Xiao}\
  \bibnamefont {Wang}}, \ and\ \bibinfo {author} {\bibfnamefont {Edward~F.}\
  \bibnamefont {Valeev}},\ }\bibfield  {title} {\enquote {\bibinfo {title}
  {Robust {P}ipek-{M}ezey orbital localization in periodic solids},}\ }\href
  {\doibase 10.1021/acs.jctc.1c00238} {\bibfield  {journal} {\bibinfo
  {journal} {J. Chem. Theory Comput.}\ }\textbf {\bibinfo {volume} {17}},\
  \bibinfo {pages} {7406--7415} (\bibinfo {year} {2021})}\BibitemShut {NoStop}%
\bibitem [{\citenamefont {Schreder}\ and\ \citenamefont
  {Luber}(2024)}]{Schreder_JCP_2024_214117}%
  \BibitemOpen
  \bibfield  {author} {\bibinfo {author} {\bibfnamefont {Lukas}\ \bibnamefont
  {Schreder}}\ and\ \bibinfo {author} {\bibfnamefont {Sandra}\ \bibnamefont
  {Luber}},\ }\bibfield  {title} {\enquote {\bibinfo {title} {Propagated
  (fragment) {P}ipek-{M}ezey {W}annier functions in real-time time-dependent
  density functional theory},}\ }\href {\doibase 10.1063/5.0203442} {\bibfield
  {journal} {\bibinfo  {journal} {J. Chem. Phys.}\ }\textbf {\bibinfo {volume}
  {160}},\ \bibinfo {pages} {214117} (\bibinfo {year} {2024})}\BibitemShut
  {NoStop}%
\bibitem [{\citenamefont {Knizia}(2013)}]{Knizia_JCTC_2013_4834}%
  \BibitemOpen
  \bibfield  {author} {\bibinfo {author} {\bibfnamefont {Gerald}\ \bibnamefont
  {Knizia}},\ }\bibfield  {title} {\enquote {\bibinfo {title} {Intrinsic atomic
  orbitals: An unbiased bridge between quantum theory and chemical concepts},}\
  }\href {\doibase 10.1021/ct400687b} {\bibfield  {journal} {\bibinfo
  {journal} {J. Chem. Theory Comput.}\ }\textbf {\bibinfo {volume} {9}},\
  \bibinfo {pages} {4834--4843} (\bibinfo {year} {2013})}\BibitemShut {NoStop}%
\bibitem [{\citenamefont {Tew}(2019)}]{Tew_JCTC_2019_6597}%
  \BibitemOpen
  \bibfield  {author} {\bibinfo {author} {\bibfnamefont {David~P.}\
  \bibnamefont {Tew}},\ }\bibfield  {title} {\enquote {\bibinfo {title}
  {Principal domains in local correlation theory},}\ }\href {\doibase
  10.1021/acs.jctc.9b00619} {\bibfield  {journal} {\bibinfo  {journal} {J.
  Chem. Theory Comput.}\ }\textbf {\bibinfo {volume} {15}},\ \bibinfo {pages}
  {6597--6606} (\bibinfo {year} {2019})}\BibitemShut {NoStop}%
\bibitem [{\citenamefont {Alml\"{o}f}(1991)}]{Almlof_CPL_1991_319}%
  \BibitemOpen
  \bibfield  {author} {\bibinfo {author} {\bibfnamefont {Jan}\ \bibnamefont
  {Alml\"{o}f}},\ }\bibfield  {title} {\enquote {\bibinfo {title} {Elimination
  of energy denominators in m{\o}ller--plesset perturbation theory by a laplace
  transform approach},}\ }\href {\doibase
  https://doi.org/10.1016/0009-2614(91)80078-C} {\bibfield  {journal} {\bibinfo
   {journal} {Chem. Phys. Lett.}\ }\textbf {\bibinfo {volume} {181}},\ \bibinfo
  {pages} {319--320} (\bibinfo {year} {1991})}\BibitemShut {NoStop}%
\bibitem [{\citenamefont {H{\"a}ser}\ and\ \citenamefont
  {Alml{\"o}f}(1992)}]{Haeser_JCP_1992_489}%
  \BibitemOpen
  \bibfield  {author} {\bibinfo {author} {\bibfnamefont {Marco}\ \bibnamefont
  {H{\"a}ser}}\ and\ \bibinfo {author} {\bibfnamefont {Jan}\ \bibnamefont
  {Alml{\"o}f}},\ }\bibfield  {title} {\enquote {\bibinfo {title} {Laplace
  transform techniques in m{\o}ller--plesset perturbation theory},}\ }\href
  {\doibase 10.1063/1.462485} {\bibfield  {journal} {\bibinfo  {journal} {J.
  Chem. Phys.}\ }\textbf {\bibinfo {volume} {96}},\ \bibinfo {pages} {489--494}
  (\bibinfo {year} {1992})}\BibitemShut {NoStop}%
\bibitem [{\citenamefont {Vahtras}\ \emph {et~al.}(1993)\citenamefont
  {Vahtras}, \citenamefont {Alml{\"o}f},\ and\ \citenamefont
  {Feyereisen}}]{Vahtras_CPL_1993}%
  \BibitemOpen
  \bibfield  {author} {\bibinfo {author} {\bibfnamefont {O.}~\bibnamefont
  {Vahtras}}, \bibinfo {author} {\bibfnamefont {J.}~\bibnamefont {Alml{\"o}f}},
  \ and\ \bibinfo {author} {\bibfnamefont {M.~W.}\ \bibnamefont {Feyereisen}},\
  }\bibfield  {title} {\enquote {\bibinfo {title} {Integral approximations for
  {LCAO}-{SCF} calculations},}\ }\href {\doibase 10.1016/0009-2614(93)89151-7}
  {\bibfield  {journal} {\bibinfo  {journal} {Chem. Phys. Lett.}\ }\textbf
  {\bibinfo {volume} {213}},\ \bibinfo {pages} {514--518} (\bibinfo {year}
  {1993})}\BibitemShut {NoStop}%
\bibitem [{\citenamefont {Weigend}\ \emph {et~al.}(1998)\citenamefont
  {Weigend}, \citenamefont {H{\"a}ser}, \citenamefont {Patzelt},\ and\
  \citenamefont {Ahlrichs}}]{Weigend_CPL_1998_143}%
  \BibitemOpen
  \bibfield  {author} {\bibinfo {author} {\bibfnamefont {Florian}\ \bibnamefont
  {Weigend}}, \bibinfo {author} {\bibfnamefont {Marco}\ \bibnamefont
  {H{\"a}ser}}, \bibinfo {author} {\bibfnamefont {Holger}\ \bibnamefont
  {Patzelt}}, \ and\ \bibinfo {author} {\bibfnamefont {Reinhart}\ \bibnamefont
  {Ahlrichs}},\ }\bibfield  {title} {\enquote {\bibinfo {title} {{RI-MP2}:
  optimized auxiliary basis sets and demonstration of efficiency},}\ }\href
  {\doibase 10.1016/S0009-2614(98)00862-8} {\bibfield  {journal} {\bibinfo
  {journal} {Chem. Phys. Lett.}\ }\textbf {\bibinfo {volume} {294}},\ \bibinfo
  {pages} {143--152} (\bibinfo {year} {1998})}\BibitemShut {NoStop}%
\bibitem [{\citenamefont {Tew}(2018)}]{Tew_JCP_2018_011102}%
  \BibitemOpen
  \bibfield  {author} {\bibinfo {author} {\bibfnamefont {David~P.}\
  \bibnamefont {Tew}},\ }\bibfield  {title} {\enquote {\bibinfo {title}
  {Communication: Quasi-robust local density fitting},}\ }\href {\doibase
  10.1063/1.5013111} {\bibfield  {journal} {\bibinfo  {journal} {J. Chem.
  Phys.}\ }\textbf {\bibinfo {volume} {148}},\ \bibinfo {pages} {011102}
  (\bibinfo {year} {2018})}\BibitemShut {NoStop}%
\bibitem [{\citenamefont {Dunlap}(2000)}]{Dunlap_PCCP_2000_2113}%
  \BibitemOpen
  \bibfield  {author} {\bibinfo {author} {\bibfnamefont {Brett~I.}\
  \bibnamefont {Dunlap}},\ }\bibfield  {title} {\enquote {\bibinfo {title}
  {Robust and variational fitting},}\ }\href {\doibase 10.1039/B000027M}
  {\bibfield  {journal} {\bibinfo  {journal} {Phys. Chem. Chem. Phys.}\
  }\textbf {\bibinfo {volume} {2}},\ \bibinfo {pages} {2113--2116} (\bibinfo
  {year} {2000})}\BibitemShut {NoStop}%
\bibitem [{\citenamefont {Nijboer}\ and\ \citenamefont {{De
  Wette}}(1958)}]{Nijboer_Physica_1958_422}%
  \BibitemOpen
  \bibfield  {author} {\bibinfo {author} {\bibfnamefont {B.~R.~A.}\
  \bibnamefont {Nijboer}}\ and\ \bibinfo {author} {\bibfnamefont {F.~W.}\
  \bibnamefont {{De Wette}}},\ }\bibfield  {title} {\enquote {\bibinfo {title}
  {The internal field in dipole lattices},}\ }\href {\doibase
  10.1016/S0031-8914(58)95803-8} {\bibfield  {journal} {\bibinfo  {journal}
  {Physica}\ }\textbf {\bibinfo {volume} {24}},\ \bibinfo {pages} {422--431}
  (\bibinfo {year} {1958})}\BibitemShut {NoStop}%
\bibitem [{\citenamefont {Madelung}(1918)}]{Madelung_PZ_1918_524}%
  \BibitemOpen
  \bibfield  {author} {\bibinfo {author} {\bibfnamefont {E.}~\bibnamefont
  {Madelung}},\ }\bibfield  {title} {\enquote {\bibinfo {title} {Das
  elektrische feld in systemen von regelm{\"a}ßig angeordneten
  punktladungen},}\ }\href@noop {} {\bibfield  {journal} {\bibinfo  {journal}
  {Phys. Z.}\ }\textbf {\bibinfo {volume} {19}},\ \bibinfo {pages} {524--533}
  (\bibinfo {year} {1918})}\BibitemShut {NoStop}%
\bibitem [{\citenamefont {Ewald}(1921)}]{Ewald_AP_1921}%
  \BibitemOpen
  \bibfield  {author} {\bibinfo {author} {\bibfnamefont {P.~P.}\ \bibnamefont
  {Ewald}},\ }\bibfield  {title} {\enquote {\bibinfo {title} {Die berechnung
  optischer und elektrostatischer gitterpotentiale},}\ }\href {\doibase
  10.1002/andp.19213690304} {\bibfield  {journal} {\bibinfo  {journal} {Ann.
  Phys.}\ }\textbf {\bibinfo {volume} {369}},\ \bibinfo {pages} {253--287}
  (\bibinfo {year} {1921})}\BibitemShut {NoStop}%
\bibitem [{\citenamefont {Stuart}(1978)}]{Stuart_JCompP_1978_127}%
  \BibitemOpen
  \bibfield  {author} {\bibinfo {author} {\bibfnamefont {S.N}\ \bibnamefont
  {Stuart}},\ }\bibfield  {title} {\enquote {\bibinfo {title} {Depolarization
  correction for {C}oulomb lattice sums},}\ }\href {\doibase
  10.1016/0021-9991(78)90113-4} {\bibfield  {journal} {\bibinfo  {journal} {J.
  Comput. Phys.}\ }\textbf {\bibinfo {volume} {29}},\ \bibinfo {pages}
  {127--132} (\bibinfo {year} {1978})}\BibitemShut {NoStop}%
\bibitem [{\citenamefont {Harris}(1975)}]{Harris_chapter_1975}%
  \BibitemOpen
  \bibfield  {author} {\bibinfo {author} {\bibfnamefont {Frank~E.}\
  \bibnamefont {Harris}},\ }\bibfield  {title} {\enquote {\bibinfo {title}
  {Hartree-fock studies of electronic structures of crystalline solids},}\ }in\
  \href {\doibase 10.1016/B978-0-12-681901-4.50011-8} {\emph {\bibinfo
  {booktitle} {Theoretical Chemistry: Advances and Perspectives}}},\ \bibinfo
  {series} {Theoretical Chemistry}, Vol.~\bibinfo {volume} {1},\ \bibinfo
  {editor} {edited by\ \bibinfo {editor} {\bibfnamefont {Henry}\ \bibnamefont
  {Eyring}}\ and\ \bibinfo {editor} {\bibfnamefont {Doublas}\ \bibnamefont
  {Henderson}}}\ (\bibinfo  {publisher} {Academic Press},\ \bibinfo {address}
  {New York},\ \bibinfo {year} {1975})\ pp.\ \bibinfo {pages}
  {147--218}\BibitemShut {NoStop}%
\bibitem [{\citenamefont {Saunders}\ \emph {et~al.}(1992)\citenamefont
  {Saunders}, \citenamefont {{Freyria-Fava}}, \citenamefont {Dovesi},
  \citenamefont {Salasco},\ and\ \citenamefont
  {Roetti}}]{Saunders_MP_1992_629}%
  \BibitemOpen
  \bibfield  {author} {\bibinfo {author} {\bibfnamefont {V.~R.}\ \bibnamefont
  {Saunders}}, \bibinfo {author} {\bibfnamefont {C.}~\bibnamefont
  {{Freyria-Fava}}}, \bibinfo {author} {\bibfnamefont {R.}~\bibnamefont
  {Dovesi}}, \bibinfo {author} {\bibfnamefont {L.}~\bibnamefont {Salasco}}, \
  and\ \bibinfo {author} {\bibfnamefont {C.}~\bibnamefont {Roetti}},\
  }\bibfield  {title} {\enquote {\bibinfo {title} {On the electrostatic
  potential in crystalline systems where the charge density is expanded in
  {G}aussian functions},}\ }\href {\doibase 10.1080/00268979200102671}
  {\bibfield  {journal} {\bibinfo  {journal} {Mol. Phys.}\ }\textbf {\bibinfo
  {volume} {77}},\ \bibinfo {pages} {629--665} (\bibinfo {year}
  {1992})}\BibitemShut {NoStop}%
\bibitem [{\citenamefont {{De Wette}}\ and\ \citenamefont
  {Nijboer}(1958)}]{DeWette_Physica_1958_1105}%
  \BibitemOpen
  \bibfield  {author} {\bibinfo {author} {\bibfnamefont {F.~W.}\ \bibnamefont
  {{De Wette}}}\ and\ \bibinfo {author} {\bibfnamefont {B.~R.~A.}\ \bibnamefont
  {Nijboer}},\ }\bibfield  {title} {\enquote {\bibinfo {title} {The
  electrostatic potential in multipole lattices},}\ }\href {\doibase
  10.1016/S0031-8914(58)80132-9} {\bibfield  {journal} {\bibinfo  {journal}
  {Physica}\ }\textbf {\bibinfo {volume} {24}},\ \bibinfo {pages} {1105--1118}
  (\bibinfo {year} {1958})}\BibitemShut {NoStop}%
\bibitem [{\citenamefont {Kudin}\ and\ \citenamefont
  {Scuseria}(1998{\natexlab{b}})}]{Kudin_CPL_1998_611}%
  \BibitemOpen
  \bibfield  {author} {\bibinfo {author} {\bibfnamefont {Konstantin~N}\
  \bibnamefont {Kudin}}\ and\ \bibinfo {author} {\bibfnamefont {Gustavo~E}\
  \bibnamefont {Scuseria}},\ }\bibfield  {title} {\enquote {\bibinfo {title} {A
  fast multipole algorithm for the efficient treatment of the {C}oulomb problem
  in electronic structure calculations of periodic systems with {G}aussian
  orbitals},}\ }\href {\doibase 10.1016/S0009-2614(98)00468-0} {\bibfield
  {journal} {\bibinfo  {journal} {Chem. Phys. Lett.}\ }\textbf {\bibinfo
  {volume} {289}},\ \bibinfo {pages} {611--616} (\bibinfo {year}
  {1998}{\natexlab{b}})}\BibitemShut {NoStop}%
\bibitem [{\citenamefont {Kudin}\ and\ \citenamefont
  {Scuseria}(2004)}]{Kudin_JCP_2004_2886}%
  \BibitemOpen
  \bibfield  {author} {\bibinfo {author} {\bibfnamefont {Konstantin~N.}\
  \bibnamefont {Kudin}}\ and\ \bibinfo {author} {\bibfnamefont {Gustavo~E.}\
  \bibnamefont {Scuseria}},\ }\bibfield  {title} {\enquote {\bibinfo {title}
  {Revisiting infinite lattice sums with the periodic fast multipole method},}\
  }\href {\doibase 10.1063/1.1771634} {\bibfield  {journal} {\bibinfo
  {journal} {J. Chem. Phys.}\ }\textbf {\bibinfo {volume} {121}},\ \bibinfo
  {pages} {2886--2890} (\bibinfo {year} {2004})}\BibitemShut {NoStop}%
\bibitem [{\citenamefont {Redlack}\ and\ \citenamefont
  {Grindlay}(1975)}]{Redlack_JPCS_1975_73}%
  \BibitemOpen
  \bibfield  {author} {\bibinfo {author} {\bibfnamefont {A.}~\bibnamefont
  {Redlack}}\ and\ \bibinfo {author} {\bibfnamefont {J.}~\bibnamefont
  {Grindlay}},\ }\bibfield  {title} {\enquote {\bibinfo {title} {Coulombic
  potential lattice sums},}\ }\href {\doibase 10.1016/0022-3697(75)90116-X}
  {\bibfield  {journal} {\bibinfo  {journal} {J. Phys. Chem. Solids}\ }\textbf
  {\bibinfo {volume} {36}},\ \bibinfo {pages} {73--82} (\bibinfo {year}
  {1975})}\BibitemShut {NoStop}%
\bibitem [{\citenamefont {Roberts}\ and\ \citenamefont
  {Schnitker}(1994)}]{Roberts_JCP_1994_5024}%
  \BibitemOpen
  \bibfield  {author} {\bibinfo {author} {\bibfnamefont {James~E.}\
  \bibnamefont {Roberts}}\ and\ \bibinfo {author} {\bibfnamefont {Jurgen}\
  \bibnamefont {Schnitker}},\ }\bibfield  {title} {\enquote {\bibinfo {title}
  {How the unit cell surface charge distribution affects the energetics of
  ion--solvent interactions in simulations},}\ }\href {\doibase
  10.1063/1.467425} {\bibfield  {journal} {\bibinfo  {journal} {J. Chem.
  Phys.}\ }\textbf {\bibinfo {volume} {101}},\ \bibinfo {pages} {5024--5031}
  (\bibinfo {year} {1994})}\BibitemShut {NoStop}%
\bibitem [{\citenamefont {Herce}\ \emph {et~al.}(2007)\citenamefont {Herce},
  \citenamefont {Garcia},\ and\ \citenamefont
  {Darden}}]{Herce_JCP_2007_124106}%
  \BibitemOpen
  \bibfield  {author} {\bibinfo {author} {\bibfnamefont {Henry~David}\
  \bibnamefont {Herce}}, \bibinfo {author} {\bibfnamefont {Angel~Enrique}\
  \bibnamefont {Garcia}}, \ and\ \bibinfo {author} {\bibfnamefont {Thomas}\
  \bibnamefont {Darden}},\ }\bibfield  {title} {\enquote {\bibinfo {title} {The
  electrostatic surface term: ({I}) periodic systems},}\ }\href {\doibase
  10.1063/1.2714527} {\bibfield  {journal} {\bibinfo  {journal} {J. Chem.
  Phys.}\ }\textbf {\bibinfo {volume} {126}},\ \bibinfo {pages} {124106}
  (\bibinfo {year} {2007})}\BibitemShut {NoStop}%
\bibitem [{\citenamefont {Ballenegger}(2014)}]{Ballenegger_JCP_2014_161102}%
  \BibitemOpen
  \bibfield  {author} {\bibinfo {author} {\bibfnamefont {V.}~\bibnamefont
  {Ballenegger}},\ }\bibfield  {title} {\enquote {\bibinfo {title}
  {Communication: On the origin of the surface term in the {E}wald formula},}\
  }\href {\doibase 10.1063/1.4872019} {\bibfield  {journal} {\bibinfo
  {journal} {J. Chem. Phys.}\ }\textbf {\bibinfo {volume} {140}},\ \bibinfo
  {pages} {161102} (\bibinfo {year} {2014})}\BibitemShut {NoStop}%
\bibitem [{\citenamefont {{\L}azarski}\ \emph {et~al.}(2016)\citenamefont
  {{\L}azarski}, \citenamefont {Burow}, \citenamefont {Grajciar},\ and\
  \citenamefont {Sierka}}]{Lazarski_JCC_2016_2518}%
  \BibitemOpen
  \bibfield  {author} {\bibinfo {author} {\bibfnamefont {Roman}\ \bibnamefont
  {{\L}azarski}}, \bibinfo {author} {\bibfnamefont {Asbj{\"o}rn~Manfred}\
  \bibnamefont {Burow}}, \bibinfo {author} {\bibfnamefont {Luk{\' a}{\v s}}\
  \bibnamefont {Grajciar}}, \ and\ \bibinfo {author} {\bibfnamefont {Marek}\
  \bibnamefont {Sierka}},\ }\bibfield  {title} {\enquote {\bibinfo {title}
  {Density functional theory for molecular and periodic systems using density
  fitting and continuous fast multipole method: Analytical gradients},}\ }\href
  {\doibase 10.1002/jcc.24477} {\bibfield  {journal} {\bibinfo  {journal} {J.
  Comput. Chem.}\ }\textbf {\bibinfo {volume} {37}},\ \bibinfo {pages}
  {2518--2526} (\bibinfo {year} {2016})}\BibitemShut {NoStop}%
\bibitem [{\citenamefont {Burow}\ and\ \citenamefont
  {Sierka}(2011)}]{Burow_JCTC_2011_3097}%
  \BibitemOpen
  \bibfield  {author} {\bibinfo {author} {\bibfnamefont {Asbj{\"o}rn~M.}\
  \bibnamefont {Burow}}\ and\ \bibinfo {author} {\bibfnamefont {Marek}\
  \bibnamefont {Sierka}},\ }\bibfield  {title} {\enquote {\bibinfo {title}
  {Linear scaling hierarchical integration scheme for the exchange-correlation
  term in molecular and periodic systems},}\ }\href {\doibase
  10.1021/ct200412r} {\bibfield  {journal} {\bibinfo  {journal} {J. Chem.
  Theory Comput.}\ }\textbf {\bibinfo {volume} {7}},\ \bibinfo {pages}
  {3097--3104} (\bibinfo {year} {2011})}\BibitemShut {NoStop}%
\bibitem [{\citenamefont {M{\"u}ller}\ \emph {et~al.}(2020)\citenamefont
  {M{\"u}ller}, \citenamefont {Sharma},\ and\ \citenamefont
  {Sierka}}]{Mueller_JCC_2020_2573}%
  \BibitemOpen
  \bibfield  {author} {\bibinfo {author} {\bibfnamefont {Carolin}\ \bibnamefont
  {M{\"u}ller}}, \bibinfo {author} {\bibfnamefont {Manas}\ \bibnamefont
  {Sharma}}, \ and\ \bibinfo {author} {\bibfnamefont {Marek}\ \bibnamefont
  {Sierka}},\ }\bibfield  {title} {\enquote {\bibinfo {title} {Real-time
  time-dependent density functional theory using density fitting and the
  continuous fast multipole method},}\ }\href {\doibase 10.1002/jcc.26412}
  {\bibfield  {journal} {\bibinfo  {journal} {J. Comput. Chem.}\ }\textbf
  {\bibinfo {volume} {41}},\ \bibinfo {pages} {2573--2582} (\bibinfo {year}
  {2020})}\BibitemShut {NoStop}%
\bibitem [{\citenamefont {Peintinger}\ \emph {et~al.}(2013)\citenamefont
  {Peintinger}, \citenamefont {Oliveira},\ and\ \citenamefont
  {Bredow}}]{Peintinger_JCC_2013_451}%
  \BibitemOpen
  \bibfield  {author} {\bibinfo {author} {\bibfnamefont {Michael~F.}\
  \bibnamefont {Peintinger}}, \bibinfo {author} {\bibfnamefont {Daniel~Vilela}\
  \bibnamefont {Oliveira}}, \ and\ \bibinfo {author} {\bibfnamefont {Thomas}\
  \bibnamefont {Bredow}},\ }\bibfield  {title} {\enquote {\bibinfo {title}
  {Consistent gaussian basis sets of triple-zeta valence with polarization
  quality for solid-state calculations},}\ }\href {\doibase 10.1002/jcc.23153}
  {\bibfield  {journal} {\bibinfo  {journal} {J. Comput. Chem.}\ }\textbf
  {\bibinfo {volume} {34}},\ \bibinfo {pages} {451--459} (\bibinfo {year}
  {2013})}\BibitemShut {NoStop}%
\bibitem [{\citenamefont {Laun}\ and\ \citenamefont
  {Bredow}(2022)}]{Laun_JCC_2022_839}%
  \BibitemOpen
  \bibfield  {author} {\bibinfo {author} {\bibfnamefont {Joachim}\ \bibnamefont
  {Laun}}\ and\ \bibinfo {author} {\bibfnamefont {Thomas}\ \bibnamefont
  {Bredow}},\ }\bibfield  {title} {\enquote {\bibinfo {title} {{BSSE}-corrected
  consistent gaussian basis sets of triple-zeta valence with polarization
  quality of the fifth period for solid-state calculations},}\ }\href {\doibase
  https://doi.org/10.1002/jcc.26839} {\bibfield  {journal} {\bibinfo  {journal}
  {J. Comput. Chem.}\ }\textbf {\bibinfo {volume} {43}},\ \bibinfo {pages}
  {839--846} (\bibinfo {year} {2022})}\BibitemShut {NoStop}%
\bibitem [{\citenamefont {Weigend}(2006)}]{Weigend_PCCP_2006_1057}%
  \BibitemOpen
  \bibfield  {author} {\bibinfo {author} {\bibfnamefont {Florian}\ \bibnamefont
  {Weigend}},\ }\bibfield  {title} {\enquote {\bibinfo {title} {Accurate
  coulomb-fitting basis sets for {H} to {Rn}},}\ }\href {\doibase
  10.1039/B515623H} {\bibfield  {journal} {\bibinfo  {journal} {Phys. Chem.
  Chem. Phys.}\ }\textbf {\bibinfo {volume} {8}},\ \bibinfo {pages}
  {1057--1065} (\bibinfo {year} {2006})}\BibitemShut {NoStop}%
\bibitem [{\citenamefont {Sorathia}\ and\ \citenamefont
  {Tew}(2020)}]{Sorathia_JCP_2020_174112}%
  \BibitemOpen
  \bibfield  {author} {\bibinfo {author} {\bibfnamefont {Kesha}\ \bibnamefont
  {Sorathia}}\ and\ \bibinfo {author} {\bibfnamefont {David~P.}\ \bibnamefont
  {Tew}},\ }\bibfield  {title} {\enquote {\bibinfo {title} {Basis set
  extrapolation in pair natural orbital theories},}\ }\href {\doibase
  10.1063/5.0022077} {\bibfield  {journal} {\bibinfo  {journal} {J. Chem.
  Phys.}\ }\textbf {\bibinfo {volume} {153}},\ \bibinfo {pages} {174112}
  (\bibinfo {year} {2020})}\BibitemShut {NoStop}%
\bibitem [{\citenamefont {Sorathia}\ \emph {et~al.}(2024)\citenamefont
  {Sorathia}, \citenamefont {Frantzov},\ and\ \citenamefont
  {Tew}}]{Sorathia_JCTC_2024_2740}%
  \BibitemOpen
  \bibfield  {author} {\bibinfo {author} {\bibfnamefont {Kesha}\ \bibnamefont
  {Sorathia}}, \bibinfo {author} {\bibfnamefont {Damyan}\ \bibnamefont
  {Frantzov}}, \ and\ \bibinfo {author} {\bibfnamefont {David~P.}\ \bibnamefont
  {Tew}},\ }\bibfield  {title} {\enquote {\bibinfo {title} {Improved {CPS} and
  {CBS} extrapolation of {PNO-CCSD(T)} energies: The {MOBH35} and {ISOL24} data
  sets},}\ }\href {\doibase 10.1021/acs.jctc.3c00974} {\bibfield  {journal}
  {\bibinfo  {journal} {J. Chem. Theory and Comput.}\ }\textbf {\bibinfo
  {volume} {20}},\ \bibinfo {pages} {2740--2750} (\bibinfo {year}
  {2024})}\BibitemShut {NoStop}%
\bibitem [{\citenamefont {Weigend}\ and\ \citenamefont
  {H{\"a}ser}(1997)}]{Weigend_TCA_1997_331}%
  \BibitemOpen
  \bibfield  {author} {\bibinfo {author} {\bibfnamefont {Florian}\ \bibnamefont
  {Weigend}}\ and\ \bibinfo {author} {\bibfnamefont {Marco}\ \bibnamefont
  {H{\"a}ser}},\ }\bibfield  {title} {\enquote {\bibinfo {title} {{RI-MP2}:
  first derivatives and global consistency},}\ }\href {\doibase
  10.1007/s002140050269} {\bibfield  {journal} {\bibinfo  {journal} {Theor.
  Chem. Acc.}\ }\textbf {\bibinfo {volume} {97}},\ \bibinfo {pages} {331--340}
  (\bibinfo {year} {1997})}\BibitemShut {NoStop}%
\bibitem [{\citenamefont {Balasubramani}\ \emph {et~al.}(2020)\citenamefont
  {Balasubramani}, \citenamefont {Chen}, \citenamefont {Coriani}, \citenamefont
  {Diedenhofen}, \citenamefont {Frank}, \citenamefont {Franzke}, \citenamefont
  {Furche}, \citenamefont {Grotjahn}, \citenamefont {Harding}, \citenamefont
  {H{\"a}ttig}, \citenamefont {Hellweg}, \citenamefont {Helmich-Paris},
  \citenamefont {Holzer}, \citenamefont {Huniar}, \citenamefont {Kaupp},
  \citenamefont {Marefat~Khah}, \citenamefont {Karbalaei~Khani}, \citenamefont
  {M{\"u}ller}, \citenamefont {Mack}, \citenamefont {Nguyen}, \citenamefont
  {Parker}, \citenamefont {Perlt}, \citenamefont {Rappoport}, \citenamefont
  {Reiter}, \citenamefont {Roy}, \citenamefont {R{\"u}ckert}, \citenamefont
  {Schmitz}, \citenamefont {Sierka}, \citenamefont {Tapavicza}, \citenamefont
  {Tew}, \citenamefont {van W{\"u}llen}, \citenamefont {Voora}, \citenamefont
  {Weigend}, \citenamefont {Wody{\'n}ski},\ and\ \citenamefont
  {Yu}}]{Balasubramani_JCP_2020_184107}%
  \BibitemOpen
  \bibfield  {author} {\bibinfo {author} {\bibfnamefont {Sree~Ganesh}\
  \bibnamefont {Balasubramani}}, \bibinfo {author} {\bibfnamefont {Guo~P.}\
  \bibnamefont {Chen}}, \bibinfo {author} {\bibfnamefont {Sonia}\ \bibnamefont
  {Coriani}}, \bibinfo {author} {\bibfnamefont {Michael}\ \bibnamefont
  {Diedenhofen}}, \bibinfo {author} {\bibfnamefont {Marius~S.}\ \bibnamefont
  {Frank}}, \bibinfo {author} {\bibfnamefont {Yannick~J.}\ \bibnamefont
  {Franzke}}, \bibinfo {author} {\bibfnamefont {Filipp}\ \bibnamefont
  {Furche}}, \bibinfo {author} {\bibfnamefont {Robin}\ \bibnamefont
  {Grotjahn}}, \bibinfo {author} {\bibfnamefont {Michael~E.}\ \bibnamefont
  {Harding}}, \bibinfo {author} {\bibfnamefont {Christof}\ \bibnamefont
  {H{\"a}ttig}}, \bibinfo {author} {\bibfnamefont {Arnim}\ \bibnamefont
  {Hellweg}}, \bibinfo {author} {\bibfnamefont {Benjamin}\ \bibnamefont
  {Helmich-Paris}}, \bibinfo {author} {\bibfnamefont {Christof}\ \bibnamefont
  {Holzer}}, \bibinfo {author} {\bibfnamefont {Uwe}\ \bibnamefont {Huniar}},
  \bibinfo {author} {\bibfnamefont {Martin}\ \bibnamefont {Kaupp}}, \bibinfo
  {author} {\bibfnamefont {Alireza}\ \bibnamefont {Marefat~Khah}}, \bibinfo
  {author} {\bibfnamefont {Sarah}\ \bibnamefont {Karbalaei~Khani}}, \bibinfo
  {author} {\bibfnamefont {Thomas}\ \bibnamefont {M{\"u}ller}}, \bibinfo
  {author} {\bibfnamefont {Fabian}\ \bibnamefont {Mack}}, \bibinfo {author}
  {\bibfnamefont {Brian~D.}\ \bibnamefont {Nguyen}}, \bibinfo {author}
  {\bibfnamefont {Shane~M.}\ \bibnamefont {Parker}}, \bibinfo {author}
  {\bibfnamefont {Eva}\ \bibnamefont {Perlt}}, \bibinfo {author} {\bibfnamefont
  {Dmitrij}\ \bibnamefont {Rappoport}}, \bibinfo {author} {\bibfnamefont
  {Kevin}\ \bibnamefont {Reiter}}, \bibinfo {author} {\bibfnamefont {Saswata}\
  \bibnamefont {Roy}}, \bibinfo {author} {\bibfnamefont {Matthias}\
  \bibnamefont {R{\"u}ckert}}, \bibinfo {author} {\bibfnamefont {Gunnar}\
  \bibnamefont {Schmitz}}, \bibinfo {author} {\bibfnamefont {Marek}\
  \bibnamefont {Sierka}}, \bibinfo {author} {\bibfnamefont {Enrico}\
  \bibnamefont {Tapavicza}}, \bibinfo {author} {\bibfnamefont {David~P.}\
  \bibnamefont {Tew}}, \bibinfo {author} {\bibfnamefont {Christoph}\
  \bibnamefont {van W{\"u}llen}}, \bibinfo {author} {\bibfnamefont {Vamsee~K.}\
  \bibnamefont {Voora}}, \bibinfo {author} {\bibfnamefont {Florian}\
  \bibnamefont {Weigend}}, \bibinfo {author} {\bibfnamefont {Artur}\
  \bibnamefont {Wody{\'n}ski}}, \ and\ \bibinfo {author} {\bibfnamefont
  {Jason~M.}\ \bibnamefont {Yu}},\ }\bibfield  {title} {\enquote {\bibinfo
  {title} {{TURBOMOLE}: Modular program suite for ab initio quantum-chemical
  and condensed-matter simulations},}\ }\href {\doibase 10.1063/5.0004635}
  {\bibfield  {journal} {\bibinfo  {journal} {J. Chem. Phys.}\ }\textbf
  {\bibinfo {volume} {152}},\ \bibinfo {pages} {184107} (\bibinfo {year}
  {2020})}\BibitemShut {NoStop}%
\bibitem [{\citenamefont {Franzke}\ \emph {et~al.}(2023)\citenamefont
  {Franzke}, \citenamefont {Holzer}, \citenamefont {Andersen}, \citenamefont
  {Begu{\v s}i{\' c}}, \citenamefont {Bruder}, \citenamefont {Coriani},
  \citenamefont {Della~Sala}, \citenamefont {Fabiano}, \citenamefont {Fedotov},
  \citenamefont {F{\"u}rst}, \citenamefont {Gillhuber}, \citenamefont
  {Grotjahn}, \citenamefont {Kaupp}, \citenamefont {Kehry}, \citenamefont
  {Krsti{\'c}}, \citenamefont {Mack}, \citenamefont {Majumdar}, \citenamefont
  {Nguyen}, \citenamefont {Parker}, \citenamefont {Pauly}, \citenamefont
  {Pausch}, \citenamefont {Perlt}, \citenamefont {Phun}, \citenamefont
  {Rajabi}, \citenamefont {Rappoport}, \citenamefont {Samal}, \citenamefont
  {Schrader}, \citenamefont {Sharma}, \citenamefont {Tapavicza}, \citenamefont
  {Treß}, \citenamefont {Voora}, \citenamefont {Wody{\'n}ski}, \citenamefont
  {Yu}, \citenamefont {Zerulla}, \citenamefont {Furche}, \citenamefont
  {H{\"a}ttig}, \citenamefont {Sierka}, \citenamefont {Tew},\ and\
  \citenamefont {Weigend}}]{Franzke_JCTC_2023_6859}%
  \BibitemOpen
  \bibfield  {author} {\bibinfo {author} {\bibfnamefont {Yannick~J.}\
  \bibnamefont {Franzke}}, \bibinfo {author} {\bibfnamefont {Christof}\
  \bibnamefont {Holzer}}, \bibinfo {author} {\bibfnamefont {Josefine~H.}\
  \bibnamefont {Andersen}}, \bibinfo {author} {\bibfnamefont {Tomislav}\
  \bibnamefont {Begu{\v s}i{\' c}}}, \bibinfo {author} {\bibfnamefont
  {Florian}\ \bibnamefont {Bruder}}, \bibinfo {author} {\bibfnamefont {Sonia}\
  \bibnamefont {Coriani}}, \bibinfo {author} {\bibfnamefont {Fabio}\
  \bibnamefont {Della~Sala}}, \bibinfo {author} {\bibfnamefont {Eduardo}\
  \bibnamefont {Fabiano}}, \bibinfo {author} {\bibfnamefont {Daniil~A.}\
  \bibnamefont {Fedotov}}, \bibinfo {author} {\bibfnamefont {Susanne}\
  \bibnamefont {F{\"u}rst}}, \bibinfo {author} {\bibfnamefont {Sebastian}\
  \bibnamefont {Gillhuber}}, \bibinfo {author} {\bibfnamefont {Robin}\
  \bibnamefont {Grotjahn}}, \bibinfo {author} {\bibfnamefont {Martin}\
  \bibnamefont {Kaupp}}, \bibinfo {author} {\bibfnamefont {Max}\ \bibnamefont
  {Kehry}}, \bibinfo {author} {\bibfnamefont {Marjan}\ \bibnamefont
  {Krsti{\'c}}}, \bibinfo {author} {\bibfnamefont {Fabian}\ \bibnamefont
  {Mack}}, \bibinfo {author} {\bibfnamefont {Sourav}\ \bibnamefont {Majumdar}},
  \bibinfo {author} {\bibfnamefont {Brian~D.}\ \bibnamefont {Nguyen}}, \bibinfo
  {author} {\bibfnamefont {Shane~M.}\ \bibnamefont {Parker}}, \bibinfo {author}
  {\bibfnamefont {Fabian}\ \bibnamefont {Pauly}}, \bibinfo {author}
  {\bibfnamefont {Ansgar}\ \bibnamefont {Pausch}}, \bibinfo {author}
  {\bibfnamefont {Eva}\ \bibnamefont {Perlt}}, \bibinfo {author} {\bibfnamefont
  {Gabriel~S.}\ \bibnamefont {Phun}}, \bibinfo {author} {\bibfnamefont
  {Ahmadreza}\ \bibnamefont {Rajabi}}, \bibinfo {author} {\bibfnamefont
  {Dmitrij}\ \bibnamefont {Rappoport}}, \bibinfo {author} {\bibfnamefont
  {Bibek}\ \bibnamefont {Samal}}, \bibinfo {author} {\bibfnamefont {Tim}\
  \bibnamefont {Schrader}}, \bibinfo {author} {\bibfnamefont {Manas}\
  \bibnamefont {Sharma}}, \bibinfo {author} {\bibfnamefont {Enrico}\
  \bibnamefont {Tapavicza}}, \bibinfo {author} {\bibfnamefont {Robert~S.}\
  \bibnamefont {Treß}}, \bibinfo {author} {\bibfnamefont {Vamsee}\
  \bibnamefont {Voora}}, \bibinfo {author} {\bibfnamefont {Artur}\ \bibnamefont
  {Wody{\'n}ski}}, \bibinfo {author} {\bibfnamefont {Jason~M.}\ \bibnamefont
  {Yu}}, \bibinfo {author} {\bibfnamefont {Benedikt}\ \bibnamefont {Zerulla}},
  \bibinfo {author} {\bibfnamefont {Filipp}\ \bibnamefont {Furche}}, \bibinfo
  {author} {\bibfnamefont {Christof}\ \bibnamefont {H{\"a}ttig}}, \bibinfo
  {author} {\bibfnamefont {Marek}\ \bibnamefont {Sierka}}, \bibinfo {author}
  {\bibfnamefont {David~P.}\ \bibnamefont {Tew}}, \ and\ \bibinfo {author}
  {\bibfnamefont {Florian}\ \bibnamefont {Weigend}},\ }\bibfield  {title}
  {\enquote {\bibinfo {title} {Turbomole: Today and tomorrow},}\ }\href
  {\doibase 10.1021/acs.jctc.3c00347} {\bibfield  {journal} {\bibinfo
  {journal} {J. Chem. Theory Comput.}\ }\textbf {\bibinfo {volume} {19}},\
  \bibinfo {pages} {6859--6890} (\bibinfo {year} {2023})}\BibitemShut {NoStop}%
\bibitem [{\citenamefont {Meng}\ \emph {et~al.}(2010)\citenamefont {Meng},
  \citenamefont {Lun}, \citenamefont {Qi}, \citenamefont {Bi}, \citenamefont
  {Qi}, \citenamefont {Zhu}, \citenamefont {Han}, \citenamefont {Bai},
  \citenamefont {Yin},\ and\ \citenamefont {Fan}}]{Meng_EJIC_2010_3174}%
  \BibitemOpen
  \bibfield  {author} {\bibinfo {author} {\bibfnamefont {Xiang-Lin}\
  \bibnamefont {Meng}}, \bibinfo {author} {\bibfnamefont {Ning}\ \bibnamefont
  {Lun}}, \bibinfo {author} {\bibfnamefont {Yong-Qiu}\ \bibnamefont {Qi}},
  \bibinfo {author} {\bibfnamefont {Jian-Qiang}\ \bibnamefont {Bi}}, \bibinfo
  {author} {\bibfnamefont {Yong-Xin}\ \bibnamefont {Qi}}, \bibinfo {author}
  {\bibfnamefont {Hui-Ling}\ \bibnamefont {Zhu}}, \bibinfo {author}
  {\bibfnamefont {Fu-Dong}\ \bibnamefont {Han}}, \bibinfo {author}
  {\bibfnamefont {Yu-Jun}\ \bibnamefont {Bai}}, \bibinfo {author}
  {\bibfnamefont {Long-Wei}\ \bibnamefont {Yin}}, \ and\ \bibinfo {author}
  {\bibfnamefont {Run-Hua}\ \bibnamefont {Fan}},\ }\bibfield  {title} {\enquote
  {\bibinfo {title} {Low-temperature synthesis of meshy boron nitride with a
  large surface area},}\ }\href {\doibase 10.1002/ejic.201000260} {\bibfield
  {journal} {\bibinfo  {journal} {Eur. J. Inorg. Chem.}\ }\textbf {\bibinfo
  {volume} {2010}},\ \bibinfo {pages} {3174--3178} (\bibinfo {year}
  {2010})}\BibitemShut {NoStop}%
\bibitem [{\citenamefont {Cassabois}\ \emph {et~al.}(2016)\citenamefont
  {Cassabois}, \citenamefont {Valvin},\ and\ \citenamefont
  {Gil}}]{Cassabois_NP_2016_262}%
  \BibitemOpen
  \bibfield  {author} {\bibinfo {author} {\bibfnamefont {Guillaume}\
  \bibnamefont {Cassabois}}, \bibinfo {author} {\bibfnamefont {Pierre}\
  \bibnamefont {Valvin}}, \ and\ \bibinfo {author} {\bibfnamefont {Bernard}\
  \bibnamefont {Gil}},\ }\bibfield  {title} {\enquote {\bibinfo {title}
  {Hexagonal boron nitride is an indirect bandgap semiconductor},}\ }\href
  {\doibase 10.1038/nphoton.2015.277} {\bibfield  {journal} {\bibinfo
  {journal} {Nat. Photonics}\ }\textbf {\bibinfo {volume} {10}},\ \bibinfo
  {pages} {262--266} (\bibinfo {year} {2016})}\BibitemShut {NoStop}%
\bibitem [{\citenamefont {Jain}\ \emph {et~al.}(2013)\citenamefont {Jain},
  \citenamefont {Ong}, \citenamefont {Hautier}, \citenamefont {Chen},
  \citenamefont {Richards}, \citenamefont {Dacek}, \citenamefont {Cholia},
  \citenamefont {Gunter}, \citenamefont {Skinner}, \citenamefont {Ceder},\ and\
  \citenamefont {Persson}}]{Jain_APLM_2013_011002}%
  \BibitemOpen
  \bibfield  {author} {\bibinfo {author} {\bibfnamefont {Anubhav}\ \bibnamefont
  {Jain}}, \bibinfo {author} {\bibfnamefont {Shyue~Ping}\ \bibnamefont {Ong}},
  \bibinfo {author} {\bibfnamefont {Geoffroy}\ \bibnamefont {Hautier}},
  \bibinfo {author} {\bibfnamefont {Wei}\ \bibnamefont {Chen}}, \bibinfo
  {author} {\bibfnamefont {William~Davidson}\ \bibnamefont {Richards}},
  \bibinfo {author} {\bibfnamefont {Stephen}\ \bibnamefont {Dacek}}, \bibinfo
  {author} {\bibfnamefont {Shreyas}\ \bibnamefont {Cholia}}, \bibinfo {author}
  {\bibfnamefont {Dan}\ \bibnamefont {Gunter}}, \bibinfo {author}
  {\bibfnamefont {David}\ \bibnamefont {Skinner}}, \bibinfo {author}
  {\bibfnamefont {Gerbrand}\ \bibnamefont {Ceder}}, \ and\ \bibinfo {author}
  {\bibfnamefont {Kristin~A.}\ \bibnamefont {Persson}},\ }\bibfield  {title}
  {\enquote {\bibinfo {title} {The materials project: A materials genome
  approach to accelerating materials innovation},}\ }\href {\doibase
  10.1063/1.4812323} {\bibfield  {journal} {\bibinfo  {journal} {APL Mater.}\
  }\textbf {\bibinfo {volume} {1}},\ \bibinfo {pages} {011002} (\bibinfo {year}
  {2013})}\BibitemShut {NoStop}%
\bibitem [{\citenamefont {Pretzel}\ \emph {et~al.}(1960)\citenamefont
  {Pretzel}, \citenamefont {Rupert}, \citenamefont {Mader}, \citenamefont
  {Storms}, \citenamefont {Gritton},\ and\ \citenamefont
  {Rushing}}]{Pretzel_JPCS_1960_10}%
  \BibitemOpen
  \bibfield  {author} {\bibinfo {author} {\bibfnamefont {F.~E.}\ \bibnamefont
  {Pretzel}}, \bibinfo {author} {\bibfnamefont {G.~N.}\ \bibnamefont {Rupert}},
  \bibinfo {author} {\bibfnamefont {C.~L.}\ \bibnamefont {Mader}}, \bibinfo
  {author} {\bibfnamefont {E.~K.}\ \bibnamefont {Storms}}, \bibinfo {author}
  {\bibfnamefont {G.~V.}\ \bibnamefont {Gritton}}, \ and\ \bibinfo {author}
  {\bibfnamefont {C.~C.}\ \bibnamefont {Rushing}},\ }\bibfield  {title}
  {\enquote {\bibinfo {title} {Properties of lithium hydride {I}. single
  crystals},}\ }\href {\doibase 10.1016/0022-3697(60)90064-0} {\bibfield
  {journal} {\bibinfo  {journal} {Journal of Physics and Chemistry of Solids}\
  }\textbf {\bibinfo {volume} {16}},\ \bibinfo {pages} {10--20} (\bibinfo
  {year} {1960})}\BibitemShut {NoStop}%
\bibitem [{\citenamefont {Li}\ and\ \citenamefont {Zhu}(2015)}]{Li_JM_2015_33}%
  \BibitemOpen
  \bibfield  {author} {\bibinfo {author} {\bibfnamefont {Xiao}\ \bibnamefont
  {Li}}\ and\ \bibinfo {author} {\bibfnamefont {Hongwei}\ \bibnamefont {Zhu}},\
  }\bibfield  {title} {\enquote {\bibinfo {title} {Two-dimensional {MoS$_2$}:
  Properties, preparation, and applications},}\ }\href {\doibase
  10.1016/j.jmat.2015.03.003} {\bibfield  {journal} {\bibinfo  {journal} {J.
  Mater.}\ }\textbf {\bibinfo {volume} {1}},\ \bibinfo {pages} {33--44}
  (\bibinfo {year} {2015})}\BibitemShut {NoStop}%
\bibitem [{\citenamefont {Ganatra}\ and\ \citenamefont
  {Zhang}(2014)}]{Ganatra_ACSN_2014_4074}%
  \BibitemOpen
  \bibfield  {author} {\bibinfo {author} {\bibfnamefont {Rudren}\ \bibnamefont
  {Ganatra}}\ and\ \bibinfo {author} {\bibfnamefont {Qing}\ \bibnamefont
  {Zhang}},\ }\bibfield  {title} {\enquote {\bibinfo {title} {Few-layer
  {MoS$_2$}: A promising layered semiconductor},}\ }\href {\doibase
  10.1021/nn405938z} {\bibfield  {journal} {\bibinfo  {journal} {ACS Nano}\
  }\textbf {\bibinfo {volume} {8}},\ \bibinfo {pages} {4074--4099} (\bibinfo
  {year} {2014})}\BibitemShut {NoStop}%
\bibitem [{\citenamefont {{\c S}ahin}\ \emph {et~al.}(2009)\citenamefont {{\c
  S}ahin}, \citenamefont {Cahangirov}, \citenamefont {Topsakal}, \citenamefont
  {Bekaroglu}, \citenamefont {Akturk}, \citenamefont {Senger},\ and\
  \citenamefont {Ciraci}}]{Sahin_PRB_2009_155453}%
  \BibitemOpen
  \bibfield  {author} {\bibinfo {author} {\bibfnamefont {H.}~\bibnamefont {{\c
  S}ahin}}, \bibinfo {author} {\bibfnamefont {S.}~\bibnamefont {Cahangirov}},
  \bibinfo {author} {\bibfnamefont {M.}~\bibnamefont {Topsakal}}, \bibinfo
  {author} {\bibfnamefont {E.}~\bibnamefont {Bekaroglu}}, \bibinfo {author}
  {\bibfnamefont {E.}~\bibnamefont {Akturk}}, \bibinfo {author} {\bibfnamefont
  {R.~T.}\ \bibnamefont {Senger}}, \ and\ \bibinfo {author} {\bibfnamefont
  {S.}~\bibnamefont {Ciraci}},\ }\bibfield  {title} {\enquote {\bibinfo {title}
  {Monolayer honeycomb structures of group-{IV} elements and {III-V} binary
  compounds: First-principles calculations},}\ }\href {\doibase
  10.1103/PhysRevB.80.155453} {\bibfield  {journal} {\bibinfo  {journal} {Phys.
  Rev. B}\ }\textbf {\bibinfo {volume} {80}},\ \bibinfo {pages} {155453}
  (\bibinfo {year} {2009})}\BibitemShut {NoStop}%
\bibitem [{\citenamefont {Usvyat}\ \emph {et~al.}(2011)\citenamefont {Usvyat},
  \citenamefont {Civalleri}, \citenamefont {Maschio}, \citenamefont {Dovesi},
  \citenamefont {Pisani},\ and\ \citenamefont
  {Sch{\"u}tz}}]{Usvyat_JCP_2011_214105}%
  \BibitemOpen
  \bibfield  {author} {\bibinfo {author} {\bibfnamefont {Denis}\ \bibnamefont
  {Usvyat}}, \bibinfo {author} {\bibfnamefont {Bartolomeo}\ \bibnamefont
  {Civalleri}}, \bibinfo {author} {\bibfnamefont {Lorenzo}\ \bibnamefont
  {Maschio}}, \bibinfo {author} {\bibfnamefont {Roberto}\ \bibnamefont
  {Dovesi}}, \bibinfo {author} {\bibfnamefont {Cesare}\ \bibnamefont {Pisani}},
  \ and\ \bibinfo {author} {\bibfnamefont {Martin}\ \bibnamefont
  {Sch{\"u}tz}},\ }\bibfield  {title} {\enquote {\bibinfo {title} {Approaching
  the theoretical limit in periodic local {MP2} calculations with
  atomic-orbital basis sets: The case of lih},}\ }\href {\doibase
  10.1063/1.3595514} {\bibfield  {journal} {\bibinfo  {journal} {J. Chem.
  Phys.}\ }\textbf {\bibinfo {volume} {134}},\ \bibinfo {pages} {214105}
  (\bibinfo {year} {2011})}\BibitemShut {NoStop}%
\bibitem [{\citenamefont {Maschio}\ \emph {et~al.}(2010)\citenamefont
  {Maschio}, \citenamefont {Usvyat},\ and\ \citenamefont
  {Civalleri}}]{Maschio_CEC_2010_2429}%
  \BibitemOpen
  \bibfield  {author} {\bibinfo {author} {\bibfnamefont {Lorenzo}\ \bibnamefont
  {Maschio}}, \bibinfo {author} {\bibfnamefont {Denis}\ \bibnamefont {Usvyat}},
  \ and\ \bibinfo {author} {\bibfnamefont {Bartolomeo}\ \bibnamefont
  {Civalleri}},\ }\bibfield  {title} {\enquote {\bibinfo {title} {\textit{Ab
  initio} study of van der waals and hydrogen-bonded molecular crystals with a
  periodic local-mp2 method},}\ }\href {\doibase 10.1039/C002580A} {\bibfield
  {journal} {\bibinfo  {journal} {CrystEngComm}\ }\textbf {\bibinfo {volume}
  {12}},\ \bibinfo {pages} {2429--2435} (\bibinfo {year} {2010})}\BibitemShut
  {NoStop}%
\end{thebibliography}%


\end{document}